\documentclass[12pt,axodraw]{article}
\usepackage{axodraw}
\usepackage{color}
\usepackage{pstricks}

\usepackage{amsmath}

\input{epsf}
\newcommand\as{\alpha_{\mathrm{S}}}
\newcommand\g{g_{\mathrm{S}}}

\newcommand\gq{g}

\newcommand\f[2]{\frac{#1}{#2}}
\def\ep{\epsilon}

\def\beq{\begin{equation}}
\def\eeq{\end{equation}}
\def\beeq{\begin{eqnarray}}
\def\eeeq{\end{eqnarray}}
\def\cm{{\cal M}}
\def\bom#1{{\mbox{\boldmath $#1$}}}
\def\to{\rightarrow}

\newcommand{\la}{\langle}
\newcommand{\ra}{\rangle}

\def\nn{\nonumber}

\def\ID{1 \kern -.45 em 1}

\def\sp{{\bom {Sp}}}

\def\ket#1{|{#1}\ra}
\def\bra#1{\la{#1}|}

\def\ubar{{\overline u}}

\def\vep{{\varepsilon}}

\def\cbet0{b_0}

\def\bj{{\bom J}}
\def\bt{{\bom T}}
\def\btq{{\bom t}}
\def\qm{q_{1\dots m}}
\def\hbj{{\bom {\hat J}}}

\def\rs{\rm RS}

\def\cw{w}
\def\bd{{\bom D}}

\def\waij{w^{[A]}_{ij}}
\def\wabc{w^{[A]}_{BC}}
\def\waab{w^{[A]}_{AB}}
\def\waca{w^{[A]}_{CA}}

\def\w0abc{w_{\{ABC\}}}
\def\wsabc{w^{[S]}_{\{ABC\}}}
\def\waabc{w^{[A]}_{[ABC]}}

\newcommand\eqcs{\raisebox{-2mm}{\rlap{\,{\rm cs}\,}} \raisebox{.0ex}{$\,=\,$}}

\begin{document}

\begin{titlepage}
\renewcommand{\thefootnote}{\fnsymbol{footnote}}
\begin{flushright}
IFIC/21-31\\
FTUV-21-0901.5855
\end{flushright}
\par \vspace{10mm}

\begin{center}
{\Large \bf Multiple soft radiation at one-loop order
\\[1ex] 
and the emission of a soft quark--antiquark pair
}
\end{center}

\par \vspace{2mm}
\begin{center}
{\bf Stefano Catani}~$^{(a)}$ and {\bf Leandro Cieri}~$^{(a) (b)}$

\vspace{5mm}

${}^{(a)}$INFN, Sezione di Firenze and
Dipartimento di Fisica e Astronomia,\\ 
Universit\`a
di Firenze,
I-50019 Sesto Fiorentino, 
Florence, Italy \\
\vspace*{2mm}
${}^{(b)}$Instituto de F\'{\i}sica Corpuscular, Universitat de Val\`{e}ncia -- Consejo Superior de Investigaciones Cient\'{\i}ficas, Parc Cient\'{\i}fic, E-46980 Paterna, Valencia, Spain

\vspace{5mm}

\end{center}

\par \vspace{2mm}
\begin{center} {\large \bf Abstract} \end{center}
\begin{quote}
\pretolerance 10000

We consider  the radiation of two or more soft partons in QCD hard-scattering
at one-loop order.
The corresponding scattering amplitude is singular, and the singular behaviour
is controlled by a process-independent soft current. 
Using regularization in $d=4 - 2\ep$ space-time dimensions, we explicitly evaluate
the ultraviolet and infrared divergent ($\ep$-pole) terms of the one-loop soft current
for emission of an arbitrary number of soft partons in a generic hard-scattering
process.
Then we consider the specific case of soft quark--antiquark ($q{\bar q}$) emission
and we compute the one-loop current by including the finite terms.
We find that the one-loop soft-$q{\bar q}$ current exhibits a new type of 
transverse-momentum singularity, which has a quantum (absorptive) origin
and a purely non-abelian character. At the squared amplitude (cross section)
level, this transverse-momentum singularity produces contributions to
multijet production processes in hadron collisions.
The one-loop squared current also leads to charge asymmetry terms,
which are a distinctive features of soft-$q{\bar q}$ radiation.
We also extend these results to the cases of QED and mixed 
QCD$\times$QED radiative corrections 
for
soft fermion--antifermion emission.

\end{quote}

\vspace*{\fill}
\begin{flushleft}
     August 2021 

\end{flushleft}
\end{titlepage}

\renewcommand{\thefootnote}{\fnsymbol{footnote}}

\section{Introduction}
\label{sec:in}

The physics program carried out at the Large Hadron Collider (LHC) has already
produced an impressive amount of high-precision data, and similar data will
be obtained in the next runs of the LHC. Theoretical predictions are thus demanded to achieve
a corresponding high precision.

In the context of the perturbative evaluation of QCD radiative corrections, the present
high-precision frontier is represented by computations at the
next-to-next-to-next-to-leading order (N$^3$LO) in the QCD coupling $\as$.
Some N$^3$LO results for LHC processes are already available
(see, e.g., related references in Ref.~\cite{Heinrich:2020ybq}).
In the case of observables that are highly sensitive to multiple radiation of 
soft and collinear partons, the fixed-order QCD predictions have to be supplemented 
with the all-order resummed calculations of classes of large logarithmic contributions.
In few specific cases (see, e.g., 
Refs.~\cite{N3LLwithoutC3,Chen:2018pzu,Bertone:2019nxa,Bacchetta:2019sam,Ebert:2020dfc,Becher:2020ugp,Luo:2019szz, Ebert:2020unb, Billis:2021ecs,Camarda:2021ict, Re:2021con, Ju:2021lah, Neumann:2021zkb})
resummed QCD calculations have reached the
next-to-next-to-next-to-leading logarithmic (N$^3$LL) accuracy.

An important feature of QCD scattering amplitudes is the presence of singularities in soft
and collinear regions of the phase space, and the corresponding presence of infrared (IR)
divergences in virtual contributions at the loop level. The soft and collinear singularities
have a process-independent structure, and they are controlled by universal factorization formulae
and corresponding soft/collinear factors.  As briefly recalled below, these factorization properties
are relevant for both fixed-order and resummed QCD calculations.

In the computation of physical observables for hard-scattering processes, 
phase space soft/collinear singularities and virtual IR divergences cancel between themselves.
However, much technical effort is required to achieve and implement the cancellation,
and the effort highly increases by increasing the perturbative order.
Soft/collinear factorization formulae can be used to organize and greatly simplify 
the cancellation mechanism  of the IR divergences in fixed-oder calculations. 

In the evaluation of observables close to the exclusive boundary of the phase space,
real and virtual radiative corrections in the scattering amplitudes are kinematically
strongly unbalanced. As a consequence, the cancellation mechanism of
the IR divergences leaves residual effects in the form of large logarithmic
contributions. Soft/collinear factorization formulae and the corresponding singular factors
are the basic ingredients for the explicit computation and resummation of these
large logarithmic contributions.

The singular factors at ${\cal O}(\as)$ and ${\cal O}(\as^2)$  for soft and collinear 
factorization of scattering amplitudes are known since long time.
The explicit knowledge of soft/collinear factorization at ${\cal O}(\as)$ has been
essential to devise fully general (process-independent and observable-independent) methods
to carry out next-to-leading order (NLO) QCD calculations (see, e.g., 
Refs.~\cite{Frixione:1995ms, csdip}). Similarly, the knowledge of soft/collinear
factorization formulae at ${\cal O}(\as^2)$
\cite{Campbell:1997hg, Catani:1998nv, Bern:1998sc, Kosower:1999rx, Bern:1999ry,
Catani:1999ss, Catani:2000pi, Czakon:2011ve, Bierenbaum:2011gg, Catani:2011st, 
Sborlini:2013jba}
is exploited to develop methods 
(see, e.g., the review in Ref.~\cite{Heinrich:2020ybq})
at the next-to-next-to-leading order (NNLO).
Soft/collinear factorization up to ${\cal O}(\as^2)$ contributes to resummed calculations up
to next-to-next-to-leading logarithmic (NNLL) accuracy
(see, e.g., Refs.~\cite{Becher:2014oda, Luisoni:2015xha}).

Soft and collinear factorization at ${\cal O}(\as^3)$ can be used in the context of N$^3$LO
calculations and of resummed calculations at N$^3$LL accuracy.
The process-independent singular factors for the various collinear limits at  ${\cal O}(\as^3)$ 
are presented in 
Refs.~\cite{DelDuca:1999iql, Birthwright:2005ak, DelDuca:2019ggv, Catani:2003vu, 
Sborlini:2014mpa, Badger:2015cxa, Bern:2004cz, Badger:2004uk, Duhr:2014nda, 
Catani:2011st}. Soft factorization of scattering amplitudes at ${\cal O}(\as^3)$
requires the study of various tree-level and loop contributions.
Triple soft-gluon radiation at the tree level is studied in Ref.~\cite{Catani:2019nqv}.
Double soft emission at one loop level has been considered recently in Ref.~\cite{Zhu:2020ftr}.
Single soft-gluon radiation at two loop order is examined in detail in 
Refs.~\cite{Badger:2004uk, Li:2013lsa, Duhr:2013msa, Dixon:2019lnw}.

This paper is devoted to a study of soft-parton emission at ${\cal O}(\as^3)$ and
beyond this order. More precisely, we consider the singular behaviour of 
scattering amplitudes in the limit in which {\em two} or {\em more} external partons are soft.
The singularity is controlled in factorized form by a current for soft multiparton radiation 
from hard partons. At one-loop order the soft current contains IR and ultraviolet (UV)
divergent contributions that we explicitly evaluate for the emission of an arbitrary number
of soft partons. In the particular case of emission of a soft quark-antiquark ($q{\bar q})$ pair,
we explicitly compute also the finite contributions to the one-loop current. We comment on the 
related results of Ref.~\cite{Zhu:2020ftr} in the paper.

The outline of the paper is as follows. In Sect.~\ref{sec:softfact} we introduce our notation, and we recall
the soft factorization formula for scattering amplitudes and the known results on the tree-level
currents for emission of a single soft gluon and of soft-$q{\bar q}$ pair. We use analytic continuation
in $d=4 - 2\ep$ space-time dimensions to regularize IR and UV divergences in loop contributions.
 In Sect.~\ref{sec:1loop} we discuss general features of the current for multiple soft radiation at the
loop level. In particular, we present in explicit form the result of the IR and UV divergent 
($\ep$-pole) terms of the one-loop soft current. 
In Sect.~\ref{sec:qqcur} we consider the emission of a soft-$q{\bar q}$ pair and 
we compute the corresponding one-loop current by including the finite
(i.e., ${\cal O}(\ep^0)$) terms. We comment on general features of our result that is valid 
for generic multiparton scattering processes in arbitrary kinematical configurations. 
Section \ref{sec:square} is devoted to consider soft-$q{\bar q}$ radiation at the 
squared amplitude level. We first recall the results for the squared current at the tree level,
and then we explicitly compute the one-loop squared current. We discuss the structure of the
charge asymmetry contributions, which are a distinctive feature of soft-$q{\bar q}$ radiation at 
the loop level.
 In Sect.~\ref{sec:23hard} we present  simplified expressions for processes with two and three hard partons.
In Sect.~\ref{sec:qed} we generalize our QCD results for soft $q{\bar q}$ emission
to the cases of QED and mixed QCD$\times$QED radiative corrections for soft fermion-antifermion
emission. A brief summary of our results is presented in Sect.~\ref{sec:sum}.

\section{Soft factorization}
\label{sec:softfact}

We consider the amplitude (the $S$-matrix element) $\cm$ of a generic 
scattering process whose external particles (the external legs of $\cm$)
are QCD partons (quarks, antiquarks and gluons) and, possibly, additional 
non-QCD particles (i.e., partons with no colour charge
such as leptons,  Higgs and electroweak vector bosons and so forth).
We use the notation $\cm(p_1,p_2,\dots,p_n)$, where $p_i$ $(i=1,\dots,n)$
is the momentum of the QCD parton $A_i$
($A_i= g, q$ or ${\bar q}$ ). Unless otherwise specified,  
the dependence of $\cm$ on the momenta (and quantum numbers)  
of additional colourless particles is not explicitly denoted.

The external QCD
partons are
on-shell with physical spin polarizations (thus, $\cm$
includes the corresponding spin wave functions), and we always
define the external momenta $p_i$'s as {\em outgoing} momenta.
Note, however, that we do no restrict our treatment to processes with physical
partons in the final state. In particular, 
the time-component (i.e. the `energy') $p_i^0$ of the momentum vector
$p_i^\nu$ ($\nu=0, 1, \dots, d-1$) in $d$ space-time dimensions is {\em not}
positive definite. Different types of physical processes are described
by considering different kinematical regions of the parton momenta and
by simply applying crossing symmetry to the wave functions and quantum numbers
of the external partons of 
the same matrix element $\cm(p_1,p_2,\dots,p_n)$. According to our definition 
of the momenta, if $p_i$ has positive energy, $\cm(\dots, p_i,\dots)$ 
describes a physical
process that produces the {\em parton} $A_i$ in the final state; if $p_i$ has 
negative energy, 
$\cm(\dots, p_i,\dots)$ describes a physical
process produced by the collision of the {\em antiparton} 
${\overline A}_i$
in the initial state.

The scattering amplitude $\cm$ also depends on the colour indices 
$\{c_1,c_2,\dots\}$ and on the 
spin (e.g. helicity) indices $\{s_1,s_2,\dots\}$
of the external QCD partons, and we write
\beq
\label{mel}
\cm^{c_1,c_2,\dots,c_n}_{s_1,s_2,\dots,s_n}(p_1,p_2,\dots,p_n) \;\;.
\eeq
It is convenient to directly work in colour
(and spin) space, and to use the notation of Ref.~\cite{csdip}
(see also Ref.~\cite{Catani:1998bh}). We treat the colour and spin structures 
by formally introducing an orthonormal basis
$\{ \ket{c_1,c_2,\dots,c_n} \otimes \ket{s_1,s_2,\dots,s_n} \}$
in colour + spin space.
The scattering amplitude 
can 
be written as 
\beeq
\label{cmvdef}
\cm^{c_1,c_2,\dots}_{s_1,s_2,\dots}(p_1,p_2,\dots)
\equiv
\Bigl( \bra{c_1,c_2,\dots} \otimes \bra{s_1,s_2,\dots} \Bigr) \;
\ket{\cm(p_1,p_2,\dots)} \;\;.
\eeeq
Thus $\ket{\cm(p_1,p_2,\dots,p_n)}$
is a vector in colour + spin (helicity) space.

As previously stated, we 
define the external momenta $p_i$'s as {\em outgoing} momenta.
The colour indices $\{c_1,c_2,\dots\,c_n\}$ are consistently defined
as {\em outgoing} colour indices:
$c_i$ is the colour index of the parton $A_i$ with outgoing momentum  
$p_i$
(if $p_i$ has negative energy, $c_i$ is the colour index of the physical 
parton ${\overline A}_i$ that collides in the initial state).
An analogous comment applies to spin indices.

The amplitude $\cm$ can be evaluated in QCD perturbation
theory as a power series expansion 
(i.e., loop expansion) in the QCD coupling $\g$
(or, equivalently, in the strong coupling 
$\as=\g^2/(4\pi)$).
We write
\beq
\label{loopexnog}
\cm = \cm^{(0)} 
+ \cm^{(1)} + \cm^{(2)}  + \dots \;\;,
\eeq
where $\cm^{(0)}$ is the tree-level
scattering amplitude, $\cm^{(1)}$ is the one-loop contribution,
$\cm^{(2)}$ is the two-loop contribution, and so forth.
More generally, $\cm^{(0)}$ is not necessarily a tree amplitude, 
but rather the
lowest-order amplitude for that given process. Thus, $\cm^{(L)}$
($L=1,2\dots)$
is the corresponding $L$-loop correction. For instance, in the
cases of the diphoton production process $gg \to \gamma \gamma$
or the Higgs boson ($H$) production process $gg \to H$, the
corresponding amplitude $\cm^{(0)}$ involves a quark loop
(a massive-quark loop in the case of $gg \to H$). 
Note that in Eq.~(\ref{loopexnog}) we have not made explicit the dependence on
powers of $\g$. Thus, $\cm^{(0)}$ includes an integer power of $\g$ 
as overall factor,
and $\cm^{(1)}$ includes an extra factor of $\g^2$
(i.e., $\cm^{(1)}/\cm^{(0)} \propto \g^2$). 

Physical processes take place in four-dimensional space-time.
The four-dimensional evaluation of the $L$-loop amplitude $\cm^{(L)}$ 
leads to UV
and IR
divergences
that have to be properly regularized. We regularize both 
kind of divergences by performing the analytic continuation of the loop momenta
and phase-space in $d=4 - 2\ep$ space-time dimensions.
We postpone comments on different variants
of dimensional regularization.
The dimensional-regularization scale is denoted by $\mu$. After regularization,
the UV and IR divergences appears as $\ep$-poles of the Laurent series expansion
in powers of $\ep$ around $\ep=0$.
Throughout the paper we formally consider expressions for arbitrary values of
$d=4 - 2\ep$
(equivalently, in terms of $\ep$ expansions, the expressions are valid to
all orders in~$\ep$ before they are eventually truncated at some order in~$\ep$).  
We always consider unrenormalized amplitudes,
and $\g$ denotes the bare (unrenormalized) coupling constant. 

We are interested in studying the behaviour of $\cm$
in the kinematical configuration where 
{\em one} or {\em more} of the  momenta of the external {\em massless} partons 
(gluons or massless quark and antiquarks) become soft.
In this kinematical configuration, $\cm$ becomes singular. To make the notation
more explicit, the soft momenta are denoted by $q_k^\nu$, while the other parton
momenta are still denoted by $p_i^\nu$. The behaviour of 
$\cm(\dots,q_k,\dots,p_i,\dots)$ in this {\em multiparton} soft region is formally
specified by performing an overall rescaling of all soft momenta as 
$q_k \to \lambda q_k$ (the rescaling parameter $\lambda$ is the same for each soft
momentum $q_k$) and by considering the limit $\lambda \to 0$. In this limit,
if the set of soft partons has $m$ ($m \geq 1$) momenta $q_k$'s $(k=1,\dots,m)$,
the amplitude $\cm$ is singular and it behaves as
\beq
\label{msoft}
\cm(\lambda q_1,\dots,\lambda q_m,p_1\dots,p_n) \sim 
\f{1}{\left( \lambda \right)^m} \;{\rm mod}\,(\ln^r \lambda) + \dots \;\;,
\quad (\lambda \to 0) \;.
\eeq

The power-like behaviour $\left( \lambda \right)^{-m}$ that we have specified 
in the right-hand side of Eq.~(\ref{msoft}) determines the {\em dominant} 
singular
terms of $\cm$ in the multiple soft region. The logarithmic corrections
$\ln^r \lambda$ $(r=0,1,2,\dots)$ 
arise from scaling violation, since 
the na\"ive (power-like) scaling behaviour is violated by the effects of the 
UV and IR divergences of the
scattering amplitude at the loop level (see Sect.~\ref{sec:1loop}).
The dots
on the right-hand side of Eq.~(\ref{msoft}) denote the {\em subdominant}
singular behaviour of $\cm$. 
The relative suppression factor between subdominant and dominant terms
is (at least) of ${\cal O}({\sqrt \lambda})$.

The computation of physical observables eventually 
requires the
phase-space integration of the {\em squared} amplitude $|\cm|^2$.
We note that, after phase-space integration over the soft momenta, the dominant
singular behaviour of $|\cm|^2$ produces logarithmic soft (IR) divergences (i.e.,
$\ep$ poles), whereas the subdominant singular behaviour does not lead to soft
divergences. In this paper we are interested in the dominant
singular behaviour of Eq.~(\ref{msoft}).

In the soft multiparton limit,
the dominant singular behaviour of $\cm$ 
can be expressed by the following
process-independent (universal) factorization formula
\cite{Bern:1995ix,Bern:1999ry,Catani:1999ss,Catani:2000pi,Feige:2014wja}
\beq
\label{softfact}
\ket{\cm(q_1, \dots,q_m,p_1,\dots,p_n)} = \bj(q_1, \dots,q_m) \;
\ket{\cm(p_1,\dots,p_n)} + \dots
 \;\;,
\eeq
where, analogously to Eq.~(\ref{msoft}), the dots on the right-hand side denote
subdominant singular terms. The amplitude $\cm(p_1,\dots,p_n)$ on the right-hand 
side of Eq.~(\ref{softfact}) is simply obtained by removing the $m$ external legs
with soft parton momenta $q_1, \dots,q_m$ from the amplitude on the left-hand
side. The factor $\bj(q_1, \dots,q_m)$ is the {\em soft multiparton current}
that embodies the dominant singular behaviour denoted in the right-hand side of 
Eq.~(\ref{msoft}).
 
In the case of tree-level scattering amplitudes 
\cite{Bassetto:1984ik,Berends:1988zn,Catani:1999ss}, the factorization formula 
(\ref{softfact}) can
be simply derived by considering soft-parton radiation from the hard-parton
(the partons with momenta $p_1,\dots,p_n$) external legs of the amplitude
and by directly applying
the eikonal approximation for emission vertices and propagators. At the one-loop
level,
the factorization structure of Eq.~(\ref{softfact}) was worked out in
Refs.~\cite{Bern:1995ix,Bern:1999ry,Catani:2000pi}. In particular, as discussed in
detail in Ref.~\cite{Catani:2000pi}, the one-loop soft current can still be
computed by using the eikonal approximation for soft-parton radiation from the
external hard partons, and this discussion generalizes to two-loop
and higher-loop orders.
Owing to its origin by eikonal radiation from the hard partons,
the soft current can also straightforwardly be expressed in equivalent form as
matrix element of Wilson line operators \cite{Feige:2014wja}.
 
The soft current $\bj(q_1, \dots,q_m)$ depends on the soft partons, specifically
on their momenta and their quantum numbers (flavour, spin, colour), and it also 
depends on the hard partons 
(on their momenta and their quantum numbers), though we use a customary notation
in which the dependence of $\bj$ on $p_1,\dots,p_n$ is not explicitly denoted in
its argument. The current $\bj$ is an operator (a `rectangular' matrix) that acts
from the (lower-dimensional) colour+spin space of the hard partons to the
(higher-dimensional) colour+spin space of the soft and hard partons. 
We remark on the fact that the soft current
$\bj$ is simply proportional to the unit operator in the spin subspace
of the hard partons, since soft radiation is insensitive to the spin of the
hard radiating partons.

In spite of its dependence on hard partons,
the soft current $\bj$ is completely {\em universal}, namely, it does not
depend on the specific scattering amplitude $\cm$ and on its corresponding
specific physical process.
The universality of $\bj$ also implies that it is directly applicable in contexts
that do not directly refer to the soft behaviour of scattering amplitudes.
For instance, $\bj$ (or, more specifically, $\bj^\dagger \bj$)
is precisely the {\em integrand} 
of any specific soft function (see, e.g.,  Ref.~\cite{Becher:2014oda} and references
therein) that can be introduced 
through soft-collinear effective theory (SCET) \cite{scetref} methods.

The colour-space factorization formula (\ref{softfact}) does not require any
specifications about the detailed colour structure of the scattering amplitudes in
its left-hand and right-hand sides. Scattering amplitudes can be decomposed in a
form that factorizes the QCD colour from colourless kinematical coefficients,
which are colour-ordered subamplitudes (see, e.g., Ref.~\cite{Mangano:1990by}).
Colour-ordered subamplitudes fulfil soft factorization formulae that are
analogous to Eq.~(\ref{softfact}) in terms of colour-stripped 
(though colour-ordered) soft factors (see, e.g., 
Refs.~\cite{Berends:1988zn,Bern:1999ry}). The factorization properties of
colour-order subamplitudes and the corresponding soft factors can be directly
and explicitly 
derived from Eq.~(\ref{softfact}). To this purpose
it is sufficient to insert
the colour decomposition of $\cm$ and the explicit colour structure of $\bj$
in Eq.~(\ref{softfact}). Therefore, the colour-space factorization of 
Eq.~(\ref{softfact}) and soft factorization of colour-ordered subamplitudes are
equivalent formulations. The advantage of Eq.~(\ref{softfact}) is that it leads to
a more compact formulation, without the necessity of introducing the explicit
colour decomposition of $\cm$, whose actual form depends on the specific 
partonic content of the amplitude (e.g., on the number of gluons and
quark-antiquark pairs\footnote{In particular,
if the set of soft partons includes one or more quark-antiquark pairs,
the scattering amplitudes in the left-hand and right-hand sides
of Eq.~(\ref{softfact}) have different numbers of quark-antiquark pairs and,
therefore, they have different colour decompositions.}) and on the loop order.
Moreover, the colour space formulation can simplify the direct computation of
the soft limit of {\em squared} amplitudes (see, e.g., Sect.~\ref{sec:square} ).

The soft current $\bj$ in Eq.~(\ref{softfact}) can be evaluated in QCD
perturbation theory, and it can be expressed in terms of a loop expansion that is
completely analogous to that in Eq.~(\ref{loopexnog}). We write
\beq
\label{jexp}
\bj = \bj^{(0)} 
+ \bj^{(1)} + \bj^{(2)}  + \dots \;\;,
\eeq
where $\bj^{(0)}$ is the tree-level current, $\bj^{(1)}$ is the one-loop current,
and so forth. Analogously to Eq.~(\ref{loopexnog}), the loop label $L$ in
$\bj^{(L)}$ refers to the unrenormalized current. Inserting the expansions 
(\ref{loopexnog}) and (\ref{jexp}) in Eq.~(\ref{softfact}) we obtain
factorization formulae that are valid order-by-order in the number of loops.
The soft factorization formula for tree-level (lowest-order) amplitudes is
\beq
\label{treef}
\ket{\cm^{(0)}(q_1, \dots,q_m,p_1,\dots,p_n)} \simeq \bj^{(0)}(q_1, \dots,q_m) \;
\ket{\cm^{(0)}(p_1,\dots,p_n)} 
 \;\;,
\eeq
where the symbol `$\simeq$' means\footnote{The symbol `$\simeq$' is used
throughout the paper with the same meaning as in Eq.~(\ref{treef}).} 
that we are neglecting subdominant
terms in the soft limit (i.e., the terms denoted by dots in the right-hand side of
Eqs.~(\ref{msoft}) and (\ref{softfact})).
The soft factorization formula for one-loop amplitudes is
\beeq
\label{onef}
\ket{\cm^{(1)}(q_1, \dots,q_m,p_1,\dots,p_n)} &\simeq& 
\bj^{(1)}(q_1, \dots,q_m) \; \ket{\cm^{(0)}(p_1,\dots,p_n)} \\
&+& 
\bj^{(0)}(q_1, \dots,q_m) \; \ket{\cm^{(1)}(p_1,\dots,p_n)} 
 \;\;. \nn
\eeeq

The tree-level current for the emission of a {\em single} 
soft gluon of momentum $q^\nu$
is well known \cite{Bassetto:1984ik}:
\beq
\label{treeg}
\bj^{(0)}(q) = \g \,\mu^\ep \;\sum_{i \in H} \;\bt_i \;
\frac{p_i \cdot \varepsilon(q)}{p_i \cdot q} 
\equiv \bj^{(0)}_\nu(q) \varepsilon^\nu(q)\;\;,
\eeq
where the notation $i \in H$ means that the sum extends over all hard partons
(with momenta $p_i$) in $\cm$, $\varepsilon^\nu(q)$ is the spin polarization 
vector of
the soft gluon, and $\bt_i$ is the colour charge of the hard parton $i$.

The spin index $s$ and the colour index $a$ ($a=1,\dots,N_c^2-1$, for 
$SU(N_c)$ QCD
with $N_c$ colours) of the soft gluon can be specified by acting onto 
Eq.~(\ref{treeg}) in colour+spin space as in Eq.~(\ref{cmvdef}). Considering
$(\bra{a}\otimes \bra{s}\,)\, \bj^{(0)}(q) \equiv \bj^{(0)\, a}_s(q)$,
we have 
$(\bra{a}\otimes \bra{s}\,)\,\varepsilon^\nu(q)\,\bt_i = 
\varepsilon^\nu_{(s)}(q) \;T^a_i$, where $T^a_i$ denotes the generators of 
$SU(N_c)$ of the representation of the parton $i$. We have 
$\bra{c_i}\,T^a_i\,\ket{c_i^\prime} = (T^a_i)_{c_i\,c_i^\prime}$, where
$(T^a)_{cb} \equiv if_{cab}$ (colour-charge matrix in the adjoint representation)
if the parton $i$ is a gluon and $(T^a)_{\alpha \beta} \equiv t^a_{\alpha \beta}$ 
(colour-charge matrix in the fundamental representation, with $\alpha,\beta =
1,\dots,N_c$) if the parton $i$ is a quark ($(T^a)_{\alpha \beta} \equiv 
{\bar t}^a_{\alpha \beta} = - t^a_{\beta \alpha}$ if the parton $i$ is 
an antiquark).
We normalize the colour matrices such as $[t^a,t^b]= if_{abc}$ and 
${\rm Tr} (t^a t^b) = T_R \,\delta^{ab}$ with $T_R=1/2$. 
The colour-charge algebra gives $[T^a_i,T_j^b]= if_{abc} T^a_i \delta_{ij}$
and $\sum_a T_i^a T_j^a \equiv \bt_i \cdot \bt_j$ with $\bt_i^2 = C_i$, where
$C_i$ is the Casimir operator of the representation of the parton $i$,
i.e. $C_i=C_A=N_c$ if $i$ is a gluon and
$C_i=C_F=(N_c^2-1)/(2N_c)$ if $i$ is a quark or antiquark.

We note that the colour charge operators $\bt_i$ fulfil some relevant properties
related to colour conservation. For instance, we have \cite{csdip}
\beq
\label{colcon}
\sum_{i \in \cm} \bt_i \; \ket{\cm} = 0 \;\;,
\eeq
which follows from the fact that the scattering amplitude $\cm$ is a
colour-singlet state (the notation $i \in \cm$ means that the sum in 
Eq.~(\ref{colcon}) extends over all external partons of $\cm$).
In particular, Eq.~(\ref{colcon}) implies that the tree-level soft current
in Eq.~(\ref{treeg}) leads to a gauge invariant soft factor
since $q^\nu \bj^{(0)}_\nu(q) \,\ket{\cm(p_1,\dots,p_n)} \propto 
\sum_{i \in H} \bt_i \;\ket{\cm(p_1,\dots,p_n)} =0$.

A property analogous to that in Eq.~(\ref{colcon}) is fulfilled by the soft 
multiparton current $\bj$. We have
\beq
\label{flowcon}
\left(\sum_{k \in S} \bt_k + \sum_{i \in H} \bt_i \right) \bj(q_1, \dots,q_m)
= \bj(q_1, \dots,q_m) \;\sum_{i \in H} \bt_i \;\;,
\eeq
or, equivalently,
\beq
\label{flowcon2}
\sum_{k \in S} \bt_k  \;\,\bj(q_1, \dots,q_m) =
\left[ \bj(q_1, \dots,q_m) \;,\;\sum_{i \in H} \bt_i \right]\;\;,
\eeq
where the notation $k \in S$ ($i \in H$) means that the sum extends over all the
soft (hard) partons in $\bj$. The relations in Eqs.~(\ref{flowcon}) and 
(\ref{flowcon2}) express the property of colour flow conservation and follow
from the fact that the total colour charge is conserved
in the radiation process of soft partons by hard partons.
Note from Eq.~(\ref{flowcon2}) that the total charge of the soft partons
acts on $\bj$ as a rotation of its hard-parton charges.
We remark that Eqs.~(\ref{colcon})--(\ref{flowcon2}) are valid to all orders in the
loop expansion or, equivalently, order-by-order in QCD perturbation theory.
It is straightforward to explicitly check that the soft single-gluon current in
Eq.~(\ref{treeg}) fulfils the colour flow conservation property 
in Eq.~(\ref{flowcon}).

Since scattering amplitudes are colour-singlet states, the structure of soft
factorization in Eq.~(\ref{softfact})
implies that the explicit expression of the universal soft current $\bj$
necessarily involves some degrees of arbitrariness. Different expressions for
$\bj$ are indeed permitted, provided the difference is proportional to an
operator that is proportional to the total colour charge of the hard partons.
Owing to Eq.~(\ref{colcon}), this degree of arbitrariness is physically harmless
(it does not affect the soft behaviour of the scattering amplitude) and the
ensuing different expressions of $\bj$ are fully equivalent (although they are
not exactly equal at the formal level, before acting onto colour-singlet states).
The property of colour-flow conservation in Eq.~(\ref{flowcon}) does not remove
this degree of arbitrariness.

For subsequent use  
 (and similarly to Ref.~\cite{Catani:2019nqv})
we introduce the notation 
\beq
\label{csequal}
{\bf O} \;\eqcs \;{\bf O}^\prime \;\;,
\eeq
where the subscript CS in the symbol $\eqcs$ means that the equality between the 
colour operators $\bf O$ and ${\bf O}^\prime$ (e.g., soft currents or their corresponding
squared currents)
is valid if these operators act (either on the left or on the right) onto colour-singlet states.
The notation in Eq.~(\ref{csequal}) permits to directly relate (and equate)
expressions that simply differ by contributions that are due to the physically harmless
arbitrariness of the soft current $\bj$.

The soft factor for radiation of {\em two} soft gluons from tree-level
colour-ordered subamplitudes with external gluons and with external gluons
and an additional quark-antiquark pair was computed in Ref.~\cite{Berends:1988zn}.
The tree-level current $\bj^{(0)}(q_1,q_2)$ for emission of 
{\em two} soft gluons in a generic scattering amplitude was given in 
Ref.~\cite{Catani:1999ss}. 
The tree-level current for emission of {\em three} soft gluons 
was computed in Ref.~\cite{Catani:2019nqv}.

The tree-level current for emission of a {\em single} soft quark (or antiquark)
vanishes. This result is equivalent to say that the dominant singular behaviour 
in Eq.~(\ref{msoft}) is absent 
in the soft single-quark limit
(the radiation of a single soft quark only produces
a subdominant behaviour of ${\cal O}(1/{\sqrt \lambda})$ in the right-hand
side of Eq.~(\ref{msoft})).

The tree-level current for emission of a soft $q{\bar q}$  pair
was computed in Ref.~\cite{Catani:1999ss}, where the
result was explicitly reported at the level of squared amplitudes
(i.e., the result refers to $\bj^\dagger \bj$). The corresponding result for the
$q\bar q$ current is
\beq
\label{treeqq}
\bj^{(0)}(q_1,q_2) = - \left( \g \mu^\ep\right)^2 \,
\sum_{i \in H} \,\btq^c \;T^c_i
\;\frac{p_i \cdot j(1,2)}{p_i \cdot q_{12}} \;\;,
\eeq
where we have introduced the fermionic current $j^\nu(1,2)$,
\beq
\label{fercur}
j^\nu(1,2) \equiv \frac{\ubar(q_1)\, \gamma^\nu \,v(q_2)}{q_{12}^2} \;\;,
\quad \quad \;\;\; \quad q_{12} = q_1+q_2 \;\;.
\eeq
The soft quark and antiquark have momenta $q_1^\nu$ and $q_2^\nu$, respectively,
and $u(q)$ and $v(q)$ are the customary Dirac spinors. The spin indices ($s_1$ and
$s_2$) and the colour indices ($\alpha_1$ and $\alpha_2$) of the 
quark and antiquark are embodied in the colour+spin space notation of 
Eq.~(\ref{treeqq}). Analogously to Eq.~(\ref{treeg}), we can consider
$(\bra{\alpha_1,\alpha_2} \otimes \bra{s_1,s_2} \,) \,\bj(q_1,q_2)
\equiv \bj^{\alpha_1,\alpha_2}_{s_1,s_2}(q_1,q_2)$ and we have
$(\bra{\alpha_1,\alpha_2} \otimes \bra{s_1,s_2} \,) \,\btq^c \;
\ubar(q_1)\, \gamma^\nu \,v(q_2) = t^c_{\alpha_1\alpha_2} \,
\ubar_{(s_1)}(q_1)\, \gamma^\nu \,v_{(s_2)}(q_2)$.

\section{One-loop current for multiple soft emission: UV and IR divergences}
\label{sec:1loop}

The soft singular behaviour of one-loop amplitudes (see Eq.~(\ref{onef}))
is controlled by $\bj^{(0)}$
and by an additional new ingredient, the one-loop soft current $\bj^{(1)}$.

The one-loop soft limit for emission of a {\em single} soft gluon was worked out
independently by two groups \cite{Bern:1999ry,Catani:2000pi}, finding results
that are in agreement. The analysis of Ref.~\cite{Bern:1999ry} is based on the
study of colour-ordered subamplitudes, while Ref.~\cite{Catani:2000pi}
considers generic scattering amplitudes. The results of 
Refs.~\cite{Bern:1999ry,Catani:2000pi} are valid for the case of massless hard
partons. The generalization of the results of Ref.~\cite{Catani:2000pi}
to include massive hard partons (such as heavy quarks) was carried out in
Ref.~\cite{Bierenbaum:2011gg}.
In the remaining part of this paper we limit ourselves to consider scattering
amplitudes with {\em massless} hard partons.

The result of the one-loop current for {\em single} gluon emission
is \cite{Catani:2000pi} (we explicitly write $\bj^{(1)\,a} \equiv
\bra{a} \,\bj^{(1)}$, where $a$ is the colour index of the soft gluon):
\beeq
\bj^{(1)\,a} &=& - \left( \g \,\mu^\ep \right)^3 \,c_{\Gamma}
\,\frac{1}{\ep^2} \,\Gamma(1-\epsilon) \Gamma(1+\epsilon) \;\,i f_{abc}\nn \\
&\times& 
\label{oneg}
\sum_{\substack{i,j \,\in H \\ i \,\neq\, j}}
T^b_i \,T^c_j 
\left( \frac{p_i^\nu}{p_i \cdot q} - \frac{p_j^\nu}{p_j \cdot q} \right)
\vep_\nu(q) \;
\frac{\left(-2p_i \cdot q -i0\right)^{-\ep} 
\left(-2p_j \cdot q -i0\right)^{-\ep}}{\left(-2p_i \cdot p_j -i0\right)^{-\ep}}
\;\;,
\eeeq
where `$x-i0$' denotes the customary Feynman prescription for analytic
continuation in different kinematical regions ($x>0$ and $x<0$) and 
$c_{\Gamma}$  is the typical volume factor of 
$d$-dimensional one-loop integrals:
\beq
\label{cgammaf}
c_{\Gamma} \equiv \frac{\Gamma(1+\epsilon) 
\Gamma^2(1-\epsilon)}{\left(4\pi\right)^{2-\epsilon}\Gamma(1-2\epsilon)} \;\;.
\eeq
We remark that Eq.~(\ref{oneg}) gives the complete result to all orders in the $\ep$
expansion around $\ep=0$ (equivalently, the result in arbitrary $d=4-2\ep$ space-time
dimensions).

We comment on some features of Eq.~(\ref{oneg}). The one-loop current is proportional
to the structure constants $f_{abc}$ of the gauge group and, therefore, it is {\em
purely} non-abelian. This is in agreement with the absence of one-loop corrections to
the soft current for single soft-photon emission in massless QED 
\cite{Yennie:1961ad}. The current in Eq.~(\ref{oneg}) involves non-abelian colour
correlations, $i f_{abc} \,T^b_i \,T^c_j$, with two hard partons. Its kinematical
structure has a rational dependence on $p_i\cdot \vep(q)/p_i\cdot q$
(which is analogous to that in the tree-level current of Eq.~(\ref{treeg})) that is
only modified through logarithmic corrections by the one-loop interactions. The
logarithmic corrections are due to the $\ep$ expansion of the last factor in the
right-hand side of Eq.~(\ref{oneg}), and they are proportional to powers of
$\ln q^2_{\perp \, ij}$ (modulo branch-cut effects), where $q_{\perp \, ij}$,
\beq
\label{qt}
q^2_{\perp \, ij} = \frac{2 (p_i \cdot q) (p_j \cdot q)}{p_i \cdot p_j} \;\;,
\eeq
has a simple kinematical interpretation since it is the transverse component of 
the
gluon momentum $q$ with respect to the longitudinal direction singled out by the
momenta $p_i$ and $p_j$ (in a reference frame in which $p_i$ and $p_j$ are
back-to-back) of the colour-correlated partons.
The overall scaling behaviour of $\bj^{(1)}(\lambda q)$  
(with $\lambda > 0$) in the limit $\lambda \to 0$ is proportional to 
$(\lambda^2)^{-\ep}/\lambda = (1/\lambda){\rm mod}(\ep^r \ln^r \lambda)$, 
and it is 
in agreement with Eq.~(\ref{msoft}). In particular, we explicitly see that the 
$\ln
\lambda$-enhancement is produced by the use of dimensional regularization 
to avoid 
the IR and UV divergences in the one-loop contribution to $\bj$. 
Performing the
$\ep$ expansion of Eq.~(\ref{oneg}), this IR and UV behaviour produces double
($1/\ep^2$) and single ($1/\ep$) poles near $\ep=0$.

The two-loop current for {\em single} soft-gluon emission was computed in
Ref.~\cite{Badger:2004uk} up to including contributions of ${\cal O}(\ep^0)$
for the simplest case of scattering amplitudes with {\em only} two hard partons.
Subsequently this result was extended up to ${\cal O}(\ep^2)$
\cite{Li:2013lsa,Duhr:2013msa} and to all orders in $\ep$ \cite{Duhr:2013msa}.
The two-loop result of Refs.~\cite{Badger:2004uk,Li:2013lsa,Duhr:2013msa}
has a structure that is very similar to the one-loop current in Eq.~(\ref{oneg}).
More involved structures, in terms of both colour correlations and kinematical
dependence, do appear in the general case of scattering amplitudes with three
or more hard partons, and the corresponding two-loop current for 
single soft-gluon emission was considered and explicitly computed 
in Ref.~\cite{Dixon:2019lnw},
by including the finite contributions up to ${\cal O}(\ep^0)$.

We now discuss multiple soft radiation at one-loop order. The structure
of the loop-level 
current $\bj$ for multiple soft radiation is expected to be definitely
more complex (in terms of both colour and kinematical dependence) than the single
soft-gluon current in Eq.~(\ref{oneg}). The presence of two or more soft
partons and the ensuing dependence on their momenta increases the number of
relevant kinematical invariants, which drive an increased complexity of colour
and kinematical correlations (especially at high orders in the $\ep$ expansion).
In the remaining part of this Section we deal with general properties of the soft
current $\bj$ with $m \geq 2$ soft partons. In particular, we consider the UV and
IR divergences of $\bj$ and we discuss their regularization scheme dependence. 


The one-loop current for multiple soft emission has 
(analogously to Eq.~(\ref{oneg})) double
($1/\ep^2$) and single ($1/\ep$) pole contributions due to the presence of IR and
UV divergences in the four-dimensional case ($\ep=0$).
At $L$-loop order, the current $\bj^{(L)}$ has poles of the type
$1/\ep^k$ with $2L \geq k \geq 1$.
These $\ep$-pole contributions are directly related to the corresponding
contributions to the multiparton scattering amplitudes
\cite{Catani:1998bh, Sterman:2002qn, Aybat:2006mz, Gardi:2009qi, 
Becher:2009qa, Almelid:2015jia}.
The $\ep$-pole contributions to the one-loop current $\bj^{(1)}$ 
have a general structure, whose
explicit form can be directly derived from the known universal structure of the 
IR and UV divergences of {\em one-loop} scattering amplitudes
\cite{Giele:1991vf, csdip, Catani:1998bh}. Starting from the results in 
Refs.~\cite{Giele:1991vf, csdip, Catani:1998bh}, the procedure to derive the 
$\ep$-pole contributions to $\bj^{(1)}$ is completely analogous to that used in
Refs.~\cite{Catani:2003vu,Catani:2011st}
for the study of the multiparton {\em collinear} limit of scattering amplitudes
(see, in particular, Eqs.~(104)--(109) 
in the arXiv version of
Ref.~\cite{Catani:2011st} and replace the collinear splitting matrix $\sp^{(1)}$
with the soft current $\bj^{(1)}$). Moreover that procedure can be extended to
higher-loop orders and it leads to a compact representation of the $\ep$-pole 
contributions to $\bj$ at arbitrary perturbative orders (see the analogous
procedure in Sect.~6.1 and, in
particular, Eq.~(137) in the arXiv version of
Ref.~\cite{Catani:2011st} and replace $\sp$ with $\bj$).
Owing to the complete analogy with the collinear limit studied in 
Ref.~\cite{Catani:2011st}, we limit ourselves to present the
final results for the soft limit.

The general all-order representation of the $\ep$-pole contributions to $\bj$ 
is
\beq
\label{poleall}
\bj(q_1,\dots,q_m) = {\bf V}(q_1, \dots,q_m,p_1,\dots,p_n) \;
\bj^{[ {\rm no} \;\ep-{\rm poles}]}(q_1,\dots,q_m) \;
{\bf V}^{-1}(p_1,\dots,p_n) \;\;,
\eeq
where $\bj^{[ {\rm no} \;\ep-{\rm poles}]}$ is obtained from $\bj$ by properly
subtracting its $\ep$-pole part order by order in the loop expansion. Therefore
at each perturbative order $\bj^{[ {\rm no} \;\ep-{\rm poles}]}$ is finite in
the limit $\ep \to 0$ order by order in the $\ep$ expansion around $\ep = 0$.
Note that this statement refers to the loop expansion of
$\bj^{[ {\rm no} \;\ep-{\rm poles}]}$
with respect to
renormalized QCD coupling (the use of the renormalized QCD coupling 
removes $\ep$ poles  of UV origin, which cannot be absorbed in the 
${\bf V}$ factors
of Eq.~(\ref{poleall})).
Obviously, at the tree level $\bj$ and $\bj^{[ {\rm no} \;\ep-{\rm poles}]}$
coincides ($\bj^{(0)} = \bj^{(0) \,[ {\rm no} \;\ep-{\rm poles}]}$).
The colour space operator ${\bf V}(q_1, \dots,q_m,p_1,\dots,p_n)$
is the process-independent factor 
\cite{Catani:1998bh, Sterman:2002qn, Aybat:2006mz, Gardi:2009qi, 
Becher:2009qa, Almelid:2015jia}
that controls the IR $\ep$-pole contributions to the scattering amplitude 
$\ket{\cm(q_1, \dots,q_m,p_1,\dots,p_n)}$ in the left-hand side of the
factorization formula (\ref{softfact}) (at the tree level, ${\bf V}^{(0)}=1$).
The operator  ${\bf V}(p_1,\dots,p_n)$ to the right of 
$\bj^{[ {\rm no} \;\ep-{\rm poles}]}$ in Eq.~(\ref{poleall})
is the restriction of ${\bf V}$ to the scattering amplitude
$\ket{\cm(p_1,\dots,p_n)}$ in the left-hand side of Eq.~(\ref{softfact})
(i.e., the amplitude with the external soft legs removed).
Since the ${\bf V}$ factors in Eq.~(\ref{poleall}) only depends on the colour,
flavour and momentum of the soft and hard partons, their perturbative knowledge
determines in a recursive manner (i.e., order by order in the loop expansion)
the $\ep$-pole part of $\bj$ at a given order in terms of $\bj$ at lower
perturbative orders.

At the one-loop level, using Eq.~(\ref{poleall}) and the known expression
\cite{Giele:1991vf, csdip, Catani:1998bh} of the one-loop term 
${\bf V}^{(1)}$ of the operator ${\bf V}$, we obtain the following expression
for the $\ep$-pole contributions to the soft multiparton current $\bj^{(1)}$:
\beeq
\label{1divmin}
&&\!\!\!\!\!\!\!\!\!\!\!\! \bj^{(1)}(q_1,\dots,q_m) =
 - \g^2 \;
c_{\Gamma} \,
\Bigl\{
\; \sum_{k \in S} 
\left[ \frac{1}{\ep^2} C_k + \frac{1}{\ep}\left( \gamma_k - \cbet0 \right)\right]
\bj^{(0)}(q_1,\dots,q_m) 
\Bigr.
\nn \\
&&\!\!\!\!\!\!\!\!\!+ \, \frac{1}{\ep} \; \Bigl[ \,\,
\sum_{\substack{k,l \,\in S \\ k \,\neq\, l}} 
\ln\!\left( \frac{-2 q_k \cdot q_l - i0}{\mu^2} \right) \bt_k \cdot  \bt_l
+ \sum_{\substack{i \in H \\ k \in S}} 
\ln\!\left( \frac{-2 p_i \cdot q_k - i0}{\mu^2} \right)
\;2\, \bt_i \cdot  \bt_k
\, \Bigr] \bj^{(0)}(q_1,\dots,q_m) \nn \\
&&\!\!\!\!\!\!\!\!\!- \, \frac{1}{\ep} \;\sum_{\substack{i,j \,\in H \\ i \,\neq\, j}} 
\ln\!\left( \frac{-2 p_i \cdot p_j - i0}{\mu^2} \right)
\;\Bigl. 
\left[ \bj^{(0)}(q_1,\dots,q_m) \,, \;\bt_i \cdot  \bt_j \right]
\frac{}{}\Bigr\} + {\cal O}(\ep^0) \;\;,
\eeeq
where $C_k$ ($\bt_k^2=C_k$) is the Casimir coefficient of the parton $k$ and, 
analogously, the coefficient $\gamma_k$ depends on the flavour of the parton
$k$ and, explicitly, we have
\beq
\label{gacoef}
\gamma_q=\gamma_{\bar q}= \f{3}{2} \; C_F \;\;, \quad
\quad \gamma_g= \f{1}{6} \;(11 \,C_A -4 T_R \,N_f) \;\;,
\eeq
where $N_f$ is the number of flavours of massless quarks.
The coefficient $\cbet0$ 
is the first perturbative coefficient of the QCD $\beta$ function,
\beq
\label{norbeta0}
\cbet0=\f{1}{6} \;(11 \,C_A - 4 T_R \,N_f) \;\;.
\eeq
Note that, in our normalization, we have $\cbet0= \gamma_g$.

The various $\ep$-pole terms in Eq.~(\ref{1divmin}) have different origins.
The single-pole term that is proportional to $\cbet0$ is of UV origin;
it can be removed by renormalizing the soft current $\bj$
(we recall that we are considering unrenormalized scattering amplitudes
and, correspondingly, unrenormalized soft currents).
The other $\ep$-pole terms are of IR origin.  
The double-pole terms, which are proportional to the 
Casimir coefficients $C_k$ ($C_k=C_F$ and $C_A$ for quarks and gluons,
respectively),
originate from one-loop contributions in which the loop
momentum is nearly on-shell, very soft and parallel to the momentum of one of the
soft partons involved in the current.
The single-pole terms with $\gamma_k$ coefficients are produced by
contributions in which the loop
momentum is not soft, though it is nearly on-shell
and parallel to the momentum of one of the external soft partons of the current. 
The single-pole terms with logarithmic dependence on soft-parton and hard-partons
subenergies ($q_k \cdot q_l, \,p_i \cdot q_k, \,p_i \cdot p_j$) originate from 
configurations in which the loop momentum is very soft and at wide angle with
respect to the direction of the external-leg (soft and hard) partons. 
Specifically,
the radiative part of these terms (i.e., the real part of the logarithms) is due 
to a nearly on-shell virtual gluon in the loop, while the absorptive part
(i.e., the imaginary part of the logarithms) is due to the exchange of an 
off-shell
Coulomb-type gluon.

The expression in the right-hand side of Eq.~(\ref{1divmin}) is valid for 
an {\em arbitrary} number $m$ of soft partons in the current 
(and for an arbitrary number of hard partons in the scattering amplitude).
This expression is given in terms of explicit coefficients and of the
tree-level current $\bj^{(0)}$ for the corresponding parton
configuration. Once $\bj^{(0)}$ (and, in particular, its colour structure) is
explicitly known, Eq.~(\ref{1divmin}) can be directly applied to determine the 
explicit $\ep$-pole contributions to the one-loop current $\bj^{(1)}$.

In particular, in the case of a single ($m=1$) soft parton, using 
Eq.~(\ref{treeg})
it is straightforward to check that Eq.~(\ref{1divmin}) gives the $\ep$-pole 
terms of the one-loop result $\bj^{(1)}$ in Eq.~(\ref{oneg}).
In this respect, we note that the expression in the right-hand side of 
Eq.~(\ref{1divmin}) does not identically (in its precise algebraic form) 
correspond to the  $\ep$-pole terms in Eq.~(\ref{oneg}): the difference is due
to terms of ${\cal O}(1/\ep^2)$ and ${\cal O}(1/\ep)$ that are proportional to 
the total colour charge ($\sum_{i \in H} \bt_i$) of the hard partons. 
As previously discussed (see Eq.~(\ref{csequal}) and related comments)
the presence of such terms in $\bj$ is physically harmless.

The tree-level currents $\bj^{(0)}$ for emission of two $(m=2)$ soft partons
(either two gluons or a $q{\bar q}$ pair) are also explicitly known
\cite{Catani:1999ss}. Therefore Eq.~(\ref{1divmin}) can also be straightforwardly applied
to explicitly obtain the $\ep$-pole terms of the one-loop current $\bj^{(1)}$ for double
soft-parton emission. The case of a soft quark and antiquark is discussed in detail in 
Sect.~\ref{sec:qqcur}. The case of two soft gluons is studied in Ref.~\cite{Zhu:2020ftr}.
We have checked that the $\ep$-pole terms of the one-loop double-gluon current 
computed in Ref.~\cite{Zhu:2020ftr} agree with the corresponding result that is obtained 
by using Eq.~(\ref{1divmin}) and the colour conservation relation (\ref{colcon}).
To be precise about the absolute normalization of the one-loop current,
we think that the factor $e^{-\ep \gamma_E}$ has to be removed from the expansion parameter  
in Eq.~(3.2) of Ref.~\cite{Zhu:2020ftr}.

We comment on the behaviour of the one-loop current $\bj^{(1)}(q_1,\dots,q_m)$
with respect to the overall rescaling $q_k \to \lambda q_k$ of all the momenta of
the soft partons. To avoid the effects of branch-cut contributions from crossing 
different kinematical regions of soft and hard momenta, with limit ourselves to
considering the case with $\lambda > 0$.
According to Eq.~(\ref{msoft})
(see also the discussion below it) and Eq.~(\ref{softfact}),
the tree-level current $\bj^{(0)}$ behaves as
\beq
\label{la0}
\bj^{(0)}(\lambda q_1,\dots,\lambda q_m) = \frac{1}{\left(\lambda \right)^m }
\; \bj^{(0)}(q_1,\dots,q_m) \;\;, 
\eeq
and the expected one-loop behaviour is
\beq
\label{la1}
\bj^{(1)}(\lambda q_1,\dots,\lambda q_m) = 
\frac{\left(\lambda \right)^{-2\ep}}{\left(\lambda \right)^m }
\; \bj^{(1)}(q_1,\dots,q_m) \;\;, 
 \quad \quad (\lambda > 0) \, .
\eeq
The behaviour as in Eqs.~(\ref{la0}) and (\ref{la1}) is indeed observed in the
tree-level results of Eqs.~(\ref{treeg}) and (\ref{treeqq}) and in the one-loop
soft single-parton 
current of Eq.~(\ref{oneg}).
Using Eq.~(\ref{la0}) and applying the $\lambda$ rescaling to the explicit
expression in the right-hand side of Eq.~(\ref{1divmin}), we obtain the result
$\bj^{(1)}(\lambda q_1,\dots,\lambda q_m) \eqcs 
\frac{1}{\left(\lambda \right)^m } (1 -2\ep \ln \lambda) \,
\bj^{(1)}(q_1,\dots, q_m) + {\cal O}(\ep^0)$ (note that we neglect harmeless
contributions proportional to the total colour charge of the hard partons).
This result is perfectly consistent with Eq.~(\ref{la1}), 
since Eq.~(\ref{1divmin}) only embodies the correct $\ep$-pole contributions to 
$\bj^{(1)}$. In particular, in Eq.~(\ref{1divmin}) these contributions are
embodied in a `minimal' form by systematically neglecting terms of 
${\cal O}(\ep^n)$ $(n \geq 0)$, with the sole exception of terms that arise from
the $\ep$-expansion of the overall factor $c_{\Gamma}$ ($(4\pi)^2 c_{\Gamma}
= 1+ {\cal O}(\ep)$). The $\ep$-pole contributions to $\bj^{(1)}$ can be expressed
in alternative forms with respect to Eq.~(\ref{1divmin}). In particular, the
right-hand side of Eq.~(\ref{1divmin}) can be supplemented with terms of 
${\cal O}(\ep^n)$ $(n \geq 0)$ in a manner that restores the behaviour in
Eq.~(\ref{la1}) to all orders in the $\ep$ expansion.

An alternative explicit form of the $\ep$-pole contributions to $\bj^{(1)}$
for the soft multiparton $(m \geq 2)$ limit is as follows
\beeq
\label{gen1div}
&&\!\!\!\!\!\!\!\!\!\!\!\! \bj^{(1)}(q_1,\dots,q_m) \eqcs
 - \g^2 \left( \frac{-\qm^2 - i0}{\mu^2}\right)^{\!\!-\ep}
c_{\Gamma} \,
\Bigl\{
\; \sum_{k \in S} 
\left[ \frac{1}{\ep^2} C_k + \frac{1}{\ep}\left( \gamma_k - \cbet0 \right)\right]
\bj^{(0)}(q_1,\dots,q_m) 
\Bigr.
\nn \\
&&\quad + \, \frac{1}{\ep} \; \Bigl[ \,\,
\sum_{\substack{k,l \,\in S \\ k \,\neq\, l}} 
\ln\!\left( \frac{-2 q_k \cdot q_l - i0}{-\qm^2 - i0} \right) \bt_k \cdot  \bt_l
+ \sum_{\substack{i \in H \\ k \in S}} \ell_{ik}(\qm) \;2\, \bt_i \cdot  \bt_k
\, \Bigr] \bj^{(0)}(q_1,\dots,q_m) \nn \\
&&\quad + \frac{1}{\ep} \;\sum_{\substack{i,j \,\in H \\ i \,\neq\, j}} 
L_{ij}(\qm) \;\Bigl. 
\left[ \bj^{(0)}(q_1,\dots,q_m) \,, \;\bt_i \cdot  \bt_j \right]
\frac{}{}\Bigr\} + {\cal O}(\ep^0) \;\;, \quad \quad \quad (m \geq 2) \;,
\eeeq
where the total soft momentum is denoted by $\qm$,
\beq
\label{mmom}
\qm \equiv \sum_{k \in S} q_k = q_1+ \dots + q_m \;\;,
\eeq
and we have introduced the logarithmic functions 
$\ell_{i k}$ and $L_{ij}$ of hard and soft momenta:
\beq
\label{loghs}
\ell_{i k}(\qm) \equiv \ln\left( 
\frac{- p_i \cdot q_k - i0}{- p_i \cdot \qm - i0} \right) \;\;,
\eeq
\beq
\label{loghh}
L_{ij}(\qm) = L_{ji}(\qm) \equiv 
\ln\left( \frac{- p_i \cdot \qm - i0}{- p_i \cdot p_j - i0} \right) +
\ln\left( \frac{- 2p_j \cdot \qm - i0}{- \qm^2 - i0} \right) \;\;.
\eeq
It can be explicitly checked that the two expressions in Eqs.~(\ref{1divmin})
and (\ref{gen1div}) only 
differ by terms of ${\cal O}(\ep^0)$ and higher orders in $\ep$
while acting onto colour singlet quantities.
We note that the logarithmic functions $\ell_{i k}$ and $L_{ij}$ in 
Eqs.~(\ref{loghs}) and (\ref{loghh}) are invariant under the overall rescaling 
$q_k \to \lambda q_k$ $(\lambda > 0)$ of the soft momenta. Therefore,
the explicit expression in the right-hand side of Eq.~(\ref{gen1div})
exactly fulfils the scaling behaviour in Eq.~(\ref{la1}). We also note that both
expressions of Eqs.~(\ref{1divmin})
and (\ref{gen1div}) fulfil the colour flow conservation property in
Eq.~(\ref{flowcon}).

Throughout the paper we use the dimensional regularization procedure to deal
with UV and IR divergences and, therefore, the momenta (and their associated
phase space) of the virtual particles inside loops are analytically continued to
$d=4-2\ep$ space-time dimensions \cite{'tHooft:1972fi, cdr, Gastmans:1973uv}.
Different variants of dimensional regularization can be used, and each variant
defines a specific regularization scheme (RS). The RSs that are mostly used are
conventional dimensional regularization (CDR)
\cite{cdr, Gastmans:1973uv}, the 't~\!Hooft--Veltman (HV)
\cite{'tHooft:1972fi} scheme, dimensional reduction (DR) \cite{Siegel:1979wq}
and the four-dimensional helicity (4DH) scheme \cite{Bern:1991aq}.
The momenta of the external-leg particles in the scattering amplitude can be
either $d$-dimensional (CDR and DR schemes) or four-dimensional (HV and 4DH
schemes). The number of spin polarization (helicity) states of the gluon also
depends on the RS: external-leg gluons can have either $d-2=2-2\ep$
polarizations (CDR) or 2 polarizations (HV, DR, 4DH), and virtual gluons can
have either $d-2=2-2\ep$
polarizations (CDR, HV) or 2 polarizations (DR, 4DH).
Scattering amplitudes and, consequently, soft currents (as defined by the soft
limit) depend on the RS. As for the RS dependence on external-leg particles,
throughout the paper we formally express soft (tree-level and one-loop)
currents in terms of external-leg momenta $(p_i, q_k)$ and corresponding
polarization wave functions ($\varepsilon(q_k), u(q_k), v(q_k)$): these
expressions are formally RS invariant, although momenta and wave functions
implicitly embody an RS dependence (which can be regarded as a dependence of
${\cal O}(\ep)$). At the one-loop level, soft currents (and scattering
amplitudes) have a residual RS dependence that can be explicitly parametrized by
the number of polarization states $h_g$ of virtual gluons. We write
$h_g= 2 - 2\ep \delta_R$ and, therefore, we have
(this is the same notation as used, e.g., in 
Refs.~\cite{Bern:1999ry, Kosower:1999rx})
\beq
\label{deleq}
\delta_R=1   \quad ({\rm CDR,\; HV}) \;\;, 
\quad \quad \quad \quad
\delta_R=0 \quad ({\rm 4DH, \; DR}) \;\;.
\eeq
To formally express the explicit $\delta_R$ dependence of the one-loop soft
current $\bj^{(1)}$ we then define
\beq
\label{j1rsdef}
\bj^{(1)}_{\rs} \equiv \bj^{(1)} - \bigl[ \bj^{(1)} \bigr]_{\delta_R=1} \;\;,
\eeq
where 
both terms in the right-hand side are expressed through the {\em same} formal
external-leg variables (momenta and wave functions), which embody an implicit
dependence on the RS, and $\bigl[ \bj^{(1)} \bigr]_{\delta_R=1}$ is obtained 
by setting $\delta_R=1$ in the explicit expression of $\bj^{(1)}$.
Roughly speaking (e.g., modulo the implicit RS dependence due
to the number of polarizations of the external partons),
$\bj^{(1)}_{\rs}$ in Eq.~(\ref{j1rsdef}) represents the
difference of $\bj^{(1)}$ between a given RS and the CDR (or HV) scheme.

One-loop scattering amplitudes have an explicit RS dependence on $\delta_R$.
Considering the $\ep$ expansion up to including terms of ${\cal O}(\ep^0)$,
the dependence on $\delta_R$ can be written in {\em factorized} form through the
tree-level scattering amplitude and universal (process independent) coefficients
\cite{Kunszt:1993sd}. Using these scattering amplitude results, we can obtain
the explicit $\delta_R$ dependence of the one-loop current up to the same order
in the $\ep$ expansion. In particular, the $\delta_R$ dependence of the one-loop
scattering amplitude can be controlled through an ensuing $\delta_R$ dependence 
of the one-loop expression ${\bf V}^{(1)}$ 
\cite{Kunszt:1993sd, Catani:1998bh, Catani:2000ef}
of the operator ${\bf V}$ in Eq.~(\ref{poleall}) and, therefore, we can
explicitly compute the right-hand side of Eq.~(\ref{j1rsdef}) up to 
${\cal O}(\ep^0)$. The expression of $\bj^{(1)}_{\rs}$ for $m$ soft partons is
\beq
\label{gen1rs}
\bj^{(1)}_{\rs}(q_1,\dots,q_m) =
 - \,(\g \mu^\ep)^2 \;c_{\Gamma} \;\left( \delta_R -1 \right)
\; \sum_{k \in S} 
 \left( {\widetilde \gamma}_k - {\widetilde \cbet0} \right)
\bj^{(0)}(q_1,\dots,q_m) 
+ {\cal O}(\ep) \;\;,
\eeq
where the coefficient ${\widetilde \gamma}_k$ depends on the flavour of the soft
parton $k$ and it has an IR origin, while the coefficient  
${\widetilde \cbet0}$ has an UV origin. The explicit IR coefficients 
\cite{Kunszt:1993sd} and the UV coefficient \cite{Altarelli:1980fi} are
\beq
\label{gatilcoef}
{\widetilde \gamma}_q={\widetilde \gamma}_{\bar q}= \f{1}{2} \; C_F \;\;, \quad
\quad {\widetilde \gamma_g}= {\widetilde \cbet0} = \f{1}{6} \;C_A  \;\;.
\eeq

Note that, analogously to the structure of its $\ep$-pole contributions
(see Eqs.~(\ref{1divmin}) and (\ref{gen1div})), the $\delta_R$ dependence
dependence of $\bj^{(1)}$ has a factorized structure in terms of its
corresponding tree-level current  $\bj^{(0)}$. Since 
${\widetilde \gamma_g}= {\widetilde \cbet0}$, in the case of a single soft gluon
$(m=1)$, Eq.~(\ref{gen1rs}) agrees with the explicit result in Eq.~(\ref{oneg}).
Incidentally, we recall \cite{Catani:2000pi} that the result in Eq.~(\ref{oneg})
(to all orders in the $\ep$ expansion)
is valid in any RS, and thus the expression 
of the single soft gluon current in Eq.~(\ref{oneg}) is basically RS invariant
(it does not depend on $\delta_R$, and the RS dependence is formally encoded in
the corresponding RS dependence on $\varepsilon (q)$ and the dimensionality of
the external momenta $p_i,p_j,q$). As shown in Eq.~(\ref{gen1rs}), in the soft
multiparton $(m \geq 2)$ case $\bj^{(1)}_{\rs}$ is of ${\cal O}(\ep^0)$.
Conceptually, however, the RS dependence of $\bj^{(1)}$ (and of one-loop
scattering amplitudes) starts and ${\cal O}(1/\ep)$: the effect of 
${\cal O}(1/\ep)$ is formally hidden in Eq.~(\ref{1divmin}) 
(or Eq.~(\ref{gen1div})) through the product $\bj^{(0)} \cdot 1/\ep^2$
($\bj^{(0)}$ conceptually embodies an RS dependence at ${\cal O}(\ep)$
through its external-leg momenta and polarization vectors).

Throughout the paper we explicitly consider unrenormalized scattering amplitudes
and currents. However, UV renormalization commutes with the soft limit and,
therefore, the renormalization procedure can be straightforwardly applied to all
the explicit expressions presented herein. In particular, since we are
considering amplitudes and currents with (on-shell) massless hard partons, 
the renormalization procedure
simply amounts to replace the bare coupling $\g$ (or $\as=\g^2/(4\pi)$) 
with its expression in terms of the renormalized running coupling
$\g(\mu_R)$ (or $\as(\mu_R)$) at the renormalization scale $\mu_R$.
In this respect, we recall that also the coupling renormalization is affected by
RS subtleties. For instance, renormalizing the coupling by subtraction of the sole
UV $\ep$-poles (e.g., the term proportional to $\cbet0$ in 
Eq.~(\ref{1divmin})) in a given RS does not lead to an RS 
invariant
definition
of the renormalized coupling $\g(\mu_R)$: an additional finite renormalization
shift of $\g(\mu_R)$ (whose size depends on the RS dependent coefficient
$\delta_R \,{\widetilde \cbet0}$ in Eq.~(\ref{gatilcoef})) \cite{Altarelli:1980fi}
is necessary to achieve
an RS 
independent
definition of $\g(\mu_R)$.

As discussed and presented in this Section, the $\ep$-pole contributions
(and also the RS dependent contributions at ${\cal O}(\ep^0)$) to the one-loop
current $\bj^{(1)}$ for the general case of $m \;(m \geq 2)$ soft partons and an
arbitrary number of hard partons are completely determined by 
Eqs.~(\ref{1divmin}) or (\ref{gen1div}) (and Eq.~(\ref{gen1rs})), and they are
explicitly known as soon as the corresponding tree-level current  $\bj^{(0)}$
is known. The determination of $\bj^{(1)}$ at ${\cal O}(\ep^0)$ and, possibly,
at higher orders in $\ep$ requires detailed one-loop computations and they have
a high complexity. To have a rough idea of the computational complexity, we can
simply observe that $\bj^{(1)}$ can (in principle) be determined by 
performing
the soft
limit of one-loop amplitudes according to Eq.~(\ref{onef}).
To apply Eq.~(\ref{onef}) we have to consider amplitudes with $m+n$
external legs, and the number $n$ of non-soft external legs cannot be `too
small', otherwise the amplitude on the right-hand side of Eq.~(\ref{onef})
vanishes. For example, the amplitude should have at least two external hard QCD
partons (because of colour conservation) and one additional colourless external
leg (because of momentum conservation): the soft limit of such an amplitude
with $m+3$ external legs leads to the current $\bj^{(1)}$ in the simplest case
with two hard QCD partons. To get information on the colour-correlation
structure of $\bj^{(1)}$ in the general case of several hard partons, the
amplitude should have at least $n=4$ hard QCD partons in its external legs
(owing to colour-conservation relations, the cases with $n=2$ and 3 hard partons
lead to simplified colour structures; see, e.g., Sect.~\ref{sec:23hard}).
In summary, the amplitudes to be considered should have $m+n$
external legs with $n \geq 4$: even in the case of double soft-parton radiation
$(m=2)$, this implies (at least) six external legs. As is well known, one-loop
computations of these multileg scattering amplitudes are definitely complex to
be carried out in analytic form (which is necessary to perform the soft limit).
The computation of the one-loop current $\bj^{(1)}$ can be highly simplified by
using general methods (e.g., the method of Ref.~\cite{Catani:2000pi}))
that do not require a full direct computation of scattering amplitudes. However,
despite some relevant simplification, even these methods have to deal with
multileg one-loop Feynman integrals,  whose evaluation is definitely complex,
especially at high orders in the $\ep$ expansion.

\section{Soft $q\bar q$ emission: the one-loop current}
\label{sec:qqcur}

In this Section we present and discuss the results of our explicit computation of
the QCD one-loop current for soft 
$q {\bar q}$
radiation.
In Sect.~\ref{sec:qed} we also generalize the results to the cases of QED and
mixed QCD$\times$QED radiative corrections.

\begin{figure}[htb]
\begin{center}

\begin{picture}(400,230)(0,150)

\SetWidth{1.5}

\SetOffset(20,200)

\Vertex(32,50){4.8}
\DashLine(0,50)(30,50){4}
\ArrowLine(30,50)(95,-25)
\ArrowLine(30,50)(95,125)
\Line(80,70)(80,30)
\ArrowLine(80,70)(100,70)
\ArrowLine(80,30)(100,30)
\Gluon(48,70)(80,70){3}{4}
\Gluon(48,30)(80,30){3}{4}
\Text(70,108)[]{$p_i$}
\Text(70,-9)[]{$p_j$}
\Text(120,70)[]{$q(q_1)$}
\Text(120,30)[]{$\bar q(q_2)$}
\Text(70,-50)[]{$(a)$}


\SetOffset(240,200)
\Vertex(32,50){4.8}
\DashLine(0,50)(30,50){4}
\ArrowLine(30,50)(95,-25)
\ArrowLine(30,50)(95,125)
\ArrowLine(80,50)(100,70)
\ArrowLine(80,50)(100,30)
\Gluon(52,50)(80,50){3}{4}
\Gluon(50,72)(50,28){3}{5}
\Text(70,108)[]{$p_i$}
\Text(70,-9)[]{$p_j$}
\Text(120,70)[]{$q(q_1)$}
\Text(120,30)[]{$\bar q(q_2)$}
\Text(70,-50)[]{$(b)$}

\end{picture}
\end{center}
\caption{\label{fig:1loop}
{\em Example of two one-loop Feynman diagrams that contribute to the one-loop
current $\bj^{(1)}$ for soft $q {\bar q}$ emission. The external-leg hard
partons with momenta $p_i$ and $p_j$ are coupled to virtual gluons by using the
eikonal approximation. The dashed line symbolically denotes the colour indices
and momenta of the additional external legs. The effect of the tree-level
scattering amplitude $\cm^{(0)}(p_1,\dots,p_n)$ (see Eq.~(\ref{onef})) is given
by an effective pointlike vertex (little black circle) that only depends on the
colour structure of $\cm^{(0)}$ (on the colour indices of the hard partons).}
}
\end{figure}
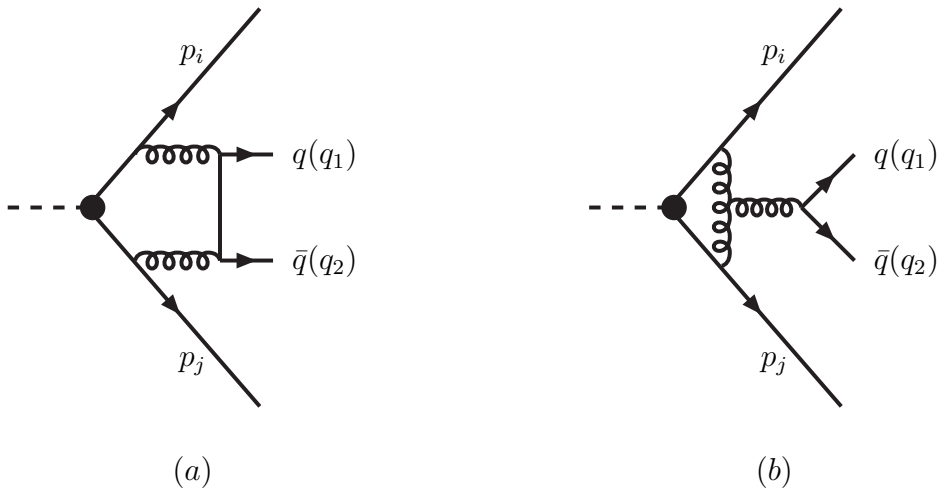

The tree-level current $\bj^{(0)}$ for emission of a soft-$q {\bar q}$ 
pair
in a scattering amplitude with an {\em arbitrary} number of hard partons
is given in Eq.~(\ref{treeqq}).
To evaluate the one-loop contribution $\bj^{(1)}$ we use the general
(process-independent) method of Ref.~\cite{Catani:2000pi}
(the same method is used in the computations of Refs.~\cite{Bierenbaum:2011gg}
and \cite{Li:2013lsa}).
The computational procedure is completely analogous to that in 
Ref.~\cite{Catani:2000pi} (though it is extended from the case of a single soft
gluon to the case of a soft-$q {\bar q}$ pair) and we do not repeat all the
details. We have to evaluate a set of one-loop Feynman diagrams 
(as example, in Fig.~\ref{fig:1loop} we show two contributing 
Feynman diagrams) in which the external-leg hard partons are coupled to virtual
gluons by using the eikonal approximation
(for both vertices and propagators), while the other vertices and propagators
are computed by using the customary QCD Feynman rules.
We perform the calculation by using both the Feynman gauge and the axial gauge
$n\cdot~\!\!A =0$, with an auxiliary light-like ($n^2=0$) gauge vector $n^\mu$.
Combining all the contributing Feynman diagrams,
the dependence on the gauge vector cancels at the integrand level
(i.e., before performing the integration over the loop momentum) and the
total axial-gauge integrand coincides with the Feynman gauge integrand: this
provides us with an explicit check of the gauge invariance of the procedure and
of the calculation. 

As usual in the context of dimensional regularization,
scaleless one-loop integrals vanish. Eventually we have to compute several 
(non-vanishing) tensor, vector and scalar one-loop Feynman integrals.
Tensor and vector integrals are expressed in terms of scalar integrals by using
customary techniques \cite{Passarino:1978jh}. One-loop integrals with five
external legs (pentagon integrals; see, e.g. the Feynman diagram in 
Fig.~\ref{fig:1loop}(a)) are expressed \cite{Bern:1993kr} in terms of 
one-loop integrals with four external legs (box integrals) plus remaining
pentagon integrals in $6-2\ep$ space-time dimensions, which only contribute at
${\cal O}(\ep)$ (and higher orders in $\ep$). We do not explicitly evaluate
these contributions at  ${\cal O}(\ep)$. We eventually express the complete
result in terms of a minimal set of basic one-loop scalar integrals. The set
involves customary two-point and three-point (with at least one on-shell leg)
Feynman integrals and some soft box integrals (box integrals with eikonal
propagators). Part of these soft box integrals was already computed in 
Ref.~\cite{Catani:2000pi} and the additional integrals are analogous to those
encountered in Ref.~\cite{Anastasiou:2015yha}. We have performed an independent
calculation of these soft box integrals and we find agreement with the results
reported in Ref.~\cite{Anastasiou:2015yha} (see `soft box 2' and `soft box 4'
in Sect.~4.2 of Ref.~\cite{Anastasiou:2015yha}). Our final result for the
one-loop current $\bj^{(1)}$ is reported below.


To present our results, we first define the tree-level and one-loop 
rescaled currents $\hbj^{(0)}$ and  $\hbj^{(1)}$ as follows
\beq
\label{treejhat}
\bj^{(0)}(q_1,q_2) = \left( \g \,\mu^\ep \right)^2  \;\hbj^{(0)}(q_1,q_2) \;\;,
\eeq
\beq
\label{onejhat}
\bj^{(1)}(q_1,q_2) = \left( \g \,\mu^\ep \right)^4 
\left(-q_{12}^2 - i0\right)^{\!-\ep} 
c_{\Gamma} \, {\hbj}^{(1)}(q_1,q_2) \;\;,
\eeq
\beq
\label{onedivfin}
 {\hbj}^{(1)}(q_1,q_2) = {\hbj}^{(1, \,{\rm div})}(q_1,q_2)
+ \hbj^{(1, \,{\rm fin})}(q_1,q_2) \;\;,
\eeq
where $\hbj^{(1)}$ is written in terms of two components, 
${\hbj}^{(1, \,{\rm div})}$ and $\hbj^{(1, \,{\rm fin})}$.
The rescaled current $\hbj^{(0)}$ can be read from comparing 
Eqs.~(\ref{treeqq}) and (\ref{treejhat}).
The component ${\hbj}^{(1, \,{\rm div})}$ of Eq.~(\ref{onedivfin}) embodies the 
$\ep$-pole contributions to $\bj^{(1)}$, while $\hbj^{(1, \,{\rm fin})}$
includes all the remaining UV/IR finite contributions at 
${\cal O}(\ep^0)$ and higher orders in $\ep$.
The explicit expressions of ${\hbj}^{(1, \,{\rm div})}$ and
$\hbj^{(1, \,{\rm fin})}$ are
\beeq
\!\!\!\!\!\!
{\hbj}^{(1, \,{\rm div})}(q_1,q_2) \!\!&=&\!\! - \,2 \left[ \frac{1}{\ep^2} \,C_F + 
\frac{1}{\ep}
\left( \frac{3}{2} \,C_F - \frac{1}{6} (11 \,C_A - 4 \,T_R\,N_f) \right) 
\right] \hbj^{(0)}(q_1,q_2) \nn \\
&-&\!\!\frac{2}{\ep} 
\,j_\nu(1,2) 
\;\btq^{a} \btq^{b}
\sum_{\substack{i,j \,\in H \\ i \,\neq\, j}} T^{a}_i \, T^{b}_j
\left(\frac{p_i^\nu}{p_i \cdot q_{12}} -
\frac{p_j ^\nu}{p_j \cdot q_{12}}
\right) \left( L_{ij} + \ell_{i1} + \ell_{j2} \right) \;\;,
\label{j1div}
\eeeq
\beeq
\!\!\!
{\hbj}^{(1, \,{\rm fin})}(q_1,q_2) \!\!\!\!&=\!&\!\!\! \left[ 
\bigl( - 8 - (\delta_R -1) \bigr) C_F  + \!
\left( \frac{76}{9} -\frac{\pi^2}{3} + \frac{1}{3}(\delta_R -1) \right) \!C_A
- \frac{20}{9} \,T_R N_f \right] \!\hbj^{(0)}(q_1,q_2) \nn \\
&+&\!\! 
\,j_\nu(1,2) 
\;\btq^{a} \btq^{b}
\sum_{\substack{i,j \,\in H \\ i \,\neq\, j}} T^{a}_i \, T^{b}_j
\left[
\left(\frac{p_i^\nu}{p_i \cdot q_{12}} -
\frac{p_j ^\nu}{p_j \cdot q_{12}}
\right) \left( L^2_{ij} + (\ell_{i1} - \ell_{j2})^2 \right)
\right. \nn \\
&+& \frac{q_{12}^2}{q_{12 \perp ij}^2}
\left. 
\left(\frac{p_i^\nu}{p_i \cdot q_{12}} +
\frac{p_j ^\nu}{p_j \cdot q_{12}}
\right) 2 \,L_{ij} \,\bigl( \ell_{i1} - \ell_{j2} \bigr) \right] + {\cal O}(\ep)
\;\;,
\label{j1fin}
\eeeq
where we have used the logarithmic functions of 
Eqs.~(\ref{loghs}) and (\ref{loghh}) and we have introduced the shorthand
notation $\ell_{i k}(q_{12}) \equiv \ell_{i k}$ (with $k=1,2$) 
and $L_{ij}(q_{12}) \equiv L_{ij}$
(i.e., we omit the explicit dependence on the argument $q_{12}$).
The kinematical variable $q^2_{12 \perp \, ij}$ that is used in 
Eq.~(\ref{j1fin}) is
\beq
\label{q12t}
q^2_{12 \perp \, ij} = 
\frac{2 (p_i \cdot q_{12}) (p_j \cdot q_{12})}{p_i \cdot p_j} - q_{12}^2 \;\;.
\eeq
We remark that the results in Eqs.~(\ref{j1div}) and (\ref{j1fin})
are valid in {\em arbitrary} kinematical regions, since the time component (`energy')
of the outgoing momenta $\{q_1, q_2, p_i, p_j \}$ of the soft and hard partons 
can have an arbitrary sign.
According to the notation in Eq.~(\ref{treeqq}) the colour indices $\alpha_1$ 
and $\alpha_2$ of the soft quark and antiquark are specified by considering
$\bra{\alpha_1,\alpha_2}\, \bj^{(1)}(q_1,q_2)
\equiv \bj^{(1) \,\alpha_1,\alpha_2}(q_1,q_2)$, and this leads to
$\bra{\alpha_1,\alpha_2}  \,\btq^a \btq^b \;= 
\left(t^a t^b\right)_{\alpha_1\alpha_2}$ in 
Eqs.~(\ref{j1div}) and (\ref{j1fin}).

The result in Eq.~(\ref{j1div}), which follows from our direct computation of 
$\bj^{(1)}$, agrees with the $\ep$-pole contributions that can straightforwardly
be obtained by applying the general results in 
Eqs.~(\ref{1divmin}) or (\ref{gen1div})
to the specific case of a soft $q {\bar q}$ pair (note that this agreement is
valid modulo harmless terms that are proportional to the total colour charge
$\sum_{i \in H} \bt_i$ of the hard partons).
The expression in Eq.~(\ref{j1div}) has a term that is directly proportional to 
$\bj^{(0)}$ and additional terms that involve colour (and kinematical)
correlations of the soft $q {\bar q}$ pair with {\em two} hard
partons. These colour correlations are produced by the colour matrix factor
$\btq^{a} \btq^{b} T^{a}_i \, T^{b}_j$. We remark that these correlations are not
purely non-abelian, but they also include a component that is still present in the
{\em abelian} limit of commuting colour matrices
(this feature has to be contrasted with the one-loop single soft-gluon case
of Eq.~(\ref{oneg}), in which correlations are purely non-abelian).
In particular, this also implies that the soft current for lepton-antilepton
radiation in massless QED has non-vanishing QED radiative corrections at one-loop
level (see Sect.~\ref{sec:qed}). The
kinematical coefficients of these colour-correlation terms 
are proportional to
the momentum function
\beq
\label{lsinglepole}
L_{ij} + \ell_{i1} + \ell_{j2} = 
\ln\left(\frac{- p_i \cdot q_1 - i0}{- p_i \cdot p_j - i0} \right)
+ \ln\left(\frac{- p_j \cdot q_2 - i0}{- q_1 \cdot q_2 - i0} \right) \;\;,
\eeq
whose real part is the logarithm of a conformally invariant cross ratio, namely,
\beq
\label{lcrossratio}
{\rm Re}(L_{ij} + \ell_{i1} + \ell_{j2}) = 
\ln\left(\frac{|p_i \cdot q_1| \, |p_j \cdot q_2|}
{|p_i \cdot p_j| \, |q_1 \cdot q_2|} \right) \;\;.
\eeq

We note that (analogously to the treatment in Sect.~\ref{sec:1loop})
in the computation of ${\hbj}^{(1)}(q_1,q_2)$ we have dressed gluon propagators
with one-loop vacuum polarization effects that are due only to massless partons.
In particular, the terms proportional to $N_f {\hbj}^{(0)}$ in the right-hand side of 
Eqs.~(\ref{j1div}) and (\ref{j1fin}) are due to the vacuum polarization of $N_f$ massless quarks.
Vacuum polarization effects of {\em massive} quarks can straightforwardly be included
in ${\hbj}^{(1)}(q_1,q_2)$, and they produce corresponding (mass-dependent) contributions that are 
proportional to ${\hbj}^{(0)}(q_1,q_2)$.

We comment on the structure of ${\hbj}^{(1, \,{\rm fin})}$.
We have explicitly computed it up to ${\cal O}(\ep^0)$ and the result is
presented in Eq.~(\ref{j1fin}). The expression of ${\hbj}^{(1, \,{\rm fin})}$
at ${\cal O}(\ep^0)$ is quite compact and
remarkably much simpler than expected. In particular,
although it involves momentum functions of trascendentality equal to two, they
are only powers of logarithmic functions with no additional dependence on
dilogarithms ${\rm Li}_2$. Dilogarithms do appear in the computation of
individual Feynman diagrams and loop integrals at ${\cal O}(\ep^0)$, but they
cancel in the complete result for ${\hbj}^{(1, \,{\rm fin})}$.
The finite component ${\hbj}^{(1, \,{\rm fin})}$ includes a term that is
proportional to $\bj^{(0)}$ and additional correlation terms with {\em two} hard
partons whose colour structure is exactly analogous to that in 
Eq.~(\ref{j1div}) (and it embodies both abelian and non-abelian components). 
We have explicitly checked that no different
colour-correlation structures occur at {\em any} 
higher orders in the $\ep$ expansion.
The term that is proportional to $\bj^{(0)}$ explicitly depends on the RS
parameter $\delta_R$: this dependence exactly agrees with that of the general
result in Eq.~(\ref{gen1rs}).

\setcounter{footnote}{1}

We also comment on the kinematical dependence of the colour correlation terms.
At the tree level the soft-$q{\bar q}$ current $\bj^{(0)}$ has a kinematical
structure with a rational dependence on $j(1,2) \cdot p_i/p_i\cdot q_{12}$
(see Eqs.~(\ref{treeqq}) and (\ref{fercur})). In particular, this dependence
leads to a collinear singularity if $q_{12}^2 \to 0$ (i.e., if the momenta of
the soft quark and antiquark are parallel).
Exactly the same rational dependence (though possibly modified by logarithmic
factors) occurs in the one-loop contributions ${\hbj}^{(1, \,{\rm div})}$
and ${\hbj}^{(1, \,{\rm fin})}$. However, by inspection of Eq.~(\ref{j1fin})
we see that the one-loop interaction at ${\cal O}(\ep^0)$ also produces
a different type of kinematical dependence as given by the factor
$j(1,2) \cdot p_i \,q_{12}^2/( p_i\cdot q_{12} \,q_{12 \perp ij}^2)$.
This rational factor has no collinear singularity at $q_{12}^2 \to 0$, but it
potentially leads to a singularity in the limit $q_{12 \perp ij}^2 \to 0$.
This is a `transverse-momentum singularity', since the kinematical variable
$\sqrt {q_{12 \perp ij}^2}$ in Eq.~(\ref{q12t}) is the transverse component of
the momentum $q_{12}$ of the soft $q{\bar q}$ pair with respect to the momenta
$p_i$ and $p_j$ of the colour-correlated hard partons in a reference frame in
which $p_i$ and $p_j$ are back-to-back.

The transverse-momentum singularity in the current is partly screened by the
logarithmic function $L_{ij}$, and we have
\beq
\label{sing}
\frac{1}{q_{12 \perp ij}^2}L_{ij}
{\underset{\;\;q_{12 \perp ij} \to 0}{\simeq}}
\;\;\,\frac{1}{q_{12 \perp ij}^2} \left[ 2\pi i \;{\rm sign}(q_{12}^2) \;
\Theta\!\left( \frac{-p_i \cdot q_{12}}{p_i \cdot p_j}\right)
\Theta\!\left( \frac{-p_j \cdot q_{12}}{p_i \cdot p_j}\right)
+ {\cal O}\!\left( \frac{q_{12 \perp ij}^2}{q_{12}^2}\right) \right] \;\;,
\eeq
which shows that the current has a one-loop singularity of absorptive origin.
Considering the physically most relevant kinematical region in which the soft
quark and antiquark are produced in the final state
$(q_1^0 >0 , q_2^0 >0)$, Eq.~(\ref{sing}) becomes
\beq
\label{sing2}
\frac{1}{q_{12 \perp ij}^2}L_{ij}
{\underset{\;\;q_{12 \perp ij} \to 0}{\simeq}}
\;\;\,\frac{1}{q_{12 \perp ij}^2}
\;2\pi i \; \Theta(-p_i^0) \;\Theta(-p_j^0) \;\;, \quad \;\;\;\;
(q_1^0 >0 , \;q_2^0 >0) \;,
\eeq
which shows that the transverse-momentum singularity is present in the
scattering amplitude of a physical
process in which the final-state soft $q{\bar q}$ pair is produced 
by the collision of the
hard partons $i$ and $j$ in the initial state.
We remark that this singularity has a pure quantum mechanics (loop) origin,
and it occurs in the limit  $q_{12 \perp ij}^2 \to 0$ even if the transverse
momenta $q_{1 \perp ij}$ and $q_{2 \perp ij}$ 
($q_{k \perp ij}^2= 2 p_i\cdot q_k \;p_j\cdot q_k/p_i\cdot p_j ,\, k=1,2$)
of the soft quark and antiquark are separately large (i.e., they are separately
non-vanishing) and $q_{12}^2$ is large.
We also note that, setting $q_{12 \perp ij} = 0$ at fixed non-vanishing values of
$q_{12}^2$ and $q_{1 \perp ij}$ (or $q_{2 \perp ij}$), we have
$\bigl( \ell_{i1} - \ell_{j2} \bigr) =
- \bigl( \ell_{j1} - \ell_{i2} \bigr)$.
Therefore, in the limit $q_{12 \perp ij}^2 \to 0$ 
the factor $\ell_{i1} - \ell_{j2}$ is (approximately) antisymmetric
with respect to the exchange $p_i \leftrightarrow p_j$ and this implies that 
we can perform the following replacement in Eq.~(\ref{j1fin}):
\beq
\label{qtsinnonab}
\btq^{a} \btq^{b} \,T^{a}_i \, T^{b}_j
\frac{L_{ij}}{q_{12 \perp ij}^2}
\bigl( \ell_{i1} - \ell_{j2} \bigr) 
{\underset{\;\;q_{12 \perp ij} \to 0}{\;\longrightarrow}}
\,- f^{abc} \btq^{c} \,T^{a}_i \, T^{b}_j
\;\,\frac{\pi \, \Theta(-p_i^0) \,\Theta(-p_j^0)}{q_{12 \perp ij}^2}
\,\bigl( \ell_{i1} - \ell_{j2} \bigr)
\;, \quad \!
(q_1^0, q_2^0 >0)
\,,
\eeq
and it follows that, in the kinematical region with $q_1^0 > 0$ and $q_2^0 >0$,
the transverse-momentum singularity has a {\em purely} non-abelian character
(see the factor $f^{abc}$ in the right-hand side of the relation
(\ref{qtsinnonab})).

As we have just discussed, the singularity of the soft-$q{\bar q}$ current in the limit
$q_{12 \perp ij} \to 0$ originates from one-loop interactions of the two
soft partons. Therefore, we expect the presence of the transverse-momentum singularity
also in the case of double soft-gluon emission at one-loop level. The one-loop
double-gluon current computed in Ref.~\cite{Zhu:2020ftr} indeed shows singular terms at 
$q_{12 \perp ij} \to 0$.

We have also computed the soft-$q{\bar q}$ one-loop current 
$\bj^{(1)}$ by explicitly evaluating its dependence on the RS parameter 
$\delta_R$ to {\em all} orders in $\ep$. Using the notation of Eq.~(\ref{j1rsdef})
and considering the rescaled currents in Eqs.~(\ref{treejhat}) and 
(\ref{onedivfin}), we find the result
\beq
\label{onejhatrs}
{\hbj}^{(1, \,{\rm fin})}_{\rs}(q_1,q_2) = - \,\left( \delta_R -1 \right)
\frac{1}{1-\ep} \left[ C_F - C_A \;\frac{1-4\ep+2\ep^2}{(1-2\ep) (3-2\ep)}
\right] \hbj^{(0)}(q_1,q_2) \;\;.
\eeq
Note that the $\delta_R$ dependence at one-loop order is completely factorized
with respect to $\bj^{(0)}$. We also note that this factorized structure and the
explicit expression of the $\ep$-dependent factor in Eq.~(\ref{onejhatrs})
are exactly equal to the corresponding RS dependence of the splitting function,
${\rm Split}(g \to q(q_1) {\bar q}(q_2))$, of one-loop scattering amplitudes
for radiation of a $q {\bar q}$ pair in the collinear limit
\cite{Bern:1999ry, Kosower:1999rx, Sborlini:2013jba}.

The one-loop current $\bj^{(1)}$ for 
soft-$q{\bar q}$ emission has been independently computed in Ref.~\cite{Zhu:2020ftr}, 
and the corresponding result is presented in Sect.~3.3 therein.
We first note that the one-loop result of  Ref.~\cite{Zhu:2020ftr} 
differs from our result
already at the level of $\ep$-pole contributions. However,
we also note that we can remove such 
difference
by adjusting the relative size
of the four contributions in the right-hand side of Eq.~(3.20) of Ref.~\cite{Zhu:2020ftr}.
More precisely, we modify the size of ${\cm}^{\,{\rm s.l.}}$  by applying the replacement
$\left( -\frac{4}{N_c} + \frac{N_c}{2} \right) \to \left( -C_F + \frac{N_c}{2} \right) = \frac{1}{2N_c}$
to its colour coefficient (see the line 10 of Eq.~(3.21)). We have contacted the author of 
Ref.~\cite{Zhu:2020ftr} and he agreed with this correction. Performing such replacement, 
we have explicitly checked that the expression of $\bj^{(1)}$ in Eq.(3.20) of Ref.~\cite{Zhu:2020ftr} 
agrees with our result (modulo the overall normalization of the one-loop current, which is not 
clearly specified in Ref.~\cite{Zhu:2020ftr}) for both the $\ep$-pole terms 
and the finite contributions at ${\cal O}(\ep^0)$. However, we note that this check and comparison
involve some `limitations'. The explicit result of  Ref.~\cite{Zhu:2020ftr}   only refers
to the `time-like' region, namely to the kinematical region in which the soft 
partons and all the hard partons are physically produced in the final state.
Moreover, the result of Ref.~\cite{Zhu:2020ftr} is specified for fixed (four dimensional)
helicities of the soft quark and antiquark, and the comparison with our result
requires the repeated use of the Schouten identity (which, precisely speaking,
is valid only in four space-time dimensions) for the product of helicity spinors.

\section{Soft $q{\bar q}$ radiation: squared amplitudes and current}
\label{sec:square}

Using the colour+spin space notation of Sect.~\ref{sec:softfact}, 
the squared amplitude $|\cm|^2$ (summed
over the colours and spins of its external legs) is written as follows
\beq
\label{squared}
|\cm|^2 = \langle{\cm} \ket{\,\cm} \;\;.
\eeq
Accordingly, the square of the soft-emission factorization formula in Eq.~(\ref{softfact})
gives
\beq
\label{softsquared}
| \cm(q_1, \dots,q_m,p_1,\dots,p_n) |^2 \simeq
\bra{\cm(p_1,\dots,p_n)} \;| \bj(q_1, \dots,q_m)\, |^2 \;
\ket{\cm(p_1,\dots,p_n)} 
 \;\;,
\eeq
where, analogously to Eqs.~(\ref{treef}) and (\ref{onef}),
the symbol $\simeq$ means that we have neglected contributions that are subdominant in the soft 
multiparton limit (i.e., the contributions that are denoted by the dots on the right-hand side of Eq.~(\ref{softfact})).
In the right-hand side of Eq.~(\ref{softsquared}),  $| \bj |^2$ denotes the all-loop squared current summed 
over the colours $\{ c_1,\dots,c_m \}$ and spins $\{ s_1,\dots,s_m \}$ of the soft partons:
\beeq
\label{spincolour}
| \bj(q_1,\dots,q_m) |^2 &=& \left[ J^{\,c_1, \dots, c_m}_{\,s_1, \dots, s_m}(q_1,\dots,q_m) \right]^\dagger J^{\,c_1, \dots, c_m}_{\,s_1, \dots, s_m}(q_1,\dots,q_m) 
\nonumber \\
&\equiv& \left[ \bj(q_1,\dots,q_m) \right]^\dagger \,\bj(q_1,\dots,q_m) \;\;.
\eeeq
The squared current $| \bj|^2 $ is a colour operator that depends on the colour charges
(and momenta) of the hard partons in $\cm(p_1,\dots,p_n)$. These colour charges produce colour correlations and,
therefore, the right-hand side of Eq.~(\ref{softsquared}) is not proportional to
$|\cm(p_1,\dots,p_n)|^2$ in the case of a generic scattering amplitude\footnote{Colour correlations 
can be simplified in the case of scattering amplitudes with two and three hard partons (see Sect.~\ref{sec:23hard}).}. 
As remarked on in Sect.~\ref{sec:softfact}, $\bj$ is simply proportional to the unit operator 
in the spin subspace of the hard partons. Therefore, we note that the squared current
$|\bj|^2$ of Eq.~(\ref{spincolour}) still applies to {\em spin-polarized} hard-scattering processes,
namely, to processes in which the spin polarizations of the {\em hard} partons are fixed (rather
than summed over).
Obviously, Eqs.~(\ref{squared})--(\ref{spincolour}) can also be properly generalized to the case in which
the spin polarizations of one or more soft partons are fixed.

In the following part of this Section, we only consider soft-$q{\bar q}$ radiation and the
corresponding soft current $\bj(q_1,q_2)$ (see Eq.~(\ref{treeqq}) and Sect.~\ref{sec:qqcur}). 
We define the loop expansion of the squared current as follows
\beq
\label{squaredexp}
| \bj(q_1,q_2) |^2 \equiv 
\left( \g \,\mu^\ep \right)^4 \, | \hbj(q_1,q_2) |^2_{(0 \ell)}
+ \left( \g \,\mu^\ep \right)^6 \left(| q_{12}^2 |\right)^{\!-\ep} \,
c_{\Gamma} \; | \hbj(q_1,q_2) |^2_{(1 \ell)} +{\cal O}(\g^8) \;\;,
\eeq
where $| \hbj |^2_{(0 \ell)}$ and $| \hbj |^2_{(1 \ell)}$ are the tree-level 
(0 loop) and one-loop rescaled contributions to $|\bj|^2$, respectively.

\subsection{The tree-level squared current}
\label{sec:j2tree}

The tree-level squared current in Eq.~(\ref{squaredexp}) is
\beq
\label{j2tree}
| \hbj(q_1,q_2) |^2_{(0 \ell)} = 
\left[ \hbj^{(0)}(q_1,q_2) \right]^\dagger \,\hbj^{(0)}(q_1,q_2)
\;\;,
\eeq
where $\hbj^{(0)}$ is the rescaled current in Eqs.~(\ref{treeqq}) and (\ref{treejhat}).
The computation of the right-hand side of Eq.~(\ref{j2tree}) is straightforward and 
the explicit result was first presented in Sect.~3.2 of Ref.~\cite{Catani:1999ss}.
We have 
\beq
\label{j2cg}
| \hbj(q_1,q_2) |^2_{(0 \ell)} = T_R \, \sum_{i,j \in H} \; \bt_i \cdot \bt_j
\;\; {\cal I}_{ij}(q_1,q_2) \,
\;\;,
\eeq
where the momentum-dependent function ${\cal I}_{ij}(q_1,q_2)$ is
(see Eq.~(96) in Ref.~\cite{Catani:1999ss})
\beq
\label{Iij1}
{\cal I}_{ij}(q_1,q_2)= \f{(p_i \cdot q_1)\, (p_j \cdot q_2)
+ (p_j \cdot q_1)\, (p_i \cdot q_2) - (p_i \cdot p_j) 
\,(q_1 \cdot q_2)}{(q_1 \cdot q_2)^2 
\,(p_i\cdot q_{12})\, (p_j \cdot q_{12})} \;\;.
\eeq

Using colour charge conservation (see Eq.~(\ref{colcon})), 
the tree-level squared current $ | \hbj|^2_{(0 \ell)}$
can be recast in the following different form
\beq
\label{hj2tree}
| \hbj(q_1,q_2) |^2_{(0 \ell)} \;\eqcs\, - \frac{1}{2} \,T_R \, 
\sum_{\substack{i,j \,\in H \\ i \,\neq\, j}}
\; \bt_i \cdot \bt_j
\; \;\cw_{ij}(q_1,q_2) \,
\;\;,
\eeq
where the soft function $\cw_{ij}$ is
\beq
\label{wij}
\cw_{ij}(q_1,q_2) = {\cal I}_{ii}(q_1,q_2) + {\cal I}_{jj}(q_1,q_2)
-2 \,{\cal I}_{ij}(q_1,q_2) \;\;.
\eeq
The expressions in the right-hand side of Eqs.~(\ref{j2cg}) and (\ref{hj2tree})
are not identical at the algebraic level, but they are fully equivalent by acting onto
scattering amplitudes (or, generically, colour-singlet states).
The expression in Eq.~(\ref{hj2tree})  has a more straightforward physical interpretation,
since the function $\cw_{ij}(q_1,q_2)$ is directly related (see Sect.~\ref{sec:2hard})
to the intensity of soft-$q{\bar q}$ radiation from two hard partons, $i$ and $j$,
in a colour-singlet configuration.

The tree-level squared current in Eqs.~(\ref{j2cg}) or (\ref{hj2tree}) produces 
{\em two-particle} correlations between the hard partons. Their colour structure
has the form of {\em dipole} contributions $\bt_i \cdot \bt_j$. We note that the 
momentum-dependent functions ${\cal I}_{ij}(q_1,q_2)$ and $\cw_{ij}(q_1,q_2)$
are {\em symmetric} with respect to the exchange $q_1 \leftrightarrow q_2$
(they are also symmetric with respect to $p_i \leftrightarrow p_j$).
In contrast, our result for the one-loop squared current (see Sect.~\ref{sec:j2one})
produces both two-particle and three-particle correlations and, moreover,
it involves also an antisymmetric dependence on the momenta $q_1$ and $q_2$.

\subsection{The one-loop squared current}
\label{sec:j2one}

The one-loop squared current in Eq.~(\ref{squaredexp}) is
\beq
\label{j2one}
| \hbj(q_1,q_2) |^2_{(1 \ell)} = 
\left( \frac{-q_{12}^2 - i0}{| q_{12}^2 |} \right)^{\!\!-\ep}
\left[ \hbj^{(0)}(q_1,q_2) \right]^\dagger \,\hbj^{(1)}(q_1,q_2) +
\;{\rm h.c.}  \;\;,
\eeq
where `h.c' denotes the hermitian-conjugate contribution, and the rescaled
currents $\hbj^{(0)}$ and $\hbj^{(0)}$ are defined in Eqs.~(\ref{treejhat}) and (\ref{onejhat}).

The explicit computation of Eq.~(\ref{j2one}) produces some contributions
that involve the fully-symmetric colour tensor $d^{abc}$,
\beq
d^{abc} = \frac{1}{T_R} \; {\rm Tr}\left(\{t^a, t^b\}\, t^c \right) \;\;.
\eeq
with indices $\{a,b,c\}$ in the adjoint representation of $SU(N_c)$.
The presence of $d^{abc}$ is a distinctive feature of (squared) currents
for radiation of soft quarks and antiquarks. 

Using $d^{abc}$ we also define the $d$-conjugated (quadratic) charge operator
${\widetilde \bd}_i$  of  the parton $i$ as follows
\beq
\label{dtildei}
{\widetilde D}_i^a \equiv d^{abc} \,T^b_i \,T^c_i \;\;.
\eeq
Performing the $SU(N_c)$ colour algebra, we explicitly find
\beeq
\label{dtildeq}
i=q &:& \quad {\widetilde D}_i^a = \frac{1}{2} \, 
d_A
\,T^a_i \;\;,\\
\label{dtildeqbar}
i={\bar q} &:& \quad {\widetilde D}_i^a = - \frac{1}{2} \, 
d_A
\,T^a_i \;\;,\\
\label{dtildeg}
i=g &:& \quad {\widetilde D}_i^a = \frac{1}{2} \,C_A \, 
D^a_i \;\;, \;\;\;\;\;\;\; \bra{b} D^a \ket{c} = d^{bac} \;\;,
\eeeq
where we have used
\beq
\label{dacoef}
d^{abc} d^{dbc} = d_A \; \delta^{ad} \;\;, 
\quad \quad \quad d_A = \frac{N_c^2 - 4}{N_c} \;\;.
\eeq
Note that the tensor $d^{abc}$ is {\em odd} under charge conjugation. This fact is responsible
for the opposite overall sign 
between
the $d$-charge ${\widetilde \bd}_i$ and the colour charge $\bt_i$
of quarks and antiquarks (see Eqs.~(\ref{dtildeq}) and (\ref{dtildeqbar})). 
Analogously, in the gluon case the $d$-charge 
$\bra{b} D^a _i \ket{c}$ in Eq.~(\ref{dtildeg}) is symmetric with respect to
$b \leftrightarrow c$, while the colour charge
$\bra{b} T^a_i \ket{c}$ is antisymmetric with respect to
$b \leftrightarrow c$.

The explicit expression of the one-loop squared current $| \hbj(q_1,q_2) |^2_{(1 \ell)} $
is obtained by inserting 
$\hbj^{(0)}$ (see Eqs.~(\ref{treeqq}) and (\ref{treejhat})) and $\hbj^{(1)}$
(see Eqs.~(\ref{onedivfin})--(\ref{j1fin})) in the right-hand side of Eq.~(\ref{j2one}),
and by performing the sum over the colours and spins of the soft quark and antiquark.
We find the following result:
\beeq
&& \!\!\!\!\!\!\!\!\!\!
| \hbj(q_1,q_2) |^2_{(1 \ell)}  =  - \frac{1}{2} \,T_R \, 
\sum_{\substack{i,j \,\in H \\ i \,\neq\, j}} \left[
\; \bt_i \cdot \bt_j \;\; \cw^{[S]}_{ij}(q_1,q_2)
+  {\widetilde \bd}_i \cdot \bt_j \;\; \waij(q_1,q_2)
\right] \nn \\
&& \!\!\!\!\!\!\!\!\!\!
 - \, T_R \!\!\sum_{\substack{i,j,k \,\in H \\ {\rm dist.} \{i,j,k\}}}\!
T^a_i \,T^b_j \,T^c_k
 \left[ 
f^{abc} \,F^{[S]}_{ijk}(q_1,q_2) + 
d^{abc} \Bigl( F^{[A]}_{ijk}(q_1,q_2) -\frac{1}{2} F^{[A]}_{iji}(q_1,q_2) -\frac{1}{2} F^{[A]}_{ijj}(q_1,q_2)\Bigr)\! 
\right] \,, \nn \\
&& \!\!\!\!\!\!\!\!
\label{j1squared}
\eeeq
which is valid to arbitrary orders in the $\ep$ expansion. The $\ep$ dependence is embodied
in the $c$-number functions $\cw^{[S]}$, $\cw^{[A]}$, $F^{[S]}$ and $F^{[A]}$.
The dependence on the colours of the hard partons is due to the colour charges $T^a_i$ and
${\widetilde D}^a_i$.
The structure of Eq.~(\ref{j1squared}) involves contributions with both {\em two} hard-parton correlations
and {\em three} hard-parton correlations. In the case of three hard-parton correlations,
the subscript `${\rm dist.} \{i,j,k\}$' in 
$\sum_{\substack{i,j,k \,\in H \\ {\rm dist.} \{i,j,k\}}}$
denotes the sum over {\em distinct} hard-parton indices $i,j$ and $k$
(i.e., $i \neq j, j\neq k, k\neq i$).

The functions $\cw^{[S]}_{ij}$, $\waij$,  $F^{[S]}_{ijk}$ and $F^{[A]}_{ijk}$ in 
Eq.~(\ref{j1squared}) depend on the momenta of the hard partons and on the momenta 
$q_1$ and $q_2$ of the soft quark and antiquark. The superscript $[S]$ in 
$\cw^{[S]}_{ij}$ and  $F^{[S]}_{ijk}$ denotes the fact that these functions are
{\em symmetric} under the exchange $q_1 \leftrightarrow q_2$ of the momenta of the soft quark and antiquark:
\beq
\cw^{[S]}_{ij}(q_1,q_2) = \cw^{[S]}_{ij}(q_2,q_1) \;, \;\;\;\;\;
F^{[S]}_{ijk}(q_1,q_2) = F^{[S]}_{ijk}(q_2,q_1) \;\;.
\eeq
Analogously, the superscript $[A]$ in $\waij$ and  $F^{[A]}_{ijk}$ highlights
the fact that these functions are {\em antisymmetric} under the exchange $q_1 \leftrightarrow q_2$:
\beq
\waij(q_1,q_2) = - \waij(q_2,q_1) \;, \;\;\;\;\;
F^{[A]}_{ijk}(q_1,q_2) = - F^{[A]}_{ijk}(q_2,q_1) \;\;.
\eeq
Therefore, $\cw^{[A]}$ and $F^{[A]}$ produce a quark--antiquark {\em charge asymmetry}
in the one-loop squared current. We note that the charge-asymmetry contributions appear 
in Eq.~(\ref{j1squared}) with the associated colour factors  
${\widetilde \bd}_i \cdot \bt_j =  d^{abc} T^a_i \,T^b_i \,T^c_j$ and 
$d^{abc} T^a_i \,T^b_j \,T^c_k$ that have a linear dependence on the colour tensor
$d^{abc}$ (which is odd under charge conjugation).
The charge-asymmetry contributions to $| \bj(q_1,q_2) |^2$ have a {\em quantum} origin
and are characteristic of the radiation of soft quark--antiquark pairs
(the squared current $| \bj(q_1,\dots,q_m) |^2$ for radiation of $m$ soft gluons is instead
fully symmetric with respect to the soft-gluon momenta $q_1,\dots,q_m$).

We present the explicit result of the $\ep$ expansion of the functions 
$\cw^{[S]}$, $\cw^{[A]}$, $F^{[S]}$ and $F^{[A]}$ up to ${\cal O}(\ep^0)$. More precisely,
we limit ourselves to presenting the expressions of these functions in the kinematical region
where $q_1^0 >0$ and $q_2^0 > 0$ (i.e., the soft quark and antiquark are produced in the
physical final state), which is the most relevant physical 
region\footnote{Expressions in other kinematical regions can be obtained by
using the fully general one-loop current in Eqs.~(\ref{j1div}) and (\ref{j1fin}).}.
In this region, the squared current depends (see Eqs.~(\ref{j1div}) and (\ref{j1fin}))
on the logarithms $\ell_{i1} \pm \ell_{j2}$ (which are purely real, independently of whether
the momenta $p_i$ and $p_j$ are physically incoming or outgoing)
and on the real part  $L_{ijR}$ and discontinuity $\Theta^{({\rm in})}_{ij}$
of the logarithm $L_{ij}$. We have (see Eqs.~(\ref{loghs}) and (\ref{loghh}))
\beq
\ell_{i1} + \ell_{j2}= \ln 
\frac{(p_i \cdot q_1) (p_j \cdot q_2)}{(p_i \cdot q_{12}) (p_j \cdot q_{12})} \,,
\; \ell_{i1} - \ell_{j2}=  \ln 
\frac{(p_i \cdot q_1) (p_j \cdot q_{12})}{(p_i \cdot q_{12}) (p_j \cdot q_{2})} 
\,,
\;
L_{ij} = L_{ijR} + 2 i \pi \,\Theta^{({\rm in})}_{ij} \;,
\eeq
where
\beq
\label{lijr}
L_{ijR} = \ln
\frac{(p_i \cdot q_{12}) (p_j \cdot q_{12})}{(p_i \cdot p_j) (q_1 \cdot q_2)}
= \ln \left(1+ \frac{q^2_{12 \perp ij}}{q_{12}^2} \right)
\;, \;\;\;\Theta^{({\rm in})}_{ij} \equiv \Theta(-p_i^0) \,\Theta(-p_j^0)
\;\;.
\eeq

The function $\cw^{[S]}_{ij}$ has the following expression in the region where
$q_1^0 >0$ and $q_2^0 > 0$:
\beeq
&& \!\!\!\!
\cw^{[S]}_{ij}(q_1,q_2) = \Bigl\{
\cw_{ij}(q_1,q_2) \; \Bigl[
- C_F \Bigl( \frac{2}{\ep^2} + \frac{3}{\ep} - \pi^2 +  8 + (\delta_R -1) \Bigr)
- \frac{4}{3} \, T_R \, N_f \,\Bigl( \frac{1}{\ep} +\frac{5}{3} \Bigr) \Bigr. \Bigr. \nn \\
&& +\, \Bigl. 
\frac{1}{3} \,C_A \Bigl( \frac{11}{\ep} + \frac{76}{3} - \pi^2 + (\delta_R -1) \Bigr)
+ \frac{1}{2} \,C_A 
\Bigl( \,\frac{2}{\ep} \left( L_{ijR} + \ell_{i1} + \ell_{j2} \right)
- L^2_{ijR} - (\ell_{i1} - \ell_{j2})^2 \Bigr)
\Bigr] \nn \\
&&  - \Bigl. \, C_A \, \bigl[ \, {\cal I}_{ii}(q_1,q_2) - {\cal I}_{jj}(q_1,q_2) \, \bigr] \,
\frac{q_{12}^2}{q_{12 \perp ij}^2} \,L_{ijR} \,\bigl( \ell_{i1} - \ell_{j2} \bigr) 
+ {\cal O}(\ep) \Bigr\} + (q_1 \leftrightarrow q_2) \;\;.
\label{w1s}
\eeeq
This function (which is {\em symmetric} under the exchange $i \leftrightarrow j$) 
controls the size of the one-loop radiative corrections to the tree-level
colour dipole correlations $\bt_i \cdot \bt_j$.

We note that $\cw^{[S]}_{ij}$ also depends on colour coefficients, while
$F^{[S]}_{ijk}$, $\waij$ and $F^{[A]}_{ijk}$ only depends on parton momenta.

The function $F^{[S]}_{ijk}$ is associated with non-abelian three-particle
correlations with colour charge factor $f^{abc} T^a_i \,T^b_j \,T^c_k$.
In the region where $q_1^0 >0$ and $q_2^0 > 0$, we have the explicit result
\beeq
F^{[S]}_{ijk}(q_2,q_1) &=& 2 \pi \,{\cal I}_{ki}(q_1,q_2) \;
\Bigl\{
L_{ijR} + \ell_{i1} + \ell_{j2} 
\Bigr. \nn \\
&+& \Bigl.
\Theta^{({\rm in})}_{ij}
\Bigl[ \,
2 \Bigl( \frac{1}{\ep} - L_{ijR} \Bigr) 
- 2 \frac{q_{12}^2}{q^2_{12 \perp ij}} (\ell_{i1} - \ell_{j2})
\Bigr]
+ {\cal O}(\ep)
\Bigr\}
 + (q_1 \leftrightarrow q_2) \;\;.
\label{f1s}
\eeeq

The charge-asymmetry contributions to Eq.~(\ref{j1squared}) can be expressed through
the function $F^{[A]}_{ijk}$. In the region where $q_1^0 >0$ and $q_2^0 > 0$ we have
\beeq
F^{[A]}_{ijk}(q_2,q_1) &=& \Bigl\{ \,
{\cal I}_{ki}(q_1,q_2) \; \Bigl[ \,-\frac{2}{\ep} \;(\ell_{i1} + \ell_{j2})
+ (\ell_{i1} - \ell_{j2})^2
+ 2 \, \frac{q_{12}^2}{q_{12 \perp ij}^2} \,L_{ijR} \,\bigl( \ell_{i1} - \ell_{j2} \bigr)
\Bigr]
\Bigr. \nn \\
&+&  \Bigl. {\cal O}(\ep) \Bigr\} - (q_1 \leftrightarrow q_2) \;\;.
\label{fant}
\eeeq
At arbitrary orders in the $\ep$ expansion,
the two-particle correlation function $\waij$ is directly related to
$F^{[A]}_{ijk}$ as follows
\beq
\label{wagen}
\waij(q_1,q_2)
=  \left[ \, F^{[A]}_{iji}(q_1,q_2) 
+ F^{[A]}_{jii}(q_1,q_2)
\right] - \bigl(  i \leftrightarrow j \bigr)  \;\;.
\eeq
In contrast to $\cw^{[S]}_{ij}$, we note that $\waij$ is {\em antisymmetric}
under the exchange $i \leftrightarrow j$ of the hard-parton momenta.
In particular, this antisymmetry of $\waij$ implies that in the sum over $i$ and $j$
of Eq.~(\ref{j1squared}) we can replace ${\widetilde \bd}_i \cdot \bt_j$ by its 
antisymmetric component, namely, ${\widetilde \bd}_i \cdot \bt_j \to 
\bigl(  {\widetilde \bd}_i \cdot \bt_j - {\widetilde \bd}_j \cdot \bt_i \bigr)/2$.
Inserting Eq.~(\ref{fant}) in Eq.~(\ref{wagen}), 
$\waij$ has the following expression:
\beeq
\label{w1a}
\waij(q_1,q_2)
&=& 
\Bigl\{
\cw_{ij}(q_1,q_2) \; \Bigl[ 
- \frac{2}{\ep} \left( \ell_{i1} + \ell_{j2} \right) + (\ell_{i1} - \ell_{j2})^2 
\Bigr] \Bigr.  \\
&+& \Bigl. \,  \bigl[ \, {\cal I}_{ii}(q_1,q_2) - {\cal I}_{jj}(q_1,q_2) \, \bigr] \,
\frac{2 \,q_{12}^2}{q_{12 \perp ij}^2} \,L_{ijR} \,\bigl( \ell_{i1} - \ell_{j2} \bigr) 
+ {\cal O}(\ep) \Bigr\} - (q_1 \leftrightarrow q_2) \;\;. \nn
\eeeq


By inspection of Eqs.~(\ref{w1s})--(\ref{w1a}) we note that only the function
$F^{[S]}_{ijk}$ exhibits a discontinuity with respect to the momenta of the hard 
partons (see $\Theta^{({\rm in})}_{ij}$ in Eqs.~(\ref{lijr}) and (\ref{f1s})).
The discontinuity contributes in the kinematical region where two hard-parton
momenta $i$ and $j$ have negative time component ($p_i^0 < 0$ and  $p_j^0 < 0$),
namely, the partons $i$ and $j$ collide in the physical initial state.
This discontinuity term of the squared current in Eq.~(\ref{j1squared}) 
originates as interference between a one-loop absorptive (imaginary) contribution
and the antihermitian colour factor $i f^{abc} \,T^a_i \,T^b_j \,T^c_k$
(we recall that $i,j$ and $k$ refer to three distinct partons). Actually,
the entire term proportional to $f^{abc} \,T^a_i \,T^b_j \,T^c_k$ in Eq.~(\ref{j1squared}) 
has this origin\footnote{The one-loop squared current for single soft-gluon radiation
\cite{Catani:2000pi} has three-particle correlations of the type $f^{abc} \,T^a_i \,T^b_j \,T^c_k$,
which have an analogous origin as absorptive/colour interference.}
as absorptive/colour interference (the absorptive term being related to the 
kinematical region where $q_1^0 >0$ and $q_2^0 > 0$).

As discussed in Sect.~\ref{sec:qqcur} (see Eqs.~(\ref{sing})--(\ref{qtsinnonab})
and accompanying comments) the one-loop current of soft-$q{\bar q}$ emission has a
transverse-momentum singularity at $q_{12 \perp ij} \to 0$. This singularity has a
non-abelian character and an absorptive origin. At the level of the one-loop
squared current, this singularity does appear in the function $F^{[S]}_{ijk}$
(see the term 
$(q_{12 \perp ij}^2)^{-1} \,\Theta^{({\rm in})}_{ij} $
in Eq.~(\ref{f1s})), while it is absent in all the other contributions
(in Eqs.~(\ref{w1s}), (\ref{fant}) and (\ref{w1a})  we see the term
$(q_{12 \perp ij}^2)^{-1} \,L_{ijR} \to (q_{12}^2)^{-1}$, which is not singular
at $q_{12 \perp ij} \to 0$). Therefore, the transverse-momentum
singularity at $q_{12 \perp ij} \to 0$ contributes 
through colour correlation $f^{abc} \,T^a_i \,T^b_j \,T^c_k$
to one-loop squared amplitudes for the class of processes with
initial-state colliding partons $i$ and $j$ and {\em two} or {\em more} 
final-state hard partons (as recalled below in Eq.~(\ref{ft3}), the colour correlation vanishes
if there is only one final-state hard parton). This class of processes includes, for instance,
dijet (or heavy-quark pair) production in hadron--hadron collisions
and the transverse-momentum singularity is directly related to the transverse momentum
of the dijet system (heavy-quark pair). Interestingly, we note that this is the
same class of processes that is sensistive to effects due to the violation of 
strict collinear factorization \cite{Catani:2011st}. However, we remark on the fact
that the transverse-momentum singularity at $q_{12 \perp ij} \to 0$
and violation of strict collinear factorization are independent phenomena
(e.g., the singularity at $q_{12 \perp ij} \to 0$ is not due to
violation of strict factorization in the one-loop collinear limit of three partons, such as
the soft quark and antiquark and a hard parton $i$ or $j$).

\setcounter{footnote}{2}
Regarding three-particle correlations of the type $f^{abc} \,T^a_i \,T^b_j \,T^c_k$
with three distinct hard partons, we also recall two general features.  As first noticed
in Ref.~\cite{Catani:2000pi}, such colour correlations vanish by acting onto
scattering amplitudes with only three hard partons (plus additional colourless
external particles). Indeed, we have \cite{Catani:2000pi}
\beq
\label{ft3}
f^{abc} \,T^a_i \,T^b_j \,T^c_k \;\; \ket{i \, j \, k} = 0 \;\;, 
\quad  \quad \quad {\rm dist.} \{i \, j \, k \} \;,
\eeq
where $\ket{i \, j \, k}$ denotes a generic colour singlet state of three distinct
hard partons $i, j$ and $k$ (the result in Eq.~(\ref{ft3}) simply follows from the colour 
conservation relation $(\bt_i + \bt_j + \bt_k) \ket{i \, j \, k} = 0$).
As pointed out in Refs.~\cite{Forshaw:2008cq, Forshaw:2012bi}, such colour correlations vanish by considering their
expectation value onto {\em pure} QCD amplitudes at the {\em tree} level.
Namely, we have \cite{Forshaw:2008cq, Seymour:2008xr}
\beq
\label{ft3ex}
\bra{\cm^{(0)}(p_1,\dots,p_n)} \; f^{abc} \,T^a_i \,T^b_j \,T^c_k \;
\ket{\cm^{(0)}(p_1,\dots,p_n)} = 0 \;\;,
\quad \quad \quad {\rm dist.} \{i \, j \, k \} 
 \;\;,
\eeq
where $\cm^{(0)}$ is a generic scattering amplitude with {\em only} quark and gluons
external lines (and no additional colourless external particles) as obtained
by {\em tree-level} QCD interactions.
Therefore, the three-particle correlations $f^{abc} \,T^a_i \,T^b_j \,T^c_k$
contributes to the one-loop squared current in Eq.~(\ref{j1squared}) 
only (see Eq.~(\ref{ft3})) for processes with four or more hard partons and
only (see Eq.~(\ref{ft3ex})) through the introduction of either QCD loop
corrections or electroweak interactions (see Refs.~\cite{Catani:2000pi, Forshaw:2012bi})
in the hard-parton scattering amplitude $\cm(p_1,\dots,p_n)$.


We present some general comments on the charge-asymmetry contributions to
the one-loop squared current of Eq.~(\ref{j1squared}). Such contributions produce
non-vanishing effects only for specific classes of scattering amplitudes
(see the discussion below and in Sect.~\ref{sec:23hard}) and quantities that are
not invariant under charge conjugation. Obviously, due to their antisymmetry under the exchange $q_1 \leftrightarrow q_2$, the charge-asymmetry contributions in
Eq.~(\ref{j1squared}) give vanishing effects after phase-space symmetric integration
over the momenta $q_1$ and $q_2$ of the soft quark and antiquark.
At the cross section level, the charge-asymmetry contributions can give
non-vanishing effects to quantities in which the soft quark (or antiquark) is triggered,
either directly (as it can be done for bottom or charm quark) or indirectly (e.g., through its fragmentation), in the final state.
For instance, we recall that the Altarelli--Parisi splitting functions
for collinear evolution of parton densities and fragmentation functions
have a quark--antiquark charge asymmetry
\cite{Catani:2003vu, Catani:2004nc, Moch:2004pa, Mitov:2006ic},
which starts at ${\cal O}(\as^3)$ (the same perturbative order of
the soft-$q{\bar q}$ one-loop squared current) and which does not vanish
in the soft limit.

Considering $\bra{\cm(p_1,\dots,p_n)} \;| \bj(q_1, q_2)\, |^2 \;\ket{\cm(p_1,\dots,p_n)} $,
the charge-asymmetry contributions vanish if $\cm(p_1,\dots,p_n)$ is a pure multigluon
scattering amplitude, namely, if it has only gluon external lines (with no additional external
$q{\bar q}$ pairs or colourless particles). This is a general consequence of the fact that
the original (i.e., before performing the soft-$q{\bar q}$ limit) squared amplitude
$| \cm(q_1, q_2,p_1,\dots,p_n) |^2$ is charge-conjugation invariant, since its external legs are
gluons and a single $q{\bar q}$ pair (one cannot distinguish between the quark and the antiquark
at the squared amplitude level). At the purely technical level, it turns out (as it can be verified)
that the colour charge operators 
${\widetilde \bd}_i \cdot \bt_j$ and $d^{abc} \,T^a_i \,T^b_j \,T^c_k$ in Eq.~(\ref{j1squared}) 
have vanishing expectation value onto pure multigluon amplitudes.

We also note that the three-particle correlations of the type $d^{abc} \,T^a_i \,T^b_j \,T^c_k$
in Eq.~(\ref{j1squared}) contribute only for processes with {\em four} or {\em more}
hard partons. Indeed, in the case of only three hard partons we have
\beq
\label{dt3}
d^{abc} \,T^a_i \,T^b_j \,T^c_k \;\; \ket{i \, j \, k} = 0 \;\;, 
\quad  \quad \quad
( \{i \, j \, k \}=\{g \, g \, g \}, \{g \, q \, {\bar q} \})  \;,
\eeq
where the three distinct
hard partons ($i, j$ and $k$) in the colour singlet state $\ket{i \, j \, k}$ are {\em either} three gluons {\em or} a gluon and a $q{\bar q}$ pair.
The proof of Eq.~(\ref{dt3}) is given in Sect.~\ref{sec:3hard}  
(see Eqs.~(\ref{dt3cas}), (\ref{ditjcas}) and related comments).

\subsection{Processes with two and three hard partons}
\label{sec:23hard}

The soft-emission factorization formula (\ref{softsquared}) 
for squared amplitudes embodies colour correlations 
produced by the squared current $| \bj |^2$.
In the case of scattering amplitudes with two or three hard partons
(plus, necessarily, additional colourless external particles)
the colour correlations have a simplified structure.
A related discussion and some general results for multiple soft-gluon radiation can be found in Ref.~\cite{Catani:2019nqv}.
The radiation of soft-$q{\bar q}$ pairs produces additional colour correlations from charge-asymmetry contributions: their main features are discussed in this Section.

\subsubsection{Processes with two hard partons}
\label{sec:2hard}

We consider a generic scattering amplitude $\cm_{BC}(q_1,q_2,p_B,p_C)$ whose external legs are two hard partons (denoted as $B$ and $C$), a soft $q{\bar q}$ pair and additional colourless particles (which are never explicitly denoted). 
The two hard partons can be either a $q{\bar q}$ pair 
(note that we specify $B=q$ and $C={\bar q}$)
or two gluons ($\{ BC \} = \{ gg \}$).
The corresponding scattering amplitude $\ket{\cm_{BC}(p_B,p_C)}$ without the 
soft-$q{\bar q}$ pair
is a colour singlet state. There is only {\em one} colour singlet configuration of the two hard partons, and the corresponding one-dimensional colour space is generated by a single colour state vector that we denote as $\ket{B C}$.

Since the soft-$q{\bar q}$ squared current $| \bj(q_1,q_2) |^2$ conserves the colour charge of the hard partons, the state 
$| \bj |^2 \,\ket{B C}$ is also proportional to $\ket{B C}$. We write
\beq
\label{jbc}
| \bj(q_1, q_2) |^2 \;\,\ket{B C} = \ket{B C} 
\;| \bj(q_1, q_2) |^{2}_{\; BC} \;\;, 
\eeq
where $| \bj |^{2}_{\; BC}$ is a $c$-number (it is the eigenvalue
of the operator $| \bj |^2$ onto the colour state $\ket{B C}$).
Therefore, the soft-factorization formula (\ref{softsquared}) 
has the following factorized $c$-number form:
\beq
\label{sbc}
| \cm_{BC}(q_1, q_2, p_B, p_C) |^2 \simeq 
| \bj(q_1,q_2) |^{2}_{\; BC} \;\;
| \cm_{BC}(p_B, p_C) |^2  \;\;,
\eeq
with no residual correlation effects in colour space
(the dependence on $SU(N_c)$ colour coefficients is embodied in 
the $c$-number factors $| \bj |^{2}_{\; BC}$ and $| \cm_{BC} |^2$).
In this respect, the structure of Eq.~(\ref{sbc}) is similar to that of soft-photon factorization formulae in QED. We recall that a $c$-number factorization formula analogous to Eq.~(\ref{sbc}) is equally valid for multiple soft-gluon radiation
from two hard partons \cite{Catani:2019nqv}.

We note that 
Eqs.~(\ref{jbc}) and (\ref{sbc}) are valid at {\em arbitrary} loop orders in the perturbative expansion of both the squared amplitude and the squared current.
Therefore, by considering Eq.~(\ref{jbc}) 
and the loop expansion in Eq.~(\ref{squaredexp}),
we can limit ourselves to evaluate the eigenvalues ($c$-numbers) 
$| \hbj |^2_{(0 \ell) BC}$ and $| \hbj |^2_{(1 \ell) BC}$, which are
the tree-level and one-loop contributions to $| \bj |^{2}_{\; BC}$.

The tree-level squared current $| \hbj |^2_{(0 \ell)}$ in Eqs.~(\ref{squaredexp}) and (\ref{hj2tree})
depends on the colour dipole factor $\bt_B \cdot \bt_C$ and, by simply using
charge conservation ($\bt_C \,\ket{B C}= - \bt_B \,\ket{B C}$), we have  
$\bt_B \cdot \bt_C \,\ket{B C} = - \ket{B C} \,\bt_B^2$ $(\bt_B^2= C_B)$.
This leads to the tree-level result first presented in Ref.~\cite{Catani:1999ss}:
\beq
\label{jbctree}
| \hbj(q_1,q_2) |^2_{(0 \ell) BC}
 \;= T_R \; C_B \;\,\cw_{BC}(q_1, q_2) \;\;,
\eeq
where $\cw_{ij}(q_1, q_2)$ is given in Eq.~(\ref{wij})
and $C_B$ is the quadratic Casimir coefficient of the hard parton 
(either $C_B=C_F$  for $\{B C\}=\{q {\bar q}\}$, or
$C_B=C_A$  for $\{B C\}=\{gg\}$).  

The one-loop squared current in Eq.~(\ref{j1squared})
depends on the colour dipole $\bt_B \cdot \bt_C$ (as at the tree level)
and on charge-asymmetry colour correlations.

By using Eqs.~(\ref{dtildei})--(\ref{dtildeg}),
we have the following colour algebra results:
\beeq
\label{c3q}
i=q &:& \quad {\widetilde \bd}_i \cdot \bt_i = \frac{1}{2} \, 
d_A
\,C_F \;\;,\\
\label{c3qbar}
i={\bar q} &:& \quad {\widetilde \bd}_i \cdot \bt_i = - \,\frac{1}{2} \, 
d_A
\,C_F \;\;,\\
\label{c3gg}
i=g &:& \quad {\widetilde \bd}_i \cdot \bt_i = 0 \;\;.
\eeeq
We note that the operators 
${\widetilde \bd}_i \cdot \bt_i = d^{abc}T_i^a T_i^b T_i^c$ in 
Eqs.~(\ref{c3q})--(\ref{c3gg}) are proportional to the unit operator in colour space. This proportionality is actually valid for ${\widetilde \bd}_i \cdot \bt_i$
in any colour (irreducible) representation $T_i^a$, and the proportionality factor is known as cubic Casimir coefficient of $SU(N_c)$.

The action onto $\ket{B C}$ of the charge-asymmetry colour correlations in 
Eq.~(\ref{j1squared}) can be explicitly evaluated by using colour conservation
(we have ${\widetilde \bd}_B \cdot \bt_C \,\ket{B C} =
 - {\widetilde \bd}_B \cdot \bt_B \,\ket{B C}$ and 
${\widetilde \bd}_C \cdot \bt_B \,\ket{B C} =
 - {\widetilde \bd}_C \cdot \bt_C \,\ket{B C}$)
and the cubic Casimir coefficients in Eqs.~(\ref{c3q})--(\ref{c3gg}).

Combining all the contributions in Eq.~(\ref{j1squared}), we find the following final result:
\beq
\label{jbcloopq}
| \hbj(q_1,q_2) |^2_{(1 \ell) BC}
 \;= T_R \; C_F \left[ \,\cw^{[S]}_{BC}(q_1, q_2) + \frac{1}{2} \;d_A \;
\wabc(q_1, q_2) \right]
\;, \;\;\;(\{ B=q, C={\bar q} \} ) \;,
\eeq
\beq
\label{jbcloopg}
| \hbj(q_1,q_2) |^2_{(1 \ell) BC}
 \;= T_R \; C_A \;\cw^{[S]}_{BC}(q_1, q_2) 
\;, \;\;\;(\{ BC \} = \{ gg\}) \;,
\eeq
where the functions $\cw^{[S]}_{ij}$ and $\waij$ are given in 
Eqs.~(\ref{w1s}) and (\ref{w1a}), respectively

In the case of soft-$q {\bar q}$ radiation from the hard partons 
$\{ B C \} =\{ q {\bar q} \}$ (see Eq.~(\ref{jbcloopq})), we do find 
charge-asymmetry contributions in $| \hbj(q_1,q_2) |^2_{(1 \ell) BC}$.
We recall that the function $\waij(q_1, q_2)$
is antisymmetric with respect to the separate exchanges $q_1 \leftrightarrow q_2$
and $i \leftrightarrow j$. Therefore, in Eq.~(\ref{jbcloopq}) the asymmetry
in the momenta of the soft-$q {\bar q}$ pair is correlated with a corresponding asymmetry in the momenta $p_B$ and $p_C$ of the hard $q$ and $\bar q$. In particular, $| \hbj(q_1,q_2) |^2_{(1 \ell) BC}$ is invariant under the overall
exchange of fermions and antifermions (i.e. 
$\{ q_1, p_B \} \leftrightarrow \{q_2, p_C\}$), consistently with charge-conjugation invariance.

In the case of soft-$q {\bar q}$ radiation from two hard gluons, the 
one-loop result in Eq.~(\ref{jbcloopg}) shows no charge-asymmetry effects. We state that this feature persists at arbitrary orders in the QCD loop expansion.
The absence of charge-asymmetry effects follows from the fact that the $c$-number squared current $| \bj(q_1,q_2) |^{2}_{\; BC}$ for $\{ B C \} =\{ g g \}$
is entirely controlled by QCD interactions, with absolutely no dependence
(both explicitly and implicitly) on the production mechanism of the two hard gluons. Therefore, such squared current is charge-conjugation invariant
(similarly to the squared amplitude for the process $gg \to  q {\bar q}$) and
one cannot distinguish between the soft quark and antiquark.

\subsubsection{Processes with three hard partons}
\label{sec:3hard}

Before considering the explicit evaluation of the soft-$q {\bar q}$ squared
current $| \bj(q_1,q_2) |^{2}$ for processes with three hard partons,
we recall and derive some general algebraic relations for the action of colour charge correlations operators onto a generic colour singlet state $\ket{i j k}$
formed by three distinct partons $i, j$ and $k$ ($i\neq j, j \neq k, k \neq i$)
in arbitrary representations of the gauge group $SU(N_c)$. We consider
the correlations operators that appear in $| \bj(q_1,q_2) |^{2}$ up to one-loop
level, namely, $\bt_i \cdot \bt_j, f^{abc} T^a_i T^b_j T^c_k,    
{\widetilde \bd}_i \cdot \bt_j$ and $d^{abc} T^a_i T^b_j T^c_k$.

As is well known, the action of dipole factors onto $\ket{i j k}$ can be evaluated in terms of quadratic Casimir coefficients $\bt_i^2 = C_i$
(see the Appendix~A of Ref.~\cite{csdip}).
We have
\beq
\label{dipijk}
2 \;\bt_i \cdot \bt_j \;\ket{i j k} = \ket{i j k} \;\bigl( C_k - C_i - C_j \bigr)
\;\;,
\eeq
and related permutations of $i, j, k$. In particular, any generic colour singlet state $\ket{i j k}$ is an eigenstate of $\bt_i \cdot \bt_j$ or, equivalently,
the action of $\bt_i \cdot \bt_j$ onto $\ket{i j k}$ is always proportional to the unit operator in colour space. The result in Eq.~(\ref{dipijk}) simply follows from the charge conservation relation $(\bt_i + \bt_j + \bt_k) \ket{i j k} = 0$,
which also leads to the result in Eq.~(\ref{ft3}) for the operator
$f^{abc} T^a_i T^b_j T^c_k$.

Considering charge-asymmetry correlations and using colour conservation
( $T^a_k \ket{i j k} = - (T^a_i + T^a_j) \ket{i j k}$ ), we have the following relations
\beq
\label{dtcons}
{\widetilde \bd}_k \cdot \bt_k \;\ket{i j k} =
- \left( {\widetilde \bd}_k \cdot \bt_i + {\widetilde \bd}_k \cdot \bt_j \right)
\;\ket{i j k} \;\;,
\eeq
\beq
\label{dt3cons}
d^{abc} \,T^a_i \,T^b_j \,T^c_k \;\; \ket{i \, j \, k} = 
- \left( {\widetilde \bd}_i \cdot \bt_j + {\widetilde \bd}_j \cdot \bt_i \right)
\;\ket{i j k} \;\;,
\eeq
and related permutations of $i, j, k$. We note that we are dealing with seven
colour correlations operators (six two-particle correlations of the type
${\widetilde \bd}_i \cdot \bt_j$, and the three-particle correlation
$d^{abc} \,T^a_i \,T^b_j \,T^c_k$) whose action onto $\ket{i j k}$ is
`non-trivial', while the action of the three operators ${\widetilde \bd}_i \cdot \bt_i$ is directly worked out in $c$-number form in terms of cubic Casimir coefficients (see Eqs.~(\ref{c3q})--(\ref{c3gg})). Since colour conservation leads to the six linear relations (exploiting permutations) in Eqs.~(\ref{dtcons}) and (\ref{dt3cons}), all the non-trivial colour correlations can be expressed in terms of a {\em single} correlation operators. To explicitly show this, we derive the following relations. The three-particle correlation is directly related to 
cubic Casimir coefficients as follows
\beq
\label{dt3cas}
d^{abc} \,T^a_i \,T^b_j \,T^c_k \;\; \ket{i \, j \, k} = \frac{1}{3}
\;\left( {\widetilde \bd}_i \cdot \bt_i + {\widetilde \bd}_j \cdot \bt_j 
+ {\widetilde \bd}_k \cdot \bt_k \right)
\;\ket{i j k} \;\;.
\eeq
The three symmetric (with respect to $i \leftrightarrow j$) two-particle correlations are equal and directly related to 
cubic Casimir coefficients as follows
\beeq
&& \!\!\!\!\!\!\!\!\!
\left( {\widetilde \bd}_i \cdot \bt_j + {\widetilde \bd}_j \cdot \bt_i \right)
\;\ket{i j k} = 
\left( {\widetilde \bd}_j \cdot \bt_k + {\widetilde \bd}_k \cdot \bt_j \right)
\;\ket{i j k} =
\left( {\widetilde \bd}_k \cdot \bt_i + {\widetilde \bd}_i \cdot \bt_k \right)
\;\ket{i j k} \nn \\
&& = - \frac{1}{3}
\;\left( {\widetilde \bd}_i \cdot \bt_i + {\widetilde \bd}_j \cdot \bt_j 
+ {\widetilde \bd}_k \cdot \bt_k \right)
\;\ket{i j k} \;\;.
\label{dtsymcas}
\eeeq
The three antisymmetric (with respect to $i \leftrightarrow j$) two-particle
correlations fulfil two independent linear relations (which are related
through $i \leftrightarrow j$) as follows
\beq
\label{dtajk}
\Bigl[ ( {\widetilde \bd}_j \cdot \bt_k - {\widetilde \bd}_k \cdot \bt_j )
- ( {\widetilde \bd}_i \cdot \bt_j - {\widetilde \bd}_j \cdot \bt_i )
+ \frac{2}{3}\,
( {\widetilde \bd}_i \cdot \bt_i + {\widetilde \bd}_k \cdot \bt_k )
- \frac{4}{3} \,{\widetilde \bd}_j \cdot \bt_j \Bigr]
\; \ket{i j k} = 0 \;\;,
\eeq
\beq
\label{dtaki}
\Bigl[ ( {\widetilde \bd}_k \cdot \bt_i - {\widetilde \bd}_i \cdot \bt_k )
- ( {\widetilde \bd}_i \cdot \bt_j - {\widetilde \bd}_j \cdot \bt_i )
- \frac{2}{3}\,
( {\widetilde \bd}_j \cdot \bt_j + {\widetilde \bd}_k \cdot \bt_k )
+ \frac{4}{3}\, {\widetilde \bd}_i \cdot \bt_i \Bigr] 
\; \ket{i j k} = 0 \;\;.
\eeq

The derivation of Eqs.~(\ref{dt3cas})--(\ref{dtaki}) from 
Eqs.~(\ref{dtcons}) and (\ref{dt3cons}) is relatively straightforward.
For instance, Eq.~(\ref{dt3cas}) is derived by first summing Eq.~(\ref{dt3cons})
and its two independent permutations to obtain
$3 d^{abc} \,T^a_i \,T^b_j \,T^c_k \ket{i j k} = 
- [\, {\widetilde \bd}_k \cdot (\bt_i + \bt_j) + (k \leftrightarrow i)
+ (k \leftrightarrow j)\, ] \,\ket{i j k}$, and then by using Eq.~(\ref{dtcons}).
Similar algebraic operations lead to Eqs.~(\ref{dtsymcas})--(\ref{dtaki}).

In the specific cases in which $i, j, k$ are either three gluons or a gluon and a $q{\bar q}$ pair, we can use the explicit results for ${\widetilde \bd}_i \cdot \bt_i$
in Eqs.~(\ref{c3q})--(\ref{c3gg})
and, consequently, Eq.~(\ref{dt3cas}) gives 
$d^{abc} \,T^a_i \,T^b_j \,T^c_k \ket{i j k} = 0$ (this proves Eq.~(\ref{dt3}))
and from Eq.~(\ref{dtsymcas}) we obtain
\beq
\label{ditjcas}
\left( {\widetilde \bd}_i \cdot \bt_j + {\widetilde \bd}_j \cdot \bt_i \right)
\;\ket{i j k} = 0 \;\;,
\quad  \quad \quad  
( \{i \, j \, k \}=\{g \, q \, {\bar q} \}, \{g \, g \, g \})  \;,
\eeq
and related permutations of $i, j, k$.

Regarding the vanishing value of the correlations 
$f^{abc} T^a_i T^b_j T^c_k$ and $d^{abc} T^a_i T^b_j T^c_k$
in 
Eqs.~(\ref{ft3}) and (\ref{dt3}), a comment is in order. The result in 
Eq.~(\ref{ft3}) applies to arbitrary colour representations of $\{i \, j \, k \}$,
while Eq.~(\ref{dt3}) is valid (as we have specified in its derivation from
Eq.~(\ref{dt3cas})) for some types of colour representations. For instance,
in the case of $SU(N_c)$ with $N_c=3$, the colour singlet state $\ket{i j k}$
can be formed by three quarks and in such case 
$d^{abc} \,T^a_i \,T^b_j \,T^c_k \,\ket{i j k}$ does not vanish.

We summarize our general discussion on colour correlations for processes with three hard partons $i, j, k$ in arbitrary colour representations of $SU(N_c)$.
The charge-symmetric component of $| \bj(q_1,q_2) |^{2}$ up to one-loop order
is proportional to the unit operator in colour space, and it can be 
expressed in $c$-number form, in terms of quadratic Casimir coefficients
(see Eqs.~(\ref{ft3}) and (\ref{dipijk})).
The charge-asymmetry component of $| \bj(q_1,q_2) |^{2}$ at one-loop order
can eventually be expressed (see Eqs.~(\ref{dt3cas})--(\ref{dtaki}))
in terms of cubic Casimir coefficients ($c$-numbers) and a single operator
(e.g., ${\widetilde \bd}_i \cdot \bt_j$) whose action onto the colour singlet state
$\ket{i j k}$ has to be explicitly computed (the result depends on the specific state $\ket{i j k}$).

We come to explicitly discuss soft-$q{\bar q}$ radiation from scattering amplitudes
with three hard partons in the specific cases that are relevant within
perturbative QCD.
We consider a generic scattering amplitude $\cm_{ABC}(q_1,q_2,p_A,p_B,p_C)$ whose external legs are colourless particles (which are not explicitly denoted), a
soft-$q{\bar q}$ pair and three hard partons (denoted as $A, B, C$) that can be
either a gluon and a $q{\bar q}$ pair ($\{ ABC \} = \{ gq{\bar q} \}$)
or three gluons ($\{ ABC \} = \{ ggg \}$).
The corresponding scattering amplitude 
$\ket{\cm_{ABC}(p_A,p_B,p_C)}$ without the soft-$q{\bar q}$ pair
is a colour singlet state formed by the three hard partons $A, B$ and $C$.
We consider the cases $\{ ABC \} = \{ gq{\bar q} \}$ and $\{ ABC \} = \{ ggg \}$
in turn.

\noindent {\underline {\large $g\, q\, {\bar q}$ {\it case}}}

We specifically set $A=g, B=q$ and $C={\bar q}$.

There is only {\em one} colour singlet configuration of the three hard partons, 
$g q {\bar q}$, and the corresponding one-dimensional colour space is generated by a single colour state vector that we denote as $\ket{A B C}$. Therefore, we are in a situation in which we can apply the same reasoning of Sect.~\ref{sec:2hard}
(see Eqs.~(\ref{jbc}) and (\ref{sbc}) and the accompanying discussion).
The state $\ket{A B C}$ is an eigenstate of the soft-$q{\bar q}$ squared current
$| \bj(q_1, q_2) |^2$,
\beq
\label{j2abc}
| \bj(q_1, q_2) |^2 \;\,\ket{A B C} = \ket{A B C} 
\;| \bj(q_1, q_2) |^{2}_{\; ABC} \;\;, 
\;\;\; \quad (\{ ABC \} = \{ gq{\bar q} \}) \;,
\eeq
and the soft-factorization formula (\ref{softsquared}) has the following factorized form:
\beq
\label{sabc}
| \cm_{ABC}(q_1, q_2, p_A, p_B, p_C) |^2 \simeq 
| \bj(q_1,q_2) |^{2}_{\; ABC} \;\;
| \cm_{ABC}(p_A, p_B, p_C) |^2  \;,
\, \quad (\{ ABC \} = \{ gq{\bar q} \}) \;,
\eeq
where $| \bj(q_1, q_2) |^{2}_{\; ABC}$ is the $c$-number eigenvalue in 
Eq.~(\ref{j2abc}). Analogously to Eq.~(\ref{sbc}),
Eq.~(\ref{sabc}) has a $c$-number factorized form 
with no residual correlation effects in colour space
(the dependence on $SU(N_c)$ colour coefficients is embodied in 
the $c$-number factors $| \bj |^{2}_{\; ABC}$ and $| \cm_{ABC} |^2$).
We also recall that a $c$-number factorized formula analogous to 
Eq.~(\ref{sabc}) applies \cite{Catani:2019nqv}
to multiple soft-gluon radiation from the three hard partons 
$\{ ABC \} = \{ gq{\bar q} \}$.

Equations~(\ref{j2abc}) and (\ref{sabc})
are valid at {\em arbitrary} loop orders in the perturbative expansion 
of both the squared amplitude and the squared current. Therefore, considering 
Eq.~(\ref{j2abc}) and the loop expansion in Eq.~(\ref{squaredexp}),
we can directly evaluate the eigenvalues $| \hbj |^2_{(0 \ell) ABC}$
and $| \hbj |^2_{(1 \ell) ABC}$, which are
the tree-level and one-loop contributions to $| \bj |^{2}_{\; ABC}$.

The tree-level squared current $| \hbj |^2_{(0 \ell)}$ in Eqs.~(\ref{squaredexp})
and (\ref{hj2tree})
involves colour dipole correlations. Using Eq.~(\ref{dipijk}),
dipole correlations can be expressed in terms of quadratic Casimir coefficients
and this leads to the tree-level result first presented in Ref.~\cite{Catani:1999ss}:
\beeq
\label{j2abctreeq}
| \hbj(q_1,q_2) |^2_{(0 \ell) ABC}
 = && \!\!\!\!\!\!\!\!\!
T_R \,\Bigl\{ C_F \,\cw_{BC}(q_1, q_2) + \frac{1}{2} \, C_A
\left[ \cw_{AB}(q_1, q_2) + \cw_{AC}(q_1, q_2) - \cw_{BC}(q_1, q_2)\right]
\Bigr\} , \nn \\
&&\quad \quad~~~~~~~~~~~~ \quad~~~~~~~ \quad~~~~~~~~~~~~~~~~~~~~~ 
(\{ ABC \} = \{ gq{\bar q} \}) \;,
\eeeq
where $\cw_{ij}(q_1, q_2)$ is given in Eq.~(\ref{wij}).
Note that the result in Eq.~(\ref{j2abctreeq}) is symmetric under the exchange
$p_B \leftrightarrow p_C$ of the momenta of the hard quark and antiquark.

The one-loop squared current $| \hbj(q_1,q_2) |^2_{(1 \ell)}$
in Eq.~(\ref{j1squared}) has contributions with and without charge asymmetry.
Owing to Eq.~(\ref{ft3}), the charge-symmetric contributions only involve colour dipole correlations, as at the tree level.
As discussed in Eqs.~(\ref{dt3cas})--(\ref{dtaki}), the charge-asymmetry contributions require the explicit evaluation of a single correlation operator
of the type ${\widetilde \bd}_i \cdot \bt_j$. We consider the operator 
${\widetilde \bd}_B \cdot \bt_C$, whose action onto $\ket{ABC}$ can be related to the action of the dipole operator $\bt_B \cdot \bt_C$. Indeed, we have
\beq
\label{dtabcq}
{\widetilde \bd}_B \cdot \bt_C \,\ket{ABC} = 
\frac{1}{2}\, d_A \, \bt_B \cdot \bt_C \,\ket{ABC}=
\ket{ABC} \;\frac{1}{4}\, d_A (C_A - 2 C_F) \;\;, 
\quad (\{ ABC \} = \{ gq{\bar q} \}) \;,
\eeq
where we have used first Eq.~(\ref{dtildeq}) and then Eq.~(\ref{dipijk}).
We note that $\ket{ABC}$ is an eigenstate of ${\widetilde \bd}_B \cdot \bt_C$,
as expected one the basis of the general relation in Eq.~(\ref{j2abc}).
Using Eq.~(\ref{dtabcq}) and the cubic Casimir coefficients in 
Eqs.~(\ref{c3q})--(\ref{c3gg}), we can express all the charge-asymmetry colour
correlations in $c$-number form (see Eqs.~(\ref{dt3cas})--(\ref{dtaki})).
We find the following result for the eigenvalue 
$| \hbj(q_1,q_2) |^2_{(1 \ell) ABC}$ of the one-loop squared current for 
soft-$q{\bar q}$ radiation:
\beeq
\label{j2abcloopq}
&& \!\!\!\!\!\!\!\!\!\!\!\!\!
| \hbj(q_1,q_2) |^2_{(1 \ell) ABC}
 =  
T_R \,\Bigl\{ \Bigl[ C_F \,\cw^{[S]}_{BC}(q_1, q_2) + \frac{1}{2} \, C_A
\bigl( \cw^{[S]}_{AB}(q_1, q_2) + \cw^{[S]}_{AC}(q_1, q_2) - \cw^{[S]}_{BC}(q_1, q_2)\bigr)
\Bigr] \Bigr.  \nn \\
&& \!\!\!\!\!\!\!\!\!\!\!\!\! + 
\frac{1}{2}\, d_A \,
\Bigl. \Bigl[ C_F \,\wabc(q_1, q_2) - \frac{1}{2} \, C_A
\bigl( \waab(q_1, q_2) + \waca(q_1, q_2) + \wabc(q_1, q_2)\bigr)
\Bigr] \Bigr\}  \;, \\
&&\quad~~~~~~~~~ \quad~~~~~~~~~~~~~ \quad~~~~~~~~~~~~~~~~~ \quad~~~~~~~~~~~~~~~~~~~ 
(\{ ABC \} = \{ gq{\bar q} \}) \,\;, \nn
\eeeq
where the functions $\cw^{[S]}_{ij}(q_1, q_2)$ and $\waij(q_1, q_2)$ are given in Eqs.~(\ref{w1s}) and (\ref{w1a}), respectively.
We note that the charge symmetric contribution in Eq.~(\ref{j2abcloopq})
is symmetric under the exchange $p_B \leftrightarrow p_C$ of the hard quark and antiquark. The charge-asymmetry contribution in Eq.~(\ref{j2abcloopq}) is instead
{\em antisymmetric} under the exchange $p_B \leftrightarrow p_C$, in complete analogy with the corresponding contribution for soft-$q{\bar q}$ radiation
from two hard partons (see Eq.~(\ref{jbcloopq})).

\noindent {\underline {\large $g\, g\, g$ {\it case}}}

We now consider the case in which the three hard partons  $A, B$ and $C$ are gluons.
The colour singlet space spanned by the three hard gluons is two-dimensional. It is convenient to choose the basis 
formed by the orthogonal colour state vectors $\ket{(ABC)_f \,}$ and
$\ket{(ABC)_d \,}$ that are defined as follows
\beq
\label{gggstates}
\bra{\,abc\,} \left(ABC\right)_f \,\rangle \equiv i f^{abc}
\,,
\;\;\;
\bra{\,abc\,} \left(ABC\right)_d \,\rangle \equiv  d^{abc}
\,,
\;\;\; \quad (\{ ABC \} = \{ ggg \}) \;\;,
\eeq
where $a, b, c$ are the colour indices of the three gluons.
We note that the two states in Eq.~(\ref{gggstates}) have different charge conjugation.
The 
scattering amplitude $\ket{\cm_{ABC}(p_A,p_B,p_C)}$ is, in general, a linear combination 
of the colour antisymmetric state $\ket{\left( ABC \right)_f}$
and the colour symmetric state  $\ket{\left( ABC \right)_d}$, and we write
\beq
\label{gggampl}
\ket{\cm_{ABC}(p_A,p_B,p_C)} = \ket{\left( ABC \right)_f} \;\;\cm_{f}(p_A,p_B,p_C) \,+
\ket{\left( ABC \right)_d} \;\;\cm_{d}(p_A,p_B,p_C) \;\;, 
\eeq
where $\cm_f$ and $\cm_d$ are colour stripped amplitudes.
Owing to the Bose symmetry of $\ket{\cm_{ABC}}$ with respect to the three gluons,
the amplitude $\cm_{f}(p_A,p_B,p_C)$ is antisymmetric under the exchange of two gluon momenta
(e.g., $p_A \leftrightarrow p_B$), while $\cm_{f}(p_A,p_B,p_C)$ has a symmetric dependence on
$p_A,p_B,p_C$.

As examples of the scattering amplitude $\ket{\cm_{ABC}(p_A,p_B,p_C)}$, we can mention the 
three scattering processes $H \to ggg$, $\gamma \to ggg$ and $Z \to ggg$.
In the Higgs boson process $H \to ggg$ (see, e.g., Ref.~\cite{Ellis:1987xu}) the amplitude component $\cm_d$
of Eq.~(\ref{gggampl}) vanishes, while in the photon process $\gamma \to ggg$ (see, e.g., Ref.~\cite{vanderBij:1988ac})
we have $\cm_f=0$. In the case of the $Z$ boson process $Z \to ggg$ both components $\cm_f$ and $\cm_d$ are
not vanishing (see, e.g., Ref.~\cite{vanderBij:1988ac}). We also note that all these scattering amplitudes are produced through QCD interactions
involving quark loops (within the Standard Model, gluons 
have tree-level interactions only with quarks and, consequently, 
$\cm_{ABC}$ vanishes at the tree level).

We have previously discussed the case of the three hard partons $\{ ABC \} = \{ gq{\bar q} \}$, which generate
a one-dimensional colour singlet space. The fact that the colour singlet space is two-dimensional for 
$\{ ABC \} = \{ ggg \}$ is an essential difference. In particular, in the case $\{ ABC \} = \{ ggg \}$
the all-order soft-factorization formula (\ref{softsquared}) for squared amplitudes cannot be recast
in the factorized $c$-number form of Eq.~(\ref{sabc}).
The action of the squared current $| \bj |^2 $ onto $\ket{\cm_{ABC}}$ of Eq.~(\ref{gggampl})
is colour conserving, but it can produce colour correlations between the two colour singlet states
$\ket{\left( ABC \right)_f}$ and $\ket{\left( ABC \right)_d}$ of the three hard gluons. In general,
the squared soft current $| \bj |^2 $ can be represented as a $2 \times 2$ correlation matrix that
acts onto the two-dimensional space generated by $\ket{\left( ABC \right)_f}$ and $\ket{\left( ABC \right)_d}$.
The all-order structure of this correlation matrix is discussed in Ref.~\cite{Catani:2019nqv}
for the case of multiple soft-gluon radiation. In the following we explicitly consider soft-$q{\bar q}$
radiation at the tree level and one-loop order.

The tree-level squared current $| \hbj |^2_{(0 \ell)}$ in Eqs.~(\ref{squaredexp})
and (\ref{hj2tree}) only involves colour dipole correlations, whose action onto both 
$\ket{\left( ABC \right)_f}$ and $\ket{\left( ABC \right)_d}$ is proportional to the unit matrix in colour space
(see Eq.~(\ref{dipijk})). Therefore, the contribution of $| \hbj |^2_{(0 \ell)}$ to the factorization
formula (\ref{softsquared}) can be expressed in factorized $c$-number form and,
using Eq.~(\ref{dipijk}), we have (see also Ref.~\cite{Catani:1999ss})
\beeq
&&\bra{\cm_{ABC}(p_A,p_B,p_C)} \;| \hbj(q_1,q_2) |^2_{(0 \ell)} \;\ket{\cm_{ABC}(p_A,p_B,p_C)}  \nn
\\
&&= | \cm_{ABC}(p_A,p_B,p_C) |^2 \;\;\frac{T_R \,C_A}{2} \;\w0abc(q_1,q_2) \;\;,
\;\;\;\;\; \quad (\{ ABC \} = \{ ggg \}) \;\;,
\label{j03g}
\eeeq
where
\beq
\w0abc(q_1,q_2) = \cw_{AB}(q_1, q_2) + \cw_{BC}(q_1, q_2) + \cw_{CA}(q_1, q_2) \;\;,
\eeq
and $\cw_{ij}(q_1, q_2)$ is given in Eq.~(\ref{wij}). Since $\cw_{ij}$ is symmetric under
the exchange
$p_i \leftrightarrow p_j$, we note that the function $\w0abc$ has a completely symmetric 
dependence on the gluon momenta $p_A,p_B,p_C$ (as required by Bose symmetry).
We also note that Eq.~(\ref{j03g}) is valid at arbitrary orders in the loop expansion of the amplitude
$\cm_{ABC}(p_A,p_B,p_C)$.

The action of the one-loop squared current $| \hbj(q_1,q_2) |^2_{(1 \ell)}$
in Eqs.~(\ref{squaredexp}) and 
(\ref{j1squared}) onto $\ket{\cm_{ABC}}$ involves charge symmetric and 
charge-asymmetry contributions. As summarized in the discussion below Eq.~(\ref{ditjcas}),
the charge symmetric contributions are proportional to the unit matrix in colour space,
while the charge-asymmetry contributions can be expressed in terms of a single colour correlation 
operator. Specifically, by using Eq.~(\ref{dipijk})) and Eqs.~(\ref{dt3cas})--(\ref{dtaki}),
we explicitly find
\beeq
\label{j1ggg}
&&\!\!\! \!\!\!\!\!\!\!\!\!  | \hbj(q_1,q_2) |^2_{(1 \ell)} \;\ket{ABC} = T_R \,\Bigl\{ \frac{C_A}{2} \;\wsabc(q_1,q_2)
+ {\widetilde \bd}_B\cdot \bt_A \;\waabc(q_1,q_2) \Bigr\}  \ket{ABC} \;,  \\
&&\quad~~~~~~ \quad~~~~~~~~~~~~~~~~ \quad~~~~~~~~~~~~~~~~~~~ \quad~~~~~~~~~~~~~~~~~~~~~ 
(\{ ABC \} = \{ ggg \}) \;, \nn
\eeeq
where
\beq
\wsabc(q_1,q_2) = \cw^{[S]}_{AB}(q_1, q_2) + \cw^{[S]}_{BC}(q_1, q_2) + \cw^{[S]}_{CA}(q_1, q_2) \;\;,
\eeq
\beq
\waabc(q_1,q_2) = \waab(q_1, q_2) + \wabc(q_1, q_2) + \waca(q_1, q_2) \;\;,
\eeq
and the functions $\cw^{[S]}_{ij}(q_1, q_2)$ and $\waij(q_1, q_2)$ are given in Eqs.~(\ref{w1s}) and (\ref{w1a}), respectively.
We note that the charge symmetric contribution to Eq.~(\ref{j1ggg}) depends on the function
$\wsabc$ that has a fully symmetric dependence on the hard-gluon momenta $p_A,p_B,p_C$.
The charge-asymmetry function $\waabc$ is instead antisymmetric under the exchange of two gluon momenta
(e.g., $p_A \leftrightarrow p_B$).

The charge-asymmetry operator ${\widetilde \bd}_B\cdot \bt_A$ in the right-hand side of Eq.~(\ref{j1ggg})
acts differently onto the two colour states $\ket{\left( ABC \right)_f}$ and $\ket{\left( ABC \right)_d}$ of 
Eq.~(\ref{gggampl}). By explicitly performing the $SU(N_c)$ colour algebra, we find the following result:
\beq
\label{fdaction}
{\widetilde \bd}_B\cdot \bt_A \,\ket{\left( ABC \right)_f} =  \frac{C_A^2}{4}  \;\ket{\left( ABC \right)_d} \;,
\;\;\;\;
{\widetilde \bd}_B\cdot \bt_A \,\ket{\left( ABC \right)_d} = \frac{C_A \,d_A}{4}  \;\ket{\left( ABC \right)_f} \;,
\eeq
and we note that the operator ${\widetilde \bd}_B\cdot \bt_A$ produces `pure' transitions between the colour 
symmetric and colour antisymmetric states $\ket{\left( ABC \right)_f}$ and $\ket{\left( ABC \right)_d}$,
which have different charge conjugation.

Using Eqs.~(\ref{gggampl}), (\ref{j1ggg}) and (\ref{fdaction}), we obtain the final result for the contribution
of the one-loop soft-$q{\bar q}$ squared current to squared amplitudes with three hard gluons. We find
\beeq
&&
\!\!\!\!\!\!\!\!\!
\!\!\!\!\!\!\!\!\!
\bra{\cm_{ABC}(p_A,p_B,p_C)} \;| \hbj(q_1,q_2) |^2_{(1 \ell)} \;\ket{\cm_{ABC}(p_A,p_B,p_C)}  
= \frac{T_R \,C_A}{2} \nn
\\
&&
\!\!\!\!\!
\times
\;\Bigl\{ \wsabc(q_1,q_2) \; | \cm_{ABC}(p_A,p_B,p_C) |^2 \Bigr. \nn \\
\label{s3gloop}
&&+ \Bigl. \waabc(q_1,q_2) \;\frac{1}{2} C_A d_A (N_c^2-1) 
\bigl[ \, \cm^\dagger_d(p_A,p_B,p_C)  \cm_f(p_A,p_B,p_C) + {\rm h.c.} \bigr]
 \Bigr\}\;\;, \\
&& \quad~~~~~~~~~~~~~~~~~~~~~~~~~~~~~~
 ~~~~~~~~~~~~~~~~~~~~~~~~~~~~~~~~~~~~~\, (\{ ABC \} = \{ ggg \}) \;\;, \nn
\eeeq
which is not simply proportional to $| \cm_{ABC} |^2$ (unlike the corresponding result in Eq.~(\ref{sabc})
for $\{ ABC \} = \{ gq{\bar q} \})$). In contrast with the case of scattering amplitudes with two hard gluons
(see Eq.~(\ref{jbcloopg})),
we note that the expression in Eq.~(\ref{s3gloop}) involves a charge-asymmetry contribution that is not vanishing,
provided the hard-scattering amplitude includes non-vanishing components $\cm_f$ and $\cm_d$
(i.e., $\cm_{ABC}$ has no definite charge conjugation). Such feature of $\cm_{ABC}$ depends on the specific production 
mechanism of the three hard gluons.
The functions $\waabc$ and $(\cm_d^\dagger \cm_f + {\rm h.c.})$ 
are separately antisymmetric under the exchange of two gluon momenta and, consequently, their
product is symmetric. Therefore, the right-hand side of Eq.~(\ref{s3gloop})
(including its charge-asymmetry contribution) is fully symmetric under permutations of the three
hard gluons, as expected and required by Bose symmetry.

\section{Soft fermion-antifermion radiation in QED and \\
mixed QCD$\times$QED}
\label{sec:qed}

Our results in Sects.~\ref{sec:qqcur} and \ref{sec:square} for soft-$q{\bar q}$
emission can be generalized to consider the emission of a soft fermion--antifermion
($f{\bar f}$) pair through QED (photon) interactions and mixed QCD$\times$QED
(gluon and photon) interactions. Before presenting the results, we precisely specify
our framework.

The soft fermions can be either massless quarks ($f=q$) or electrically-charged 
massless leptons ($f= \ell$). We consider generic scattering amplitudes, $\cm$, 
whose external particles are massless quarks and gluons, massless leptons and, 
additionally, particles that carry no colour charge and no electric charge
(i.e., photons, Higgs and $Z$ bosons in the context of Standard Model).
The external particles (i.e., their momenta and quantum numbers) of $\cm$ are treated
as outgoing particles (as already specified in Sect.~\ref{sec:softfact} for the pure QCD case).
The internal legs of $\cm$ can include massless (photons, gluons) and massive
(e.g., heavy quarks and/or $W^\pm$ bosons) particles. If an external $f{\bar f}$ pair
becomes soft, the scattering amplitude $\cm$ is singular and the singular behaviour is due
to the production of the soft-$f{\bar f}$ pair through QCD (gluon) and QED (photon)
interactions. We formally treat QCD, QED and mixed QCD$\times$QED interactions
on equal footing. Therefore, the scattering amplitude $\cm$ has a generalized
perturbative (loop) expansion in powers of two unrenormalized couplings:
the QCD coupling $\g$ and the QED coupling $\gq$ 
($\gq^2/(4\pi) = \alpha$ is the fine structure constant at the unrenormalized level).
Regarding the RS of the UV and IR divergences, photons and charged leptons are treated
in the same way (see Sect.~\ref{sec:1loop}) as gluons and massless quarks, respectively.

\subsection{The soft-$\!f{\bar f}$ current}
\label{sec:ffcur}

The dominant singular behaviour of $\ket{\cm}$ for emission of a
soft-$f{\bar f}$ pair is given by the factorization formula in Eq.~(\ref{softfact})
through the replacement $\bj(q_1, \dots,q_m) \rightarrow \bj_{f{\bar f}}(q_1,q_2)$.
Here $\bj_{f{\bar f}}(q_1,q_2)$ is the soft current for emission of a fermion
$f$ and an antifermion $\bar f$ with momenta $q_1$ and $q_2$, respectively.
Analogously to the scattering amplitude $\cm$, the current $\bj_{f{\bar f}}$ is perturbatively 
computable by performing a loop expansion, and we write
\beq
\label{jff}
\bj_{f{\bar f}}(q_1,q_2) = \bj^{(0)} _{f{\bar f}}(q_1,q_2)
+ \bj^{(1)}_{f{\bar f}}(q_1,q_2) + \bj^{(2)}_{f{\bar f}}(q_1,q_2)  + \dots \;\;.
\eeq
Since we formally treat QCD and QED interactions on equal footing, the
$k$-th loop term $\bj^{(k)} _{f{\bar f}}$ include contributions that are proportional to powers
of both coupling constants $\g$ and $\gq$. The pure-QCD and pure-QED cases are recovered 
by setting $\{\gq=0, f=q \} $ and $\{\g=0, f=\ell \}$, respectively.

The lowest-order (tree-level) term $\bj^{(0)} _{f{\bar f}}$ of Eq.~(\ref{jff}) is
\beq
\label{treejff}
\bj^{(0)}_{f{\bar f}}(q_1,q_2) = \left( \g \,\mu^\ep \right)^2  \;\hbj^{(0)}(q_1,q_2)
+ \left( \gq \,\mu^\ep \right)^2  \;\hbj^{(0)}_{(1\gamma)}(q_1,q_2)
 \;\;,
\eeq
where $\hbj^{(0)}(q_1,q_2)$ is the rescaled current in Eqs.~(\ref{treeqq}) and (\ref{treejhat})
for soft-$q{\bar q}$ emission in QCD
(note that $\hbj^{(0)}$ vanishes if $f=\ell$). The term $\hbj^{(0)}_{(1\gamma)}$
has the following explicit expression:
\beq
\label{tree1gamma}
\hbj^{(0)}_{(1\gamma)}(q_1,q_2) = - \,e_f \; {\bf \Delta}_f \,
\sum_{i \in H} \,e_i
\;\frac{p_i \cdot j(1,2)}{p_i \cdot q_{12}} \;\;,
\eeq
where  $j^\nu(1,2)$ is the fermionic current in Eq.~(\ref{fercur}).
The current $\hbj^{(0)}_{(1\gamma)}$ is due to a single-photon interaction between the soft
fermion $f$ (with electric charge $e_f$) and the other external charged particles 
(with electric charges $e_i$), $i \in H$, of $\cm$. The charges
$e_f$ and $e_i$ are expressed in units of the positron charge 
(e.g., for the up-quark $u$ we have $e_u=+2/3$).
The factor ${\bf \Delta}_f$ in the right-hand side of Eq.~(\ref{tree1gamma}) 
is a colour operator that depends on the type
of soft fermion $f$. If $f=\ell$, we simply have ${\bf \Delta}_f= 1$. If $f=q$, 
${\bf \Delta}_f$ is the projection operator onto the colour singlet
state of the $f{\bar f}$ pair, namely, by using the colour space notation of Sect.~\ref{sec:softfact} we have
$\bra{\alpha_1,\alpha_2}  \,{\bf \Delta}_f = \delta_{\alpha_1\alpha_2}$.

The soft-$f{\bar f}$ current $\bj^{(1)}_{f{\bar f}}$ in Eq.~(\ref{jff}) is due to the one-loop
corrections (with respect to both $\g$ and $\gq$) to the tree-level current $\bj^{(0)}_{f{\bar f}}$.
We can write
\beq
\label{onejffhat}
\bj^{(1)}_{f{\bar f}}(q_1,q_2) = \left( \,\mu^\ep \right)^4 
\left(-q_{12}^2 - i0\right)^{\!-\ep} 
c_{\Gamma} \,\left[ \, \g^4  \,{\hbj}^{(1)}(q_1,q_2) +
\g^2 \,\gq^2 \,\hbj^{(1)}_{(1\gamma)} (q_1,q_2) \, 
+ \gq^4 \, \hbj^{(1)}_{(2\gamma)}(q_1,q_2)
\right] \;,
\eeq
where the rescaled currents ${\hbj}^{(1)}$, ${\hbj}^{(1)}_{(1\gamma)}$ and 
${\hbj}^{(1)}_{(2\gamma)}$ are introduced similarly to Eq.~(\ref{onejhat}).
The term ${\hbj}^{(1)}(q_1,q_2)$ in the right-hand side of Eq.~(\ref{onejffhat})
is exactly the soft-$q{\bar q}$ current of Eqs.~(\ref{onejhat})--(\ref{j1fin})
for the QCD case. The results for the terms ${\hbj}^{(1)}_{(1\gamma)}$ and 
${\hbj}^{(1)}_{(2\gamma)}$ are obtained by properly modifying the QCD result
in Eqs.~(\ref{j1div}) and (\ref{j1fin}).

The rescaled current ${\hbj}^{(1)}_{(2\gamma)}$ in Eq.~(\ref{onejffhat})
is entirely due to QED interactions, and it has the following expression:
\beeq
\!\!\!\!\!\!
{\hbj}^{(1)}_{(2\gamma)}(q_1,q_2) \!\!&=&\!\! \left[ 
- \, e_f^2 
\left(  \frac{2}{\ep^2} + \frac{3}{\ep} + 8 + (\delta_R -1) \right)   
- \frac{4}{3} N_{\rm ch.}  
\left( \frac{1}{\ep} + \frac{5}{3}  \right) \right] \hbj^{(0)}_{(1\gamma)}(q_1,q_2) \nn \\
&& {} \!\!\!\!\!\!    \!\!\!\!\!\!  \!\!\!\!\!\!    \!\!\!\!\!\!  
\!\!\!\!\!\!    \!\!\!\!\!\!  
+
\,j_\nu(1,2) 
\;e_f^2\; {\bf \Delta}_f
\sum_{\substack{i,j \,\in H \\ i \,\neq\, j}} 
e_i \,e_j  
\left[
\left(\frac{p_i^\nu}{p_i \cdot q_{12}} -
\frac{p_j ^\nu}{p_j \cdot q_{12}}
\right) \left( - \frac{2}{\ep } (\ell_{i1}+\ell_{j2}) + (\ell_{i1} - \ell_{j2})^2 \right)
\right. \nn \\
&&  {} \!\!\!\!\!\!    \!\!\!\!\!\!  \!\!\!\!\!\!    \!\!\!\!\!\!  
\!\!\!\!\!\!    \!\!\!\!\!\!  
+ \frac{q_{12}^2}{q_{12 \perp ij}^2}
\left. 
\left(\frac{p_i^\nu}{p_i \cdot q_{12}} +
\frac{p_j ^\nu}{p_j \cdot q_{12}}
\right) 2 \,L_{ij} \,\bigl( \ell_{i1} - \ell_{j2} \bigr) \right] + {\cal O}(\ep)
\;\;,
\label{j12gamma}
\eeeq
where $\hbj^{(0)}_{(1\gamma)}$ is given in Eq.~(\ref{tree1gamma}).
In Eq.~(\ref{j12gamma})
the soft fermion $f$ can be either a quark or a lepton and, similarly,
the  charged hard particles $i,j \in H$ can include quarks and leptons. 
The factor $N_{\rm ch.}$ in the right-hand side of Eq.~(\ref{j12gamma})
is analogous
to the factor $T_R N_f$ of the QCD expressions
in the right-hand side 
of Eqs.~(\ref{j1div}) and (\ref{j1fin}).
The coefficient $N_{\rm ch.}$
depends on the squared electric charges of the massless\footnote{Analogously to the QCD
case, in the one-loop current ${\hbj}^{(1)}_{(2\gamma)}(q_1,q_2)$ we have not considered
and included vacuum polarization effects due to {\em massive}
particles (charged leptons, quarks and $W^\pm$).} quarks and leptons in the theory,
and we have
\beq
N_{\rm ch.}= \sum_\ell e_{\ell}^2 + N_c \sum_q e_q^2 \;\;.
\eeq

The one-loop term $\hbj^{(1)}_{(1\gamma)}$ in Eq.~(\ref{onejffhat}) is due
to mixed QCD$\times$QED interactions. It has the following explicit expression:
\beeq
\!\!\!\!\!\!
{\hbj}^{(1)}_{(1\gamma)}(q_1,q_2) \!\!&=&\!\!  \delta_{fq} 
\left\{
\left(  \frac{2}{\ep^2} + \frac{3}{\ep} + 8 + (\delta_R -1) \right)   
\left[ \,- \,C_F \,\hbj^{(0)}_{(1\gamma)}(q_1,q_2) - e_f^2  \,\hbj^{(0)}(q_1,q_2)
\right] \right. \nn \\
&& {} \!\!\!\!\!\!    \!\!\!\!\!\!  \!\!\!\!\!\!    \!\!\!\!\!\!  
\!\!\!\!\!\!    \!\!\!\!\!\!  
+
\,j_\nu(1,2) 
\;e_f \,\btq^{c}
\sum_{\substack{i,j \,\in H \\ i \,\neq\, j}} 
\left( e_i \, T^{c}_j + e_j \, T^{c}_i \right)
\left[
\left(\frac{p_i^\nu}{p_i \cdot q_{12}} -
\frac{p_j ^\nu}{p_j \cdot q_{12}}
\right) \left( - \frac{2}{\ep } (\ell_{i1}+\ell_{j2}) + (\ell_{i1} - \ell_{j2})^2 \right)
\right. \nn \\
&&  {} \!\!\!\!\!\!    \!\!\!\!\!\!  \!\!\!\!\!\!    \!\!\!\!\!\!  
\!\!\!\!\!\!    \!\!\!\!\!\!  
\left. + \frac{q_{12}^2}{q_{12 \perp ij}^2}
\left. 
\left(\frac{p_i^\nu}{p_i \cdot q_{12}} +
\frac{p_j ^\nu}{p_j \cdot q_{12}}
\right) 2 \,L_{ij} \,\bigl( \ell_{i1} - \ell_{j2} \bigr) \right] + {\cal O}(\ep) \right\}
\;\;,
\label{oneloopj1gamma}
\eeeq
where $\hbj^{(0)}$ and $\hbj^{(0)}_{(1\gamma)}$ are the tree-level currents
in the right-hand side of Eq.~(\ref{treejff}). We note that $\hbj^{(1)}_{(1\gamma)}$
is entirely proportional to the Kronecker delta symbol $\delta_{fq}$ and,
consequently, it is not vanishing only if the soft fermion $f$ is a quark.
Therefore, if the soft fermion $f$ is a charged lepton, the total one-loop current
$\bj^{(1)}_{f{\bar f}}$ in Eq.~(\ref{onejffhat}) receives a non-vanishing contribution
only from the QED interaction term $\hbj^{(1)}_{(2\gamma)}$.

In Sect.~\ref{sec:qqcur} we have discussed the singularity at
$q_{12 \perp ij} \to 0$
of the current $\hbj^{(1)}$ for soft-$q{\bar q}$ QCD radiation 
at the one-loop level, and we have concluded that it has a purely non-abelian character.
The results for $\hbj^{(0)}_{(2\gamma)}$ and $\hbj^{(0)}_{(1\gamma)}$ are consistent
with this conclusion, since the expressions in Eqs.~(\ref{j12gamma}) and (\ref{oneloopj1gamma}) 
do not have the transverse-momentum singularity. Although
the right-hand side of Eqs.~(\ref{j12gamma}) and (\ref{oneloopj1gamma}) include the factor
$1/q_{12 \perp ij}^2$, its singular contribution at $q_{12 \perp ij} \to 0$
turns out to be antisymmetric under $i \leftrightarrow j$
(see Eqs.~(\ref{sing})--(\ref{qtsinnonab}) and accompanying comments) and it cancels 
by summing over $i,j \in H$.

\subsection{The square of the soft-$\!f{\bar f}$ current}
\label{sec:ffsquare}

The singular behaviour of squared amplitudes for soft-$f{\bar f}$ radiation 
is controlled by the square of the current in Eq.~(\ref{jff}). We have
\beq
\label{jffsquare}
| \bj_{f{\bar f}}(q_1,q_2) |^2 =
\left[ \bj^{(0)}_{f{\bar f}}(q_1,q_2) \right]^\dagger  \bj^{(0)}_{f{\bar f}}(q_1,q_2)
+\left\{ \left[ \bj^{(0)}_{f{\bar f}}(q_1,q_2) \right]^\dagger \bj^{(1)}_{f{\bar f}}(q_1,q_2)
+ \,{\rm h.c.} \right\}
+ \dots \;,
\eeq
where the dots stand for higher-loop contributions (i.e., terms of
 ${\cal O}\!\left((\g^2)^{4-n} (\gq^2)^n\right)$ with $0 \leq n \leq 4)$).

Using Eq.~(\ref{treejff}), the tree-level term in Eq.~(\ref{jffsquare}) is
\beq
\label{treejffsquare}
\left[ \bj^{(0)}_{f{\bar f}}(q_1,q_2) \right]^\dagger  \bj^{(0)}_{f{\bar f}}(q_1,q_2)
\equiv 
\left( \g \,\mu^\ep \right)^4 \, | \hbj(q_1,q_2) |^2_{(0 \ell)}
+ \left( g \,\mu^\ep \right)^4 
\; | \hbj(q_1,q_2) |^2_{(0 \ell;2\gamma)} \;\;,
\eeq
where  $| \hbj(q_1,q_2) |^2_{(0 \ell)}$ is the pure QCD contribution
given in Eq.~(\ref{hj2tree}). We note that the right-hand side of Eq.~(\ref{treejffsquare})
does not include a term proportional to $\g^2 \gq^2$ (such
QCD$\times$QED interference is proportional to $[ \hbj^{(0)} ]^\dagger  \hbj^{(0)}_{(1\gamma)}$
and it leads to an overall vanishing colour factor, ${\rm Tr} ( \btq^{c} {\bf \Delta}_f) =0$).
The term $| \hbj |^2_{(0 \ell;2\gamma)}$ in Eq.~(\ref{treejffsquare}) is due
to QED interactions, and it is 
\beeq
| \hbj(q_1,q_2) |^2_{(0 \ell; 2\gamma)} &=& \left[ \hbj^{(0)}_{(1\gamma)}(q_1,q_2) \right]^\dagger  \hbj^{(0)}_{(1\gamma)}(q_1,q_2)
\nn \\
&=&
- \left( \delta_{f\ell} + N_c \,\delta_{f q} \right) e_f^2 \;
\frac{1}{2} \, 
\sum_{\substack{i,j \,\in H \\ i \,\neq\, j}}
\; e_i \, e_j
\; \;\cw_{ij}(q_1,q_2) \,
\;\;,
\label{qedsquared}
\eeeq
where the 
function $\cw_{ij}$ is given in Eq.~(\ref{wij}).

The one-loop term in the squared current of Eq.~(\ref{jffsquare}) includes 
all possible contributions that are proportional to the powers 
$(\g^2)^{3-n} (\gq^2)^n$ with $0 \leq n \leq 3$. We write it in the following form:
\beeq
\left[ \bj^{(0)}_{f{\bar f}}(q_1,q_2) \right]^\dagger \bj^{(1)}_{f{\bar f}}(q_1,q_2)
+ \,{\rm h.c.} &=&
\left( \mu^\ep \right)^6 \left(| q_{12}^2 |\right)^{\!-\ep} \,
c_{\Gamma} \;
\Bigl\{ \g^6 \;| \hbj(q_1,q_2) |^2_{(1 \ell)} \Bigr. \nn\\
&+& \Bigl. \sum_{n=1}^3 \;(\g^2)^{3-n} \, (\gq^2)^{n} 
\;| \hbj(q_1,q_2) |^2_{(1 \ell;n\gamma)} \,
\Bigr\} \;\;,
\label{loopjff}
\eeeq
where $| \hbj(q_1,q_2) |^2_{(1 \ell)}$ is the pure QCD contribution given in Eq.~(\ref{j1squared}).
Using Eqs.~(\ref{treejff}) and (\ref{onejffhat}), the other contributions in the
right-hand side of Eq.~(\ref{loopjff}) are given in terms of the rescaled currents
$\hbj^{(0)}, \hbj^{(0)}_{(1\gamma)}, \hbj^{(1)}, \hbj^{(1)}_{(1\gamma)}$ 
and $\hbj^{(1)}_{(2\gamma)}$.

The one-loop contribution $| \hbj(q_1,q_2) |^2_{(1 \ell;3\gamma)}$ is entirely due
to QED interactions, and we explicitly obtain
\beeq
&& \!\!\!\!\!\!\!\!\!\!
| \hbj(q_1,q_2) |^2_{(1 \ell;3\gamma)} =
 \left[ \hbj^{(0)}_{(1\gamma)}(q_1,q_2) \right]^\dagger  
\hbj^{(1)}_{(2\gamma)}(q_1,q_2) + {\rm h.c.} \nn \\
&&\!\!\!\!\!
= \left( \delta_{f\ell} + N_c \,\delta_{f q} \right) e_f^2 \;
\Bigl\{ - \frac{1}{2} \, 
\sum_{\substack{i,j \,\in H \\ i \,\neq\, j}}
\; e_i \, e_j
\; \;\cw_{ij}^{[S] (f)}(q_1,q_2)
- \sum_{k \in H} \;\sum_{\substack{i,j \,\in H \\ i \,\neq\, j}}
\;e_f \,e_k\, e_i\,e_j \;2 \,F_{ijk}^{[A]}(q_1,q_2)
\Bigr\} \;,\nn \\
&&
\label{j3gamma}
\eeeq
where $F_{ijk}^{[A]}(q_1,q_2)$ is given in Eq.~(\ref{fant})
and the one-loop function $\cw_{ij}^{[S] (f)}(q_1,q_2)$ is
\beeq
\label{wsf}
\cw_{ij}^{[S] (f)}(q_1,q_2) &=&
\Bigl\{ \cw_{ij}(q_1,q_2)
\left[ 
- \, e_f^2 
\left(  \frac{2}{\ep^2} + \frac{3}{\ep} - \pi^2 + 8 + (\delta_R -1) \right)   
- \frac{4}{3} N_{\rm ch.}  
\left( \frac{1}{\ep} + \frac{5}{3}  \right) \right] \Bigr. \nn \\
&+& \Bigl. {\cal O}(\ep)
\Bigr\} + \bigl( q_1 \leftrightarrow q_2 \bigr) \;\;.
\eeeq
Note that $\cw_{ij}^{[S] (f)}$ explicitly depends on the squared electric charge $e_f^2$
of the radiated soft fermion $f$.

We note that the result in Eq.~(\ref{j3gamma}) has a charge symmetric contribution
(which is proportional to the two-particle correlation function $\cw_{ij}^{[S] (f)}(q_1,q_2)$)
and an abelian charge-asymmetry contribution that is proportional to the 
momentum function $F_{ijk}^{[A]}(q_1,q_2)$. This structure is consistent with the
QCD result in Eq.~(\ref{j1squared}), since the charge symmetric three-particle correlations in
Eq.~(\ref{j1squared}) are purely non-abelian. At variance with the expression in Eq.~(\ref{j1squared}),
in the right-hand side of Eq.~(\ref{j3gamma}) we do not explicitly distinguish between two-particle
and three-particle charge-asymmetry correlations (i.e., the summed index $k$ can also be equal
to either $i$ or $j$). In the QCD case, we also noticed that three-particle correlations do not
contribute to the squared of the soft-$f{\bar f}$ current for emission from three hard partons
(see Eqs.~(\ref{ft3}) and (\ref{dt3})). 
A corresponding observation 
does not apply to the one-loop contribution in Eq.~(\ref{j3gamma}). For example,
we can consider soft-$f{\bar f}$ emission from the hard-scattering process
${\bar u}d \to W^- \to {\bar \nu}_e e^-$
(the charges of the outgoing hard particles are $\{ +2/3,+1/3,-1\}$)
and we see that the product $e_k e_i e_j$ of three distinct charges in Eq.~(\ref{j3gamma})
does not vanish.

The terms $| \hbj |^2_{(1 \ell;1\gamma)}$ and $| \hbj |^2_{(1 \ell;2\gamma)}$ in 
Eq.~(\ref{loopjff}) are due to mixed QCD$\times$QED interactions.

The contribution $| \hbj |^2_{(1 \ell;1\gamma)}$ can be regarded as a one-loop QED correction to the
QCD radiation of the soft fermion--antifermion pair. We obtain the following result:
\beeq
&& \!\!\!\!\!\!\!\!\!\!
| \hbj(q_1,q_2) |^2_{(1 \ell;1\gamma)} =
\Bigl\{
 \left[ \hbj^{(0)}(q_1,q_2) \right]^\dagger  
\hbj^{(1)}_{(1\gamma)}(q_1,q_2) 
+  \left[ \hbj^{(0)}_{(1\gamma)}(q_1,q_2) \right]^\dagger  
\hbj^{(1)}(q_1,q_2) \,
\Bigr\}
+ {\rm h.c.} \nn \\
&&\!\!\!\!\!\!\!\!\!\!
=  \delta_{f q} \,T_R \,\Bigl\{ \;\frac{1}{2}
\sum_{\substack{i,j \,\in H \\ i \,\neq\, j}}
 \bt_i \cdot \bt_j
\Bigl[ \cw_{ij}(q_1,q_2) \,
e_f^2
\left(  \frac{2}{\ep^2} + \frac{3}{\ep} - \pi^2 + 8 + (\delta_R -1) 
+ {\cal O}(\ep) \right)   
+ ( q_1 \leftrightarrow q_2 )
\Bigr] 
\Bigr. \nn \\
&&\!\!\!\!\!\!\!\!\!\!
- \; e_f \sum_{k \in H} \;\sum_{\substack{i,j \,\in H \\ i \,\neq\, j}} \,
\Bigl( e_i\, \bt_k \cdot \bt_j + e_j\, \bt_i \cdot \bt_k + e_k\, \bt_i \cdot \bt_j\Bigr) \;2 \,F_{ijk}^{[A]}(q_1,q_2) \,\Bigr\}
\;\;,
\label{squarej1gamma}
\eeeq
where the charge symmetric function $\cw_{ij}(q_1,q_2)$ 
and the charge-asymmetry function $F_{ijk}^{[A]}(q_1,q_2)$ are given in Eqs.~(\ref{wij})
and (\ref{fant}), respectively. We note that the one-loop term in Eq.~(\ref{squarej1gamma})
is not vanishing only if the soft fermion is a quark. Similarly to Eq.~(\ref{j3gamma}),
the summed index $k$ in Eq.~(\ref{squarej1gamma})  can also be equal
to either $i$ or $j$.

The term $| \hbj |^2_{(1 \ell;2\gamma)}$ can be regarded as a one-loop QCD correction to the
tree-level QED radiation (see Eq.~(\ref{qedsquared})) of the soft fermion--antifermion pair.
Its explicit expression is
\beeq
&& \!\!\!\!\!\!\!\!\!\!\!
| \hbj(q_1,q_2) |^2_{(1 \ell;2\gamma)} =
 \left[ \hbj^{(0)}_{(1\gamma)}(q_1,q_2) \right]^\dagger  
\hbj^{(1)}_{(1\gamma)}(q_1,q_2) 
+ {\rm h.c.} \nn \\
&&\!\!\!\!\!\!\!\!\!\!\!
=  \delta_{f q} \,N_c\, e_f^2 \,\frac{1}{2}
\sum_{\substack{i,j \,\in H \\ i \,\neq\, j}}
 e_i \, e_j
\Bigl\{ \cw_{ij}(q_1,q_2) \!
\left[ 
C_F \!
\left(  \frac{2}{\ep^2} + \frac{3}{\ep} - \pi^2 + 8 + (\delta_R -1) \right) \!   
+ {\cal O}(\ep) \right] \! + ( q_1 \leftrightarrow q_2 )
\Bigr\} ,\nn \\
&& \!\!
\label{squarej2gamma}
\eeeq
where the function $\cw_{ij}(q_1,q_2)$ is given in Eq.~(\ref{wij}).
We note that, analogously to Eq.~(\ref{squarej1gamma}), the term $| \hbj |^2_{(1 \ell;2\gamma)}$
is not vanishing only if the soft fermion is a quark.
Unlike the cases of the one-loop terms in Eqs.~(\ref{j3gamma}) and (\ref{squarej1gamma}),
charge-asymmetry contributions do not appear in $| \hbj |^2_{(1 \ell;2\gamma)}$.

By direct inspection of the tree-level and one-loop results in 
Eqs.~(\ref{qedsquared}) , (\ref{j3gamma}) , (\ref{squarej1gamma}) and (\ref{squarej2gamma}),
we see that the charge symmetric (charge-asymmetry) contributions are proportional
to even (odd) powers of $e_f$, as expected from charge conjugation symmetry.

\section{Summary}
\label{sec:sum}

We have considered the radiation of two or more soft partons in 
QCD hard scattering. In this soft limit the scattering amplitude is singular, and the singular behaviour is controlled in factorized form by a multiparton soft current, which has a process-independent structure.
At loop level, the scattering amplitudes and the soft current have UV and IR divergences, which we regularize in the form of $\ep$ poles by analytic continuation
in $d=4-2\ep$ space-time dimensions.

We have discussed the general structure of the $\ep$-pole divergences of the multiparton soft current. We have considered the soft current at one-loop order and we have presented the explicit form of its $\ep$-pole (divergent) contributions.
We have also discussed the RS dependence of the one-loop soft 
current.

In the remaining part of the paper we have considered the specific case of
soft $q{\bar q}$ radiation, by presenting a detailed study at one-loop order.
Considering arbitrary kinematical regions of the soft-parton and hard-parton momenta, we have explicitly computed the one-loop current by including the finite terms at ${\cal O}(\ep^0)$. We find a relatively simple expression, which, for instance, includes powers of logarithmic functions but no dilog functions.

We find that the one-loop current produces a new type of singularity if the 
soft-$q{\bar q}$ pair is radiated with a vanishing transverse momentum with respect to the direction of two colliding hard partons in the initial state.
This new transverse-momentum singularity has a quantum (more precisely, absorptive) origin and a purely non-abelian character.
Owing to its dynamical origin, the transverse-momentum singularity can appear also in the one-loop current for double soft-gluon emission.

We have computed the one-loop contribution of the squared current for 
soft-$q{\bar q}$ emission and the ensuing colour correlations for squared amplitudes
of generic multiparton hard-scattering processes. We have also explicitly
considered the specific cases of processes with two or three hard partons, in which the colour correlation structure can be partly simplified.  

We find that, despite its absorptive origin, the new one-loop transverse-momentum singularity contributes to squared amplitudes (and, hence, cross sections) 
of scattering processes with two initial-state colliding partons (hadrons)
and two or more hard partons (jets) in the final state.

At variance with the case of multiple soft-gluon radiation, the emission of soft fermions and antifermions lead to charge asymmetry effects. We have discussed in details the charge asymmetry contributions of the one-loop squared current
for soft $q{\bar q}$ radiation.

We have finally generalized our QCD study of soft $q{\bar q}$ emission 
to the study of QED and mixed QCD$\times$QED radiative corrections in the context of soft fermion--antifermion radiation. We have presented the corresponding one-loop
results for the soft current and its square.

\vspace*{3mm}
{\bf Acknowledgements.} 
This project has received funding from the European Union's Horizon 2020 research and innovation programme
under the Marie Sk\l{}odowska-Curie grant agreement number 754496 and by the Generalitat Valenciana (Spain) through the plan GenT program (CIDEGENT/2020/011).


\begin{thebibliography}{99}


\bibitem{Heinrich:2020ybq}
G.~Heinrich,
Phys. Rept. \textbf{922} (2021) 1-69
[arXiv:2009.00516 [hep-ph]].

\bibitem{N3LLwithoutC3}
W.~Bizon, P.~F.~Monni, E.~Re, L.~Rottoli and P.~Torrielli,
JHEP \textbf{02} (2018), 108
[arXiv:1705.09127 [hep-ph]];
W.~Bizo\'n, X.~Chen, A.~Gehrmann-De Ridder, T.~Gehrmann, N.~Glover, A.~Huss, P.~F.~Monni, E.~Re, L.~Rottoli and P.~Torrielli,
JHEP \textbf{12} (2018), 132
[arXiv:1805.05916 [hep-ph]];
W.~Bizon, A.~Gehrmann-De Ridder, T.~Gehrmann, N.~Glover, A.~Huss, P.~F.~Monni, E.~Re, L.~Rottoli and D.~M.~Walker,
Eur. Phys. J. C \textbf{79} (2019) no.10, 868
[arXiv:1905.05171 [hep-ph]].

\bibitem{Chen:2018pzu}
X.~Chen, T.~Gehrmann, E.~W.~N.~Glover, A.~Huss, Y.~Li, D.~Neill, M.~Schulze, I.~W.~Stewart and H.~X.~Zhu,
Phys. Lett. B \textbf{788} (2019), 425-430
[arXiv:1805.00736 [hep-ph]].

\bibitem{Bertone:2019nxa}
V.~Bertone, I.~Scimemi and A.~Vladimirov,
JHEP \textbf{06} (2019), 028
[arXiv:1902.08474 [hep-ph]].

\bibitem{Bacchetta:2019sam}
A.~Bacchetta, V.~Bertone, C.~Bissolotti, G.~Bozzi, F.~Delcarro, F.~Piacenza and M.~Radici,
JHEP \textbf{07} (2020), 117
[arXiv:1912.07550 [hep-ph]].

\bibitem{Ebert:2020dfc}
M.~A.~Ebert, J.~K.~L.~Michel, I.~W.~Stewart and F.~J.~Tackmann,
JHEP \textbf{04} (2021), 102
[arXiv:2006.11382 [hep-ph]].

\bibitem{Becher:2020ugp}
T.~Becher and T.~Neumann,
JHEP \textbf{03} (2021), 199
[arXiv:2009.11437 [hep-ph]].

\bibitem{Luo:2019szz}
M.~x.~Luo, T.~Z.~Yang, H.~X.~Zhu and Y.~J.~Zhu,
Phys. Rev. Lett. \textbf{124} (2020)  092001
[arXiv:1912.05778 [hep-ph]],
JHEP \textbf{06} (2021) 115
[arXiv:2012.03256 [hep-ph]].

\bibitem{Ebert:2020unb}
M.~A.~Ebert, B.~Mistlberger and G.~Vita,
JHEP \textbf{09} (2020) 143
[arXiv:2006.03056 [hep-ph]],
JHEP \textbf{09} (2020) 146
[arXiv:2006.05329 [hep-ph]],
JHEP \textbf{07} (2021) 121
[arXiv:2012.07853 [hep-ph]].

\bibitem{Billis:2021ecs}
G.~Billis, B.~Dehnadi, M.~A.~Ebert, J.~K.~L.~Michel and F.~J.~Tackmann,
Phys. Rev. Lett. \textbf{127} (2021) no.7, 072001
[arXiv:2102.08039 [hep-ph]].

\bibitem{Camarda:2021ict}
S.~Camarda, L.~Cieri and G.~Ferrera,
[arXiv:2103.04974 [hep-ph]].

\bibitem{Re:2021con}
E.~Re, L.~Rottoli and P.~Torrielli,
[arXiv:2104.07509 [hep-ph]].

\bibitem{Ju:2021lah}
W.~L.~Ju and M.~Sch\"onherr,
JHEP \textbf{10} (2021), 088
[arXiv:2106.11260 [hep-ph]].

\bibitem{Neumann:2021zkb}
T.~Neumann,
Eur. Phys. J. C \textbf{81} (2021) no.10, 905
[arXiv:2107.12478 [hep-ph]].

\bibitem{Frixione:1995ms}
  S.~Frixione, Z.~Kunszt and A.~Signer,
  Nucl.\ Phys.\ B {\bf 467} (1996) 399
  [hep-ph/9512328];
  S.~Frixione,
  Nucl.\ Phys.\ B {\bf 507} (1997) 295
  [hep-ph/9706545].

\bibitem{csdip}
  S.~Catani and M.~H.~Seymour,
  Nucl.\ Phys.\ B {\bf 485} (1997) 291
   [Erratum:  Nucl.\ Phys.\ B {\bf 510} (1998) 503]
  [arXiv:hep-ph/9605323].


\bibitem{Campbell:1997hg}
J.~M.~Campbell and E.~W.~N.~Glover,
Nucl. Phys. B \textbf{527} (1998) 264-288
[arXiv:hep-ph/9710255 [hep-ph]].

\bibitem{Catani:1998nv}
S.~Catani and M.~Grazzini,
Phys. Lett. B \textbf{446} (1999) 143-152
[arXiv:hep-ph/9810389 [hep-ph]].

\bibitem{Bern:1998sc}
Z.~Bern, V.~Del Duca and C.~R.~Schmidt,
Phys. Lett. B \textbf{445} (1998) 168-177
[arXiv:hep-ph/9810409 [hep-ph]].

\bibitem{Kosower:1999rx}
D.~A.~Kosower and P.~Uwer,
Nucl. Phys. B \textbf{563} (1999) 477-505
[arXiv:hep-ph/9903515 [hep-ph]].

\bibitem{Bern:1999ry}
Z.~Bern, V.~Del Duca, W.~B.~Kilgore and C.~R.~Schmidt,
Phys. Rev. D \textbf{60} (1999) 116001
[arXiv:hep-ph/9903516 [hep-ph]].

\bibitem{Catani:1999ss}
  S.~Catani and M.~Grazzini,
  Nucl.\ Phys.\ B {\bf 570} (2000) 287
  [arXiv:hep-ph/9908523].

\bibitem{Catani:2000pi}
  S.~Catani and M.~Grazzini,
  Nucl.\ Phys.\ B {\bf 591} (2000) 435
  [arXiv:hep-ph/0007142].

\bibitem{Czakon:2011ve}
M.~Czakon,
Nucl. Phys. B \textbf{849} (2011) 250-295
[arXiv:1101.0642 [hep-ph]].

\bibitem{Bierenbaum:2011gg}
  I.~Bierenbaum, M.~Czakon and A.~Mitov,
  Nucl.\ Phys.\ B {\bf 856} (2012) 228
  [arXiv:1107.4384 [hep-ph]];
M.~L.~Czakon and A.~Mitov,
[arXiv:1804.02069 [hep-ph]].

\bibitem{Catani:2011st}
S.~Catani, D.~de Florian and G.~Rodrigo,
JHEP \textbf{07} (2012) 026
[arXiv:1112.4405 [hep-ph]].

\bibitem{Sborlini:2013jba}
G.~F.~R.~Sborlini, D.~de Florian and G.~Rodrigo,
JHEP \textbf{01} (2014) 018
[arXiv:1310.6841 [hep-ph]].

\bibitem{Becher:2014oda}
  T.~Becher, A.~Broggio and A.~Ferroglia,
  Lect.\ Notes Phys.\  {\bf 896} (2015) 1
  [arXiv:1410.1892 [hep-ph]].


\bibitem{Luisoni:2015xha}
  G.~Luisoni and S.~Marzani,
  J.\ Phys.\ G {\bf 42} (2015) no.10,  103101
  [arXiv:1505.04084 [hep-ph]].



\bibitem{DelDuca:1999iql}
V.~Del Duca, A.~Frizzo and F.~Maltoni,
Nucl. Phys. B \textbf{568} (2000) 211-262
[arXiv:hep-ph/9909464 [hep-ph]].

\bibitem{Birthwright:2005ak}
T.~G.~Birthwright, E.~W.~N.~Glover, V.~V.~Khoze and P.~Marquard,
JHEP \textbf{05} (2005) 013
[arXiv:hep-ph/0503063 [hep-ph]],
JHEP \textbf{07} (2005) 068
[arXiv:hep-ph/0505219 [hep-ph]].

\bibitem{DelDuca:2019ggv}
V.~Del Duca, C.~Duhr, R.~Haindl, A.~Lazopoulos and M.~Michel,
JHEP \textbf{02} (2020) 189
[arXiv:1912.06425 [hep-ph]],
JHEP \textbf{10} (2020) 093
[arXiv:2007.05345 [hep-ph]].



\bibitem{Catani:2003vu}
S.~Catani, D.~de Florian and G.~Rodrigo,
Phys. Lett. B \textbf{586} (2004) 323-331
[arXiv:hep-ph/0312067 [hep-ph]].

\bibitem{Sborlini:2014mpa}
G.~F.~R.~Sborlini, D.~de Florian and G.~Rodrigo,
JHEP \textbf{10} (2014) 161
[arXiv:1408.4821 [hep-ph]],
JHEP \textbf{03} (2015) 021
[arXiv:1409.6137 [hep-ph]].

\bibitem{Badger:2015cxa}
S.~Badger, F.~Buciuni and T.~Peraro,
JHEP \textbf{09} (2015) 188
[arXiv:1507.05070 [hep-ph]].



\bibitem{Bern:2004cz}
Z.~Bern, L.~J.~Dixon and D.~A.~Kosower,
JHEP \textbf{08} (2004)  012
[arXiv:hep-ph/0404293 [hep-ph]].

\bibitem{Badger:2004uk}
  S.~D.~Badger and E.~W.~N.~Glover,
  JHEP {\bf 07} (2004) 040
  [arXiv:hep-ph/0405236 [hep-ph]].

\bibitem{Duhr:2014nda}
C.~Duhr, T.~Gehrmann and M.~Jaquier,
JHEP \textbf{02} (2015) 077
[arXiv:1411.3587 [hep-ph]].



\bibitem{Catani:2019nqv}
  S.~Catani, D.~Colferai and A.~Torrini,
  JHEP {\bf 01} (2020) 118
  [arXiv:1908.01616 [hep-ph]].


\bibitem{Zhu:2020ftr}
Y.~J.~Zhu,
[arXiv:2009.08919 [hep-ph]].


\bibitem{Li:2013lsa}
  Y.~Li and H.~X.~Zhu,
  JHEP {\bf 11} (2013) 080
  [arXiv:1309.4391 [hep-ph]].

\bibitem{Duhr:2013msa}
  C.~Duhr and T.~Gehrmann,
  Phys.\ Lett.\ B {\bf 727} (2013) 452
  [arXiv:1309.4393 [hep-ph]].

\bibitem{Dixon:2019lnw}
 L.~J.~Dixon, E.~Herrmann, K.~Yan and H.~X.~Zhu,
 JHEP \textbf{05} (2020) 135
 [arXiv:1912.09370 [hep-ph]].


\bibitem{Catani:1998bh}
  S.~Catani,
  Phys.\ Lett.\ B {\bf 427} (1998) 161
  [arXiv:hep-ph/9802439].

\bibitem{Bern:1995ix}
  Z.~Bern and G.~Chalmers,
  Nucl.\ Phys.\  B {\bf 447} (1995) 465
  [arXiv:hep-ph/9503236].

\bibitem{Feige:2014wja}
  I.~Feige and M.~D.~Schwartz,
  Phys.\ Rev.\ D {\bf 90} (2014) 10,  105020
  [arXiv:1403.6472 [hep-ph]].

\bibitem{Bassetto:1984ik}
  A.~Bassetto, M.~Ciafaloni and G.~Marchesini,
  Phys.\ Rept.\  {\bf 100} (1983) 201. 

\bibitem{Berends:1988zn}
  F.~A.~Berends and W.~T.~Giele,
  Nucl.\ Phys.\  B {\bf 313} (1989) 595.


\bibitem{scetref}  
  C.~W.~Bauer, S.~Fleming and M.~E.~Luke,
  Phys.\ Rev.\ D {\bf 63} (2000) 014006
  [hep-ph/0005275];
  C.~W.~Bauer, S.~Fleming, D.~Pirjol and I.~W.~Stewart,
  Phys.\ Rev.\ D {\bf 63} (2001) 114020
  [hep-ph/0011336];
  C.~W.~Bauer and I.~W.~Stewart,
  Phys.\ Lett.\ B {\bf 516} (2001) 134
  [hep-ph/0107001];
  C.~W.~Bauer, D.~Pirjol and I.~W.~Stewart,
  Phys.\ Rev.\ D {\bf 65} (2002) 054022
  [hep-ph/0109045];
  C.~W.~Bauer, S.~Fleming, D.~Pirjol, I.~Z.~Rothstein and I.~W.~Stewart,
  Phys.\ Rev.\ D {\bf 66} (2002) 014017
  [hep-ph/0202088];
  M.~Beneke, A.~P.~Chapovsky, M.~Diehl and T.~Feldmann,
  Nucl.\ Phys.\ B {\bf 643} (2002) 431
  [hep-ph/0206152].

\bibitem{Mangano:1990by}
  M.~L.~Mangano and S.~J.~Parke,
  Phys.\ Rept.\  {\bf 200} (1991) 301
  [arXiv:hep-th/0509223].



\bibitem{Yennie:1961ad}
  D.~R.~Yennie, S.~C.~Frautschi and H.~Suura,
  Annals Phys.\  {\bf 13} (1961) 379;
  G.~Grammer, Jr. and D.~R.~Yennie,
  Phys.\ Rev.\ D {\bf 8} (1973) 4332.


\bibitem{Sterman:2002qn}
  G.~Sterman and M.~E.~Tejeda-Yeomans,
  Phys.\ Lett.\  B {\bf 552} (2003) 48
  [arXiv:hep-ph/0210130].

\bibitem{Aybat:2006mz}
  S.~M.~Aybat, L.~J.~Dixon and G.~Sterman,
  Phys.\ Rev.\  D {\bf 74} (2006) 074004
  [arXiv:hep-ph/0607309].

\bibitem{Gardi:2009qi}
  E.~Gardi and L.~Magnea,
  JHEP {\bf 03} (2009) 079
  [arXiv:0901.1091 [hep-ph]].

\bibitem{Becher:2009qa}
  T.~Becher and M.~Neubert,
  JHEP {\bf 06} (2009) 081
  [arXiv:0903.1126 [hep-ph]].

\bibitem{Almelid:2015jia}
 \O{}.~Almelid, C.~Duhr and E.~Gardi,
 Phys. Rev. Lett. \textbf{117} (2016)  172002
 [arXiv:1507.00047 [hep-ph]].

\bibitem{Giele:1991vf}
  W.~T.~Giele and E.~W.~N.~Glover,
  Phys.\ Rev.\  D {\bf 46} (1992) 1980;
  Z.~Kunszt, A.~Signer and Z.~Trocsanyi,
  Nucl.\ Phys.\  B {\bf 420} (1994) 550
  [arXiv:hep-ph/9401294].



\bibitem{'tHooft:1972fi}
  G.~'t Hooft and M.~J.~G.~Veltman,
  Nucl.\ Phys.\ B {\bf 44} (1972) 189.

\bibitem{cdr}
  C.~G.~Bollini and J.~J.~Giambiagi,
  Nuovo Cim.\ B {\bf 12} (1972) 20;
  J.~F.~Ashmore,
  Lett.\ Nuovo Cim.\  {\bf 4} (1972) 289;
  G.~M.~Cicuta and E.~Montaldi,
  Lett.\ Nuovo Cim.\  {\bf 4} (1972) 329.

\bibitem{Gastmans:1973uv}
  R.~Gastmans and R.~Meuldermans,
  Nucl.\ Phys.\ B {\bf 63} (1973) 277.

\bibitem{Siegel:1979wq}
  W.~Siegel,
  Phys.\ Lett.\ B {\bf 84} (1979) 193.

\bibitem{Bern:1991aq}
  Z.~Bern and D.~A.~Kosower,
  Nucl.\ Phys.\ B {\bf 379} (1992) 451.


\bibitem{Kunszt:1993sd}
  Z.~Kunszt, A.~Signer and Z.~Trocsanyi,
  Nucl.\ Phys.\ B {\bf 411} (1994) 397
  [hep-ph/9305239];
  S.~Catani, M.~H.~Seymour and Z.~Trocsanyi,
  Phys.\ Rev.\ D {\bf 55} (1997) 6819
  [hep-ph/9610553].

\bibitem{Catani:2000ef}
  S.~Catani, S.~Dittmaier and Z.~Trocsanyi,
  Phys.\ Lett.\ B {\bf 500} (2001) 149-160
  [hep-ph/0011222].

\bibitem{Altarelli:1980fi}
  G.~Altarelli, G.~Curci, G.~Martinelli and S.~Petrarca,
  Nucl.\ Phys.\ B {\bf 187} (1981) 461.


\bibitem{Passarino:1978jh}
  G.~Passarino and M.~J.~G.~Veltman,
  Nucl.\ Phys.\ B {\bf 160} (1979) 151.


\bibitem{Bern:1993kr}
  Z.~Bern, L.~J.~Dixon and D.~A.~Kosower,
  Nucl.\ Phys.\ B {\bf 412} (1994) 751
  [hep-ph/9306240];
  T.~Binoth, J.~P.~Guillet and G.~Heinrich,
  Nucl.\ Phys.\ B {\bf 572} (2000) 361
  [hep-ph/9911342].


\bibitem{Anastasiou:2015yha}
  C.~Anastasiou, C.~Duhr, F.~Dulat, E.~Furlan, F.~Herzog and B.~Mistlberger,
  JHEP {\bf 08} (2015) 051
  [arXiv:1505.04110 [hep-ph]].



\bibitem{Forshaw:2008cq}
  J.~R.~Forshaw, A.~Kyrieleis and M.~H.~Seymour,
  JHEP {\bf 09} (2008) 128
  [arXiv:0808.1269 [hep-ph]].

\bibitem{Forshaw:2012bi}
  J.~R.~Forshaw, M.~H.~Seymour and A.~Siodmok,
  JHEP {\bf 11} (2012) 066
  [arXiv:1206.6363 [hep-ph]].

\bibitem{Seymour:2008xr}
  M.~H.~Seymour and M.~Sjodahl,
  JHEP {\bf 12} (2008) 066
  [arXiv:0810.5756 [hep-ph]].

\bibitem{Catani:2004nc}
  S.~Catani, D.~de Florian, G.~Rodrigo and W.~Vogelsang,
  Phys.\ Rev.\ Lett.\  {\bf 93} (2004) 152003
[arXiv:hep-ph/0404240 [hep-ph]].
  
  
\bibitem{Moch:2004pa}
S.~Moch, J.~A.~M.~Vermaseren and A.~Vogt,
Nucl. Phys. B \textbf{688} (2004), 101-134
[arXiv:hep-ph/0403192 [hep-ph]].

\bibitem{Mitov:2006ic}
A.~Mitov, S.~Moch and A.~Vogt,
Phys. Lett. B \textbf{638} (2006), 61-67
[arXiv:hep-ph/0604053 [hep-ph]].

\bibitem{Ellis:1987xu}
  R.~K.~Ellis, I.~Hinchliffe, M.~Soldate and J.~J.~van der Bij,
  Nucl.\ Phys.\ B {\bf 297} (1988) 221.

\bibitem{vanderBij:1988ac}
  J.~J.~van der Bij and E.~W.~N.~Glover,
  Nucl.\ Phys.\ B {\bf 313} (1989) 237.


\end{thebibliography}
\end{document}